\documentclass[a4paper]{aa}

\modulolinenumbers[1]
\usepackage{graphicx}
\usepackage{capt-of}
\usepackage{amsmath}
\usepackage{amssymb}
\usepackage{natbib}
\usepackage[colorlinks=true, linkcolor=blue, citecolor=blue, urlcolor=blue]{hyperref}

\usepackage{placeins}

\bibpunct{(}{)}{;}{a}{}{,}

\begin{document}

\title{Hierarchical Bayesian full-Stokes imaging and polarimetric calibration with \texttt{Comrade}: A 25-source 15 GHz VLBA benchmark}

\titlerunning{Bayesian polarimetric imaging of a VLBA survey with Comrade}

\author{Jongho Park \inst{1,2,3} \and Paul Tiede \inst{4,5}}

\authorrunning{Park \& Tiede}

\institute{
    School of Space Research, Kyung Hee University, 1732, Deogyeong-daero, Giheung-gu, Yongin-si, Gyeonggi-do 17104, Republic of Korea
    \and
    G-LAMP NEXUS Institute, Kyung Hee University, Yongin 17104, Republic of Korea
    \and
    Institute of Astronomy and Astrophysics, Academia Sinica, P.O.\ Box 23-141, Taipei 10617, Taiwan, R.O.C.
    \and
    Black Hole Initiative at Harvard University, 20 Garden Street, Cambridge, MA 02138, USA
    \and
    Center for Astrophysics $|$ Harvard \& Smithsonian, 60 Garden Street, Cambridge, MA 02138, USA
}

\date{}

\abstract
{Bayesian forward-modeling methods have shown potential for super-resolution in horizon-scale very long baseline interferometry (VLBI), but remain largely untested for survey-scale centimeter-wavelength imaging of extended, complex jets in active galactic nuclei (AGNs). Synthetic data cannot fully reproduce calibration and non-closing errors in real observations.}
{We test whether Bayesian full-Stokes imaging reproduces conventional VLBI image and calibration products across diverse sources and establish checks for structures recovered below the CLEAN beam.}
{We applied a hierarchical \texttt{Comrade} framework to 25 AGNs observed with the Very Long Baseline Array at 15\,GHz, jointly inferring full-Stokes brightness distributions, residual antenna gains, and instrumental polarization leakages (D-terms). We compared CLEAN-beam-convolved posterior images, gain amplitudes, and time-dependent D-terms with results from an independent CLEAN$+$\texttt{GPCAL} reduction. We examined leakage consistency among sources close in sky position and observing time and compared the PolExp and Poincar\'{e}-sphere polarization representations.}
{For the fiducial PolExp representation, all 25 CLEAN-beam-convolved Stokes~$I$ reconstructions have normalized cross-correlations above 0.999 with CLEAN. The complex-polarization agreement decreases toward weakly polarized sources, indicating greater sensitivity to residual calibration errors. The inferred gain amplitudes and D-terms are consistent with CLEAN$+$\texttt{GPCAL}, and nearby sources show similar gain and leakage evolution. The R/L gain ratios are more consistent across nearby sources than the individual gain amplitudes, but remain insufficiently constrained for sub-percent Stokes~$V$ measurements. The two polarization representations are nearly indistinguishable at nominal resolution but differ at sub-beam scales. Native-resolution reconstructions reveal fine jet structure, including an edge-brightened M87 jet and counterjet consistent with higher-frequency VLBI observations, and resolved polarization features smoothed out in CLEAN images.}
{Hierarchical Bayesian full-Stokes imaging is practical and reliable for survey-scale centimeter-wavelength VLBI. Agreement at nominal resolution and independent recovery of instrumental behavior link conventional CLEAN imaging to Bayesian native-resolution reconstructions and provide real-data checks for assessing sub-beam structure.}


\maketitle
\nolinenumbers

\section{Introduction}\label{sec:intro}

A fraction of active galactic nuclei (AGNs) launch collimated relativistic jets that transport energy and momentum from sub-parsec to kiloparsec and even megaparsec scales \citep[for a review, see][]{BlandfordMeierReadhead2019}. It is widely believed that these jets are launched by large-scale magnetic fields threading the black hole ergosphere \citep{BlandfordZnajek1977} or the inner accretion disk \citep{BlandfordPayne1982}. Analytical studies and magnetohydrodynamic simulations have shown that the jets can be gradually accelerated through the conversion of Poynting flux into bulk kinetic energy while being collimated by the pressure of the surrounding medium \citep[e.g.,][]{VlahakisKonigl2004,McKinney2006,Komissarov2007,Tchekhovskoy2008,Lyubarsky2009,Nakamura2018,Nokhrina2020}. Since these processes occur on very small angular scales, very long baseline interferometry (VLBI) provides one of the most direct means of investigating the jet launching, acceleration, and collimation regions.

Recent VLBI observations have indeed resolved these regions over a wide range of physical scales. At the highest angular resolutions, the Event Horizon Telescope (EHT) resolved ring-like emission surrounding central brightness depressions in both M87$^{*}$ and Sgr~A$^{*}$, providing direct constraints on the horizon-scale emission structures around the two supermassive black holes \citep{EHTC2019d,EHT2022a}. Subsequent EHT observations of M87$^{*}$ showed that the characteristic ring diameter persisted in a later observing epoch despite changes in the brightness distribution \citep{EHTC2024_M87shadow}. At longer wavelengths, VLBI observations trace the jet downstream from the central engine. In M87, they have revealed a parabolic-to-conical transition in the jet geometry \citep{AsadaNakamura2012,NakamuraAsada2013,Hada2013,Nakamura2018}, an edge-brightened transverse structure \citep[e.g.,][]{Hada2016,Walker2018,Lu2023,Nikonov2023,Kim2025_M87_86GHz}, and gradual acceleration over a broad range of distances \citep[e.g.,][]{Mertens2016,Park2019a}. Similar acceleration and collimation regions have been found in NGC\,315 \citep{Park2021b,Park2024a,Boccardi2021,Ricci2022}, NGC\,4261 \citep{Nakahara2018,Yan2023,Yan2025}, and other radio-loud AGNs \citep[e.g.,][]{Tseng2016,Hada2018,Kovalev2020,Okino2022,Yi2024}. These results demonstrate that high-resolution imaging is essential for connecting horizon-scale physics to the downstream evolution of relativistic jets.

The restarted jet in 3C\,84 provides a particularly useful example of the information obtained from high-resolution, multi-epoch imaging. Space- and ground-based VLBI observations have resolved a wide, limb-brightened jet and a low-intensity mini-cocoon surrounding it \citep{Nagai2014,Giovannini2018,Savolainen2023}. Subsequent studies have revealed changes in the jet collimation profile, filamentary and spectral-index structures, propagation, and termination region \citep{Kam2024,Foschi2025,Benke2026,Debbrecht2026}. The deflection and spectral properties of individual jet features further provide evidence for interaction with dense ambient gas \citep{Park2024b}. These results show that high-fidelity Stokes~$I$ imaging over multiple angular scales and epochs can distinguish intrinsic jet evolution from the effects of the surrounding medium.

Polarimetric VLBI provides additional information that cannot be obtained from total-intensity imaging alone. Linear polarization constrains the ordering and projected geometry of magnetic fields, while its frequency dependence probes Faraday rotation and the magnetized plasma along the line of sight. On horizon scales, the EHT has resolved polarized ring structures in M87$^{*}$ and Sgr~A$^{*}$, providing direct constraints on the magnetic fields and plasma near the black holes, while multi-epoch observations of M87$^{*}$ have revealed substantial evolution in the linear-polarization structure \citep{EHTC2021a,EHTC2021b,EHTC2024_SgrApol,EHTC2024_SgrApol_VIII,EHTC2025_M87pol}. Extending these studies to the acceleration--collimation zone, high-sensitivity multifrequency VLBI polarimetry of M87 has revealed transverse polarization asymmetries consistent with a large-scale helical magnetic field retaining a substantial poloidal component, thereby directly testing the magnetic-field evolution predicted by magnetohydrodynamic jet models \citep{Park2026}. On sub-parsec to parsec scales, Faraday rotation measurements have constrained the magnetized ambient media around M87 and 3C\,84 \citep{Park2019b,Kam2026}, and EHT observations of 3C\,84 have revealed ordered magnetic fields close to the central engine \citep{Paraschos2024}. 

Circular polarization, quantified by the Stokes~$V$ parameter, provides complementary diagnostics of the jet plasma composition and magnetic-field topology, as it can arise from intrinsic synchrotron emission in ordered fields or from Faraday conversion of linear into circular polarization \citep{Wardle1998, WardleHoman2003}. VLBI surveys have detected parsec-scale circular polarization at the $\lesssim 1$\% level in numerous AGN \citep{HomanWardle1999, Homan2001, HomanLister2006}, and the EHT has recently reported a Stokes~$V$ signal near the event horizon of M87$^{*}$ \citep{EHTC2023_M87_StokesV}. However, Stokes~$V$ imaging remains particularly challenging: the circular polarization fraction is typically an order of magnitude below the linear polarization fraction, and its recovery is sensitive to residual calibration errors in the R/L gain ratio that couple directly to the inferred $V$ signal. Extracting all four Stokes parameters robustly therefore requires careful treatment of station-based corruptions---complex gains, R/L phase offsets, and polarization leakage terms---which couple directly to the inferred Stokes maps.

The standard VLBI workflow addresses these effects sequentially. One first uses the parallel-hand (RR, LL) visibilities to derive a Stokes~$I$ image and antenna gain solutions through iterative CLEAN imaging and self-calibration \citep{Hogbom1974, Shepherd1997}. The cross-hand (RL, LR) visibilities are then used to separate the source linear polarization from the instrumental polarization (D-terms), assuming that the gain corruptions have already been removed. LPCAL \citep{Leppanen1995} employs a similarity approximation in which the source polarization structure is assumed to be proportional to the total-intensity CLEAN model, which limits its accuracy for sources whose polarization morphology differs from Stokes~$I$. More recently, \texttt{GPCAL} \citep{Park2021} and \texttt{PolSolve} \citep{MartiVidal2021} overcome this limitation through iterative CLEAN imaging and D-term modeling, allowing calibration of sources with complex polarization structures. They can additionally combine multiple calibrators, and \texttt{GPCAL} has been extended to calibrate frequency-dependent and time-dependent leakage variations \citep{Park2023b,Park2023a}. While this sequential workflow has been extraordinarily successful in practice, it has clear limitations for modern polarimetric inference: the calibration and imaging stages are separated, uncertainty propagation into the final Stokes maps is incomplete, and the implicit image prior induced by CLEAN is not always well matched to extended or multi-scale polarized structure.

Bayesian forward-modeling approaches offer a complementary route. In such methods, the sky brightness distribution and instrumental terms are inferred jointly by modeling how they together produce the observed visibilities, so calibration uncertainty is propagated naturally into image uncertainty. A dedicated fully Bayesian implementation is the \texttt{DMC} code \citep{Pesce2021}, which performs simultaneous full-Stokes imaging and calibration---including polarimetric leakage. The \texttt{THEMIS} framework provides modular Bayesian parameter estimation for EHT analyses \citep{Broderick2020} and was subsequently extended to polarimetric imaging. \texttt{DMC} and polarimetric \texttt{THEMIS} were used as independent posterior-exploration methods in the EHT analysis of the 2017 M87$^{*}$ observations \citep{EHTC2021a}, while subsequent EHT polarimetric studies have also used \texttt{THEMIS} \citep{EHTC2024_SgrApol,EHTC2025_M87pol}. Building on this heritage, the \texttt{Comrade} software package \citep{Tiede2022} provides a composable, differentiable forward-modeling framework for VLBI data. \citet{Tiede2026} extended \texttt{Comrade} with hierarchical inference of image hyperparameters and station-based complex gains, jointly inferring the Stokes~$I$ sky brightness and gain solutions. More recently, \citet{Chavez2026} introduced Colorful Hierarchical Interferometric Bayesian Imaging (CHIBI), a hierarchical multifrequency-synthesis method implemented in \texttt{Comrade.jl} that jointly reconstructs total-intensity and spectral-index images and has been demonstrated on Very Long Baseline Array (VLBA) data. \texttt{Comrade} has been adopted for EHT analyses of M87$^{*}$, including total-intensity imaging of the persistent ring structure \citep{EHTC2024_M87shadow} and multi-epoch polarimetry \citep{EHTC2025_M87pol}. Bayesian self-calibration and imaging have also been demonstrated with \texttt{resolve}, first for Stokes~$I$ joint gain calibration and imaging \citep{Kim2024}, then for full-Stokes imaging \citep{Arras2025}, and subsequently for full-Stokes polarimetric calibration and imaging including gains and D-terms \citep{Kim2025}.

However, to date these Bayesian methods have been applied almost exclusively to EHT horizon-scale data \citep[e.g.,][]{EHTC2021a, Arras2022, EHTC2024_M87shadow, EHTC2025_M87pol} or to a small number of selected targets and data sets \citep[e.g.,][]{Kim2024, Arras2025, Kim2025, Mueller2026}. A systematic validation on cm-wavelength VLBA data---which typically feature large fields of view, complex extended jet morphologies, and a wider range of signal-to-noise ratios \citep[e.g.,][]{Jorstad2017, Pushkarev2017, Lister2018}---has not yet been carried out. Moreover, while synthetic-data tests have shown that Bayesian frameworks can recover known inputs \citep{Tiede2026, Kim2024, Kim2025}, real VLBI data may additionally contain unmodeled instrumental systematics with unknown statistical behavior, including baseline-based non-closing errors, imperfect data flagging, and residual calibration biases. Validation on real observations is therefore essential. A multi-source VLBA experiment, in which targets at similar right ascensions and declinations are observed in rapid switching cycles, provides a natural testbed for such a validation: the instrumental terms inferred independently for nearby sources can be cross-compared, and the Bayesian calibration products can be benchmarked directly against those from an established conventional pipeline.

A related open question concerns the super-resolution capability of forward-modeling methods. Both synthetic and real-data studies have demonstrated that regularized maximum-likelihood and Bayesian imaging can recover structures finer than the nominal diffraction limit in favorable cases \citep[e.g.,][]{Honma2014, Chael2016, Akiyama2017a, Tiede2026}, but validation of this super-resolution capability on real data has been limited to individual targets and data sets. At the same time, CLEAN-based imaging and self-calibration have been applied successfully to a wide variety of AGN jet morphologies across large VLBI surveys \citep{Jorstad2017, Lister2018}. If forward-modeling methods genuinely achieve higher effective resolution, then their images, once convolved with the CLEAN restoring beam, should closely reproduce the corresponding CLEAN maps, while the antenna-based calibration products---gains and D-terms---should agree regardless of the imaging resolution. A systematic comparison over a large, morphologically diverse data set is therefore needed to establish whether forward-modeling methods deliver calibration products and scientific images that are consistent with those from the conventional pipeline.

In this work, we present a systematic benchmark of Bayesian full-Stokes imaging and calibration using experiment BP266B, a 25-source VLBA data set observed at 15\,GHz. We reconstruct the sources with \texttt{Comrade} and compare the inferred gains, D-terms, and full-Stokes images with those obtained from an independent CLEAN$+$\texttt{GPCAL} reduction, thereby testing their consistency at the CLEAN resolution. We further compare the instrumental terms independently inferred with \texttt{Comrade} for neighboring sources observed at similar times. Because station-based gains are expected to vary smoothly over short time intervals and small angular separations, while D-terms should remain stable throughout the experiment, their consistency provides an independent validation of the calibration and, indirectly, of the super-resolved reconstructions. Section~\ref{sec:obs} describes the observations and comparison data products, Section~\ref{sec:imaging} presents the imaging and calibration model, Section~\ref{sec:results} presents the results and discusses their implications, and Section~\ref{sec:conclusions} summarizes the main conclusions.

\section{Observations and data reduction}\label{sec:obs}

We analyzed 25 AGN observed with all ten VLBA stations at 15\,GHz on 2024 October 4--5 in experiment BP266B. The observation covered approximately 24\,hr. The targets span a broad range of declinations and morphologies, including compact blazars, core--jet sources, and the nearby radio galaxies 3C\,84 and M87.

The raw data were correlated with the DiFX software correlator \citep{Deller2011} in Socorro, New Mexico, and pre-calibrated in AIPS using standard VLBI procedures \citep{Greisen2003}. The pre-processing included \textit{a priori} amplitude calibration, ionospheric and parallactic-angle corrections, and global fringe fitting. After fringe fitting, the visibilities were time-averaged into 10-second bins using \texttt{Difmap} \citep{Shepherd1997} with the \texttt{doscatter} option. This option derives the weight of each averaged visibility from the scatter of the samples within the bin, rather than summing the original weights, thereby accounting for any systematic variance within the bin that a simple weight sum would underestimate. The Bayesian analysis presented here starts from these pre-calibrated UVFITS data and jointly infers the full-Stokes sky brightness and residual station-based corruptions (complex gains and polarization leakage) by forward-modeling the radio interferometric measurement equation \citep[RIME;][]{Hamaker1996, Smirnov2011}. For the production \texttt{Comrade} runs described in this paper, we imaged one baseband channel\footnote{For both the Bayesian analysis and the conventional comparison, we used the first of the four 128 MHz basebands (IF~1), with a reference frequency of 14.976\,GHz.} (often referred to as intermediate frequency; IF) from the pre-calibrated data products with scan-averaged visibilities and the fractional systematic-noise settings listed in Table~\ref{tab:sources}.

Our aim is to validate both the reconstructed images and the inferred antenna gains, including large gain offsets relative to those inferred for nearby sources and adjacent scans. We therefore did not flag visibilities solely because of their low amplitudes. We kept low-amplitude data when their phases remained coherent within a scan after fringe fitting, which we used as a practical indication of reliable fringe detection. If the amplitude reductions arise from antenna-based multiplicative errors, they do not intrinsically introduce non-closing errors and should be described by the station-based gain model implemented in \texttt{Comrade}. These data therefore allow us to test the calibration model under large residual gain errors.

For comparison, we also produced conventional CLEAN images with iterative self-calibration in \texttt{Difmap} \citep{Shepherd1997}. Residual instrumental polarization for the conventional workflow was solved with \texttt{GPCAL} \citep{Park2021}, including time-dependent leakage calibration following the method of \citet{Park2023a}. The comparison between the Bayesian and conventional products is therefore performed on both the image domain and the calibration domain.

The EVPA calibration was conducted by comparing the integrated EVPAs of 0420$-$014, 0528$+$134, and 0923$+$392 measured from the CLEAN$+$\texttt{GPCAL} products with the corresponding values from near-contemporaneous 15\,GHz MOJAVE observations obtained on 2024 September 22 \citep{Lister2018}. The resulting IF-dependent global R/L phase offsets were applied to the otherwise unchanged pre-calibrated visibilities prior to the \texttt{Comrade} analysis; the CLEAN self-calibration and \texttt{GPCAL} D-term solutions themselves were not transferred to the \texttt{Comrade} input data. For IF~1 used in this work, the applied R/L phase offset was $66.369^{\circ}$, with a source-to-source standard deviation of $6.032^{\circ}$ in R/L phase, corresponding to $3.016^{\circ}$ in EVPA.

The 25 target sources and their imaging parameters are listed in Table~\ref{tab:sources}.

\begin{table*}[t]

\centering
\caption{Target sources and Comrade imaging parameters.}
\label{tab:sources}
\begin{tabular}{llrllccc}
\hline\hline
Source & Alt.\ name & $z$ & R.A.\ (J2000) & Dec.\ (J2000) & FOV ($\mu$as) & Pixels & $\sigma_{\rm sys}$ \\
\hline
0003$-$066   & J0006$-$0623    & 0.347   & 00h06m13.89s & $-$06$\degr$23$\arcmin$35$\arcsec$ & $10000\times6000$   & $250\times150$ & 0       \\
0106$+$013   & J0108$+$0135    & 2.110   & 01h08m38.77s & $+$01$\degr$35$\arcmin$00$\arcsec$ & $6000\times6000$    & $150\times150$ & 0       \\
0202$+$319   & J0205$+$3212    & 1.475   & 02h05m04.93s & $+$32$\degr$12$\arcmin$30$\arcsec$ & $4000\times4000$    & $100\times100$ & 0       \\
0234$+$285   & J0237$+$2848    & 1.206   & 02h37m52.41s & $+$28$\degr$48$\arcmin$09$\arcsec$ & $8000\times8000$    & $200\times200$ & 0       \\
0300$+$470   & J0303$+$4716    & $\cdots$& 03h03m35.24s & $+$47$\degr$16$\arcmin$16$\arcsec$ & $6000\times6000$    & $150\times150$ & 0       \\
0309$+$411   & J0313$+$4120    & 0.136   & 03h13m01.96s & $+$41$\degr$20$\arcmin$01$\arcsec$ & $6000\times6000$    & $150\times150$ & 0       \\
0316$+$413   & 3C\,84          & 0.018   & 03h19m48.16s & $+$41$\degr$30$\arcmin$42$\arcsec$ & $20000\times40000$  & $250\times500$ & 0.015   \\
0333$+$321   & NRAO\,140       & 1.259   & 03h36m30.11s & $+$32$\degr$18$\arcmin$29$\arcsec$ & $12000\times12000$  & $300\times300$ & 0       \\
0420$-$014   & PKS\,0420$-$014 & 0.914   & 04h23m15.80s & $-$01$\degr$20$\arcmin$33$\arcsec$ & $4800\times9600$    & $120\times240$ & 0       \\
0528$+$134   & J0530$+$1331    & 2.070   & 05h30m56.42s & $+$13$\degr$31$\arcmin$55$\arcsec$ & $9600\times12800$   & $240\times320$ & 0.01    \\
0552$+$398   & DA\,193         & 2.363   & 05h55m30.81s & $+$39$\degr$48$\arcmin$49$\arcsec$ & $4000\times4000$    & $100\times100$ & 0       \\
0727$-$115   & J0730$-$1141    & 1.591   & 07h30m19.11s & $-$11$\degr$41$\arcmin$13$\arcsec$ & $4000\times4000$    & $100\times100$ & 0       \\
0923$+$392   & 4C\,39.25       & 0.696   & 09h27m03.01s & $+$39$\degr$02$\arcmin$21$\arcsec$ & $8000\times6000$    & $200\times150$ & 0       \\
1219$+$044   & J1222$+$0413    & 0.966   & 12h22m22.55s & $+$04$\degr$13$\arcmin$16$\arcsec$ & $4000\times8000$    & $100\times200$ & 0       \\
1228$+$126   & M87             & 0.004   & 12h30m49.42s & $+$12$\degr$23$\arcmin$28$\arcsec$ & $38400\times19200$  & $480\times240$ & 0       \\
1236$+$077   & J1239$+$0730    & 0.400   & 12h39m24.59s & $+$07$\degr$30$\arcmin$17$\arcsec$ & $6000\times6000$    & $150\times150$ & 0       \\
1236$+$049   & J1239$+$0443    & 1.759   & 12h39m32.76s & $+$04$\degr$43$\arcmin$05$\arcsec$ & $4000\times4000$    & $100\times100$ & 0       \\
1502$+$106   & J1504$+$1029    & 1.834   & 15h04m24.98s & $+$10$\degr$29$\arcmin$39$\arcsec$ & $8000\times6000$    & $200\times150$ & 0       \\
1641$+$399   & 3C\,345         & 0.594   & 16h42m58.81s & $+$39$\degr$48$\arcmin$37$\arcsec$ & $12000\times6000$   & $300\times150$ & 0       \\
1730$-$130   & NRAO\,530       & 0.902   & 17h33m02.71s & $-$13$\degr$04$\arcmin$50$\arcsec$ & $4000\times6000$    & $100\times150$ & 0       \\
1749$+$096   & J1751$+$0939    & 0.322   & 17h51m32.82s & $+$09$\degr$39$\arcmin$01$\arcsec$ & $4000\times4000$    & $100\times100$ & 0       \\
1803$+$784   & J1800$+$7828    & 0.683   & 18h00m45.68s & $+$78$\degr$28$\arcmin$04$\arcsec$ & $8000\times4000$    & $200\times100$ & 0       \\
1928$+$738   & J1927$+$7358    & 0.302   & 19h27m48.50s & $+$73$\degr$58$\arcmin$02$\arcsec$ & $8000\times12000$   & $200\times300$ & 0.005   \\
2201$+$315   & J2203$+$3145    & 0.295   & 22h03m14.98s & $+$31$\degr$45$\arcmin$38$\arcsec$ & $6000\times6000$    & $150\times150$ & 0       \\
2230$+$114   & CTA\,102        & 1.038   & 22h32m36.41s & $+$11$\degr$43$\arcmin$51$\arcsec$ & $8000\times10000$   & $200\times250$ & 0.01    \\
\hline
\end{tabular}
\tablefoot{$\sigma_{\rm sys}$ is the fractional systematic noise added in quadrature to the thermal uncertainties. All sources were scan-averaged prior to imaging. The pixel size is 40\,$\mu$as for all sources except 3C\,84 and M87 (80\,$\mu$as).}
\end{table*}

% ============================================================
\section{Imaging and calibration}\label{sec:imaging}

\subsection{Bayesian polarimetric imaging with Comrade}\label{sec:comrade}

We performed full-polarization Bayesian imaging using the \texttt{Comrade} software package \citep{Tiede2022}, which jointly solves for all four Stokes parameters and station-based instrumental terms through forward modeling of the radio interferometric measurement equation \citep[RIME;][]{Hamaker1996, Smirnov2011}. Below we describe the sky model, instrumental model, and the inference procedure adopted in this work.

\subsubsection{Sky model}\label{sec:sky_model}

For each target we defined a source-dependent raster grid with field of view $\theta_x\times\theta_y$ and $N_x\times N_y$ pixels (Table~\ref{tab:sources}). The deterministic part of the image prior is a normalized circular Gaussian mean image $\bar{m}$ whose FWHM is set to the inverse of the maximum baseline length. To avoid forcing all flux into the fiducial template, the effective mean image is written as
\begin{equation}\label{eq:mimg_bkg}
    \bar{m}_{\rm eff} = (1-f_b)\,\bar{m} + \frac{f_b}{N_{\rm pix}},
\end{equation}
where $f_b$ is a background fraction and $N_{\rm pix}=N_xN_y$.

Stokes~$I$ is modeled as a strictly positive correlated field following the approach of \citet{Tiede2026}. A latent fluctuation field $\delta_I$ is mapped onto the simplex via the exponential transfer function (Eq.~16 of \citealt{Tiede2026}) and multiplied elementwise by $\bar{m}_{\rm eff}$, yielding a normalized image that is then rescaled to the measured compact flux density $F_0$:
\begin{equation}\label{eq:stokesI_generic}
    I_{ij} = F_0\,\frac{\bar{m}_{{\rm eff},ij}\,\exp(\delta_{I,ij})}{\sum_{kl}\bar{m}_{{\rm eff},kl}\,\exp(\delta_{I,kl})}.
\end{equation}
The latent field $\delta_I$ is drawn from a first-order Gaussian Markov random field \citep[GMRF;][]{Tiede2026} with two hyperparameters: a fluctuation amplitude $\sigma$ and a correlation length $\rho$ whose lower bound is set to 4\,pixels. Both are inferred from the data during sampling. The background fraction $f_b$ is an additional free parameter that we introduce to allow flux outside the fiducial Gaussian template. We note that this separate Stokes-$I$ construction is used as written by the Poincar\'{e}-sphere parameterization described below; in the PolExp parameterization the total intensity is instead constructed jointly with the polarization from the matrix-exponential coherency model, and Eq.~(\ref{eq:stokesI_generic}) does not enter (Section~\ref{sec:pol_model}).

\subsubsection{Polarization model}\label{sec:pol_model}

We implemented two physically admissible polarization parameterizations. Both guarantee $I>0$ and $I\geq\sqrt{Q^2+U^2+V^2}$ by construction, but they encode different prior structure.

\paragraph{Poincar\'{e} sphere representation.}
This parameterization was introduced by \citet{Pesce2021} and, with minor modifications, adopted in recent \texttt{Comrade} polarimetric analyses \citep{EHTC2025_M87pol}. The polarized fraction at each pixel is written as a logistic transform of a correlated latent field,
\begin{equation}\label{eq:poincare_pfrac}
    p_{ij} = \mathrm{logistic}\!\left(p_0 + p_\sigma\,\ell_{ij}\right)
            = \left[1 + \exp\!\left(-p_0 - p_\sigma\,\ell_{ij}\right)\right]^{-1},
\end{equation}
where $p_0$ is the mean logit-polarization level, $p_\sigma$ is its fluctuation amplitude, and $\ell_{ij}$ is a correlated GMRF realization. The polarization direction is parameterized by a unit vector on the Poincar\'{e} sphere with an isotropic prior, and the full Stokes vector is obtained through the \texttt{PoincareSphere2Map} transform, with the total intensity given by the generic construction of Eq.~(\ref{eq:stokesI_generic}).

\paragraph{Exponential polarization (PolExp) representation.}
As an alternative we implemented the matrix-exponential parameterization introduced by \citet{Arras2025}. At each pixel the coherency matrix is written as
\begin{equation}\label{eq:polexp_matrix}
    \mathbf{X}_{ij} = \exp\!\begin{pmatrix}
        a_{ij}+b_{ij} & c_{ij}+i d_{ij} \\
        c_{ij}-i d_{ij} & a_{ij}-b_{ij}
    \end{pmatrix},
\end{equation}
which yields a Hermitian positive-definite coherency matrix by construction. In particular, the total intensity follows as $I_{ij} = e^{a_{ij}}\cosh p_{ij}$ with $p_{ij} = \sqrt{b_{ij}^2 + c_{ij}^2 + d_{ij}^2}$, so that in this parameterization Stokes~$I$ is constructed jointly with the polarization from the four latent fields, rather than through the separate Stokes-$I$ field of Eq.~(\ref{eq:stokesI_generic}). In our \texttt{Comrade} implementation, the latent fields $a,b,c,d$ are independent correlated random fields with amplitudes $\sigma_a,\sigma_b,\sigma_c,\sigma_d$. The prior on $\sigma_d$ (Stokes~$V$) is set an order of magnitude tighter than those on $\sigma_{a,b,c}$ (Table~\ref{tab:priors}), encoding the expectation that circular polarization in AGN jets is typically much weaker than linear polarization. The resulting full-Stokes map is multiplied pixelwise by the normalized mean image and renormalized so that the total compact flux equals $F_0$. This differs slightly from the pure \texttt{resolve} implementation but preserves the positive-definite coherency construction while anchoring the solution to the same mean-image prior used for Stokes~$I$. 
%We note that, for implementation reasons, the global fluctuation amplitude $\sigma$ of Eq.~(\ref{eq:stokesI_generic}) was retained as a sampled parameter in the PolExp runs even though it does not enter the PolExp sky construction; it is an inert nuisance parameter that has no influence on the model prediction or, by independence, on the posterior of any other parameter.

\subsubsection{Instrumental model}\label{sec:inst_model}

We model residual station-based corruptions with a Jones formalism \citep[see, e.g.,][]{Park2021}. Because parallactic-angle corrections have already been applied during AIPS pre-processing, the feed-rotation matrix $\mathbf{R}_j$ is reintroduced in the forward model so that the per-station Jones matrix takes the form
\begin{equation}\label{eq:jones}
    \mathbf{J}_j = \mathbf{R}_j^{\dagger}\,\mathbf{G}_j\,\mathbf{D}_j\,\mathbf{R}_j,
\end{equation}
where $\mathbf{G}_j$ contains complex gains and $\mathbf{D}_j$ contains polarization leakage terms. The measured visibility on baseline $j$--$k$ is then $\mathbf{V}_{jk} = \mathbf{J}_j\,\mathbf{C}_{jk}\,\mathbf{J}_k^{\dagger}$, where $\mathbf{C}_{jk}$ is the sky coherency matrix.

\paragraph{Gain model.}
For each station $j$ and integration $t$, the RCP gain is parameterized as
\begin{equation}\label{eq:gain_R}
    g_{j,R}(t) = \exp\!\left[\mu^{\rm g}_{j} + \delta^{\rm g}_j(t) + i\,\phi_{j,R}(t)\right],
\end{equation}
where $\mu^{\rm g}_{j}$ is a track-level mean log-amplitude and $\delta^{\rm g}_j(t)$ is an integration-level fluctuation with a fixed prior width (Table~\ref{tab:priors}). This two-level (mean $+$ fluctuation) structure regularizes the per-integration gains toward a station-specific track average while keeping the amplitude solutions flexible on integration timescales. The LCP gain is written relative to the RCP gain,
\begin{equation}\label{eq:gain_L}
    g_{j,L}(t) = g_{j,R}(t)\,\exp\!\left[\mu^{\rm rat}_{j} + \delta^{\rm rat}_j(t) + i\,\phi^{\rm rat}_{j}(t)\right],
\end{equation}
where the prior width on the integration-level ratio fluctuation $\delta^{\rm rat}_j(t)$ is set a factor of two tighter than that on $\delta^{\rm g}_j(t)$ (Table~\ref{tab:priors}), reflecting the expectation that the R/L gain ratio is substantially more stable in time than the individual gain amplitudes. The absolute RCP phase gauge is fixed independently at each integration by setting the RCP phase of a reference antenna to zero. The reference antenna is selected automatically from a predefined sensitivity ranking: at each integration, the most sensitive station present is chosen, so the reference may vary with time depending on station availability. Because the input visibilities are scan-averaged, each integration corresponds to one scan in the present analysis. The EVPA gauge is constrained by assigning a tight prior ($5^{\circ}$) on $\phi_j^{\rm rat}$ at a designated reference antenna, while all other stations receive a broad prior. This approach softens the EVPA gauge with an informative prior rather than removing it by fixing all R/L phases to zero.

%We also explored a three-level variant in which the fluctuation scales are themselves inferred per station as track-level hyperparameters, following the spirit of hierarchical noise modeling; the resulting images and calibration solutions were statistically indistinguishable from the two-level model adopted here, indicating that our results are robust to the choice of gain-prior hierarchy. 

\paragraph{D-term model.}
The leakage terms are hierarchical and time dependent. For example,
\begin{equation}\label{eq:dterm}
    d_{j,R}(t) = \left[\mu^{d,\mathrm{re}}_{j,R} + \delta d^{\mathrm{re}}_{j,R}(t)\right]
    + i\left[\mu^{d,\mathrm{im}}_{j,R} + \delta d^{\mathrm{im}}_{j,R}(t)\right],
\end{equation}
with an analogous expression for $d_{j,L}(t)$. The track-level terms $\mu^d$ encode the mean leakage over an observing track, while the integration-level terms $\delta d(t)$ allow the leakage to vary from one time stamp to the next. This hierarchical time-dependent D-term model goes beyond the static leakage assumed in previous \texttt{Comrade} polarimetric analyses \citep{EHTC2025_M87pol} and is motivated by the expectation that effective leakage can change with pointing and elevation even when the underlying receiver leakage is stable \citep{Park2023a}. The tight prior on $\delta d(t)$ (Table~\ref{tab:priors}) regularizes the time-dependent component, permitting variability without allowing completely unconstrained per-integration leakage solutions.

\begin{table*}[t]
\centering
\caption{Prior distributions for the Comrade imaging model parameters.}
\label{tab:priors}
\begin{tabular}{llll}
\hline\hline
Parameter & Symbol & Prior & Timescale \\
\hline
\multicolumn{4}{l}{\textit{Sky model}} \\
Total compact flux & $F_0$ & Fixed (measured) & -- \\
Background fraction & $f_b$ & $\text{Beta}(1.0,\, 5.0)$ & -- \\
Intensity fluctuation amplitude & $\sigma$ & $\mathcal{N}^{+}(0,\, 0.5)$ & -- \\
GMRF order & -- & 1 (fixed) & -- \\
Correlation length lower bound & -- & 4 pixels (fixed) & -- \\
\hline
\multicolumn{4}{l}{\textit{Polarization model (Poincar\'{e})}} \\
Mean logit pol.\ fraction & $p_0$ & $\mathcal{N}(-1.0,\, 2.0)$ & -- \\
Pol.\ fraction fluctuation scale & $p_\sigma$ & $\mathcal{N}^{+}(0,\, 0.5)$ & -- \\
Pol.\ direction & $\hat{\boldsymbol{p}}_{ij}$ & Uniform on $S^2$ & -- \\
\hline
\multicolumn{4}{l}{\textit{Polarization model (PolExp)}} \\
Field amplitude ($a$) & $\sigma_{a}$ & $\mathcal{N}^{+}(0,\, 0.4)$ & -- \\
Field amplitudes ($b, c$) & $\sigma_{b,c}$ & $\mathcal{N}^{+}(0,\, 0.25)$ & -- \\
Field amplitude ($d$; Stokes $V$) & $\sigma_d$ & $\mathcal{N}^{+}(0,\, 0.015)$ & -- \\
\hline
\multicolumn{4}{l}{\textit{Gain model}} \\
RCP log-amplitude fluctuation & $\delta^{\rm g}_j(t)$ & $\mathcal{N}(0,\, 0.2)$ & Integration \\
RCP log-amplitude mean & $\mu^{\rm g}_j$ & $\mathcal{N}(0,\, 0.2)$ & Track \\
RCP phase & $\phi_{j,R}(t)$ & $\text{VonMises}(0,\, \pi^{-2})$ & Integration \\
L/R log-amplitude ratio fluctuation & $\delta_j^{\rm rat}(t)$ & $\mathcal{N}(0,\, 0.1)$ & Integration \\
L/R log-amplitude ratio mean & $\mu_j^{\rm rat}$ & $\mathcal{N}(0,\, 0.2)$ & Track \\
L/R phase ratio & $\phi_j^{\rm rat}(t)$ & $\text{VonMises}(0,\, [19.3^{\circ}]^{-2})$ & Integration \\
EVPA-anchor L/R phase ratio & $\phi_{\rm anchor}^{\rm rat}$ & $\text{VonMises}(0,\, [5^{\circ}]^{-2})$ & Integration \\
\hline
\multicolumn{4}{l}{\textit{D-term model}} \\
Track-level mean (re, im) & $\mu^d$ & $\mathcal{N}(0,\, 0.15)$ & Track \\
Per-integration deviation (re, im) & $\delta d(t)$ & $\mathcal{N}(0,\, 0.01)$ & Integration \\
\hline
\end{tabular}
\tablefoot{$\mathcal{N}^{+}$ denotes a half-normal distribution (truncated at zero). $\text{VonMises}(\mu,\kappa)$ denotes the von Mises distribution with mean $\mu$ and concentration parameter $\kappa$. The prior families follow the conventions of \citet{Tiede2026} and \citet{EHTC2025_M87pol}, with values adapted for cm-VLBA data. Source-dependent image geometry and fractional systematic-noise settings are listed in Table~\ref{tab:sources}.}
\end{table*}

\subsubsection{Data products and likelihood}\label{sec:likelihood}

For each source we fitted all four circular-polarization correlation products ($RR^{*}$, $RL^{*}$, $LR^{*}$, $LL^{*}$) simultaneously with a complex Gaussian visibility likelihood. The analysis was performed on scan-averaged, pre-calibrated visibilities from a single baseband channel.

The visibility uncertainties were estimated using the \texttt{doscatter} option in \texttt{Difmap} (Section~\ref{sec:obs}). For the short 10-second intervals at 15\,GHz after fringe fitting, the scatter is expected to be dominated by thermal noise, although systematic variations within each interval can also contribute. However, this estimate may not fully capture residual systematic errors that remain approximately constant within an interval. For EHT observations, the magnitude of residual non-closing errors can be estimated from intra-site baselines \citep{EHTC2024_M87shadow}, but no such baselines are available for the VLBA. We instead estimated the systematic-noise level empirically by identifying the fractional systematic-noise term $\sigma_{\rm sys}$ that brings the reduced $\chi^2$ close to unity for each source. The required $\sigma_{\rm sys}$ is source-dependent (Table~\ref{tab:sources}), as it absorbs both residual calibration errors and any model incompleteness arising from complex extended structure. This term was added in quadrature to the scatter-based visibility uncertainty before Bayesian imaging. We note that a nonzero $\sigma_{\rm sys}$ was required for only four of the 25 sources (Table~\ref{tab:sources}); for these sources the resulting near-unity reduced $\chi^2$ is partially by construction and should not be read as an independent goodness-of-fit validation, whereas for the remaining 21 sources no such tuning was applied and the $\chi^2$ statistics reflect the scatter-based visibility uncertainties without an additional systematic-noise term.

\subsubsection{Inference procedure}\label{sec:inference}

Posterior exploration proceeded in two stages. First, we obtained a maximum \textit{a posteriori} (MAP) solution through a sequence of optimization passes with progressively decreasing additional systematic noise (fractional levels of 5\%, 2.5\%, 1\%, and 0\%). At each step, the MAP solution from the previous, smoother objective serves as the starting point for the next. This staged approach guides the optimizer toward a good region of parameter space before the final pass with zero additional tempering noise (the permanent $\sigma_{\rm sys}$ of Table~\ref{tab:sources} is always retained), which provides the starting point for posterior sampling.

Second, we sampled the joint posterior with the NUTS variant of Hamiltonian Monte Carlo \citep{HoffmanGelman2014} as implemented in \texttt{AdvancedHMC.jl}, similar to the approach used in \citet{Tiede2026} and \citet{EHTC2025_M87pol}. Gradients were computed with reverse-mode automatic differentiation via \texttt{Enzyme.jl} \citep{Moses2020, Moses2021}. The sampler ran for a total of 5000 steps, of which the first 2000 constitute the adaptation (warm-up) phase during which the step size and mass matrix are tuned; every 25th sample was retained, yielding 200 stored samples per source, the first 80 of which correspond to the adaptation phase. For each retained sample we stored reconstructed Stokes images, calibration tables, visibility-domain model products, and per-product $\chi^2$ diagnostics.

\subsubsection{Burn-in determination}\label{sec:burnin}

We discarded, as burn-in, all stored samples belonging to the sampler's adaptation phase. With 2000 adaptation steps and a thinning factor of 25 (Section~\ref{sec:inference}), the first 80 of the 200 stored samples were removed before computing all posterior summaries. Adopting the adaptation length as the burn-in cut is important for quantitative posterior statistics because the step size and mass matrix are still being tuned during adaptation; late-adaptation samples can therefore appear stable in the $\chi^2$ trace while still biasing posterior widths, hyperparameter summaries, and effective-sample-size estimates.

\begin{figure*}[t!]
    \centering
    \includegraphics[width=\textwidth]{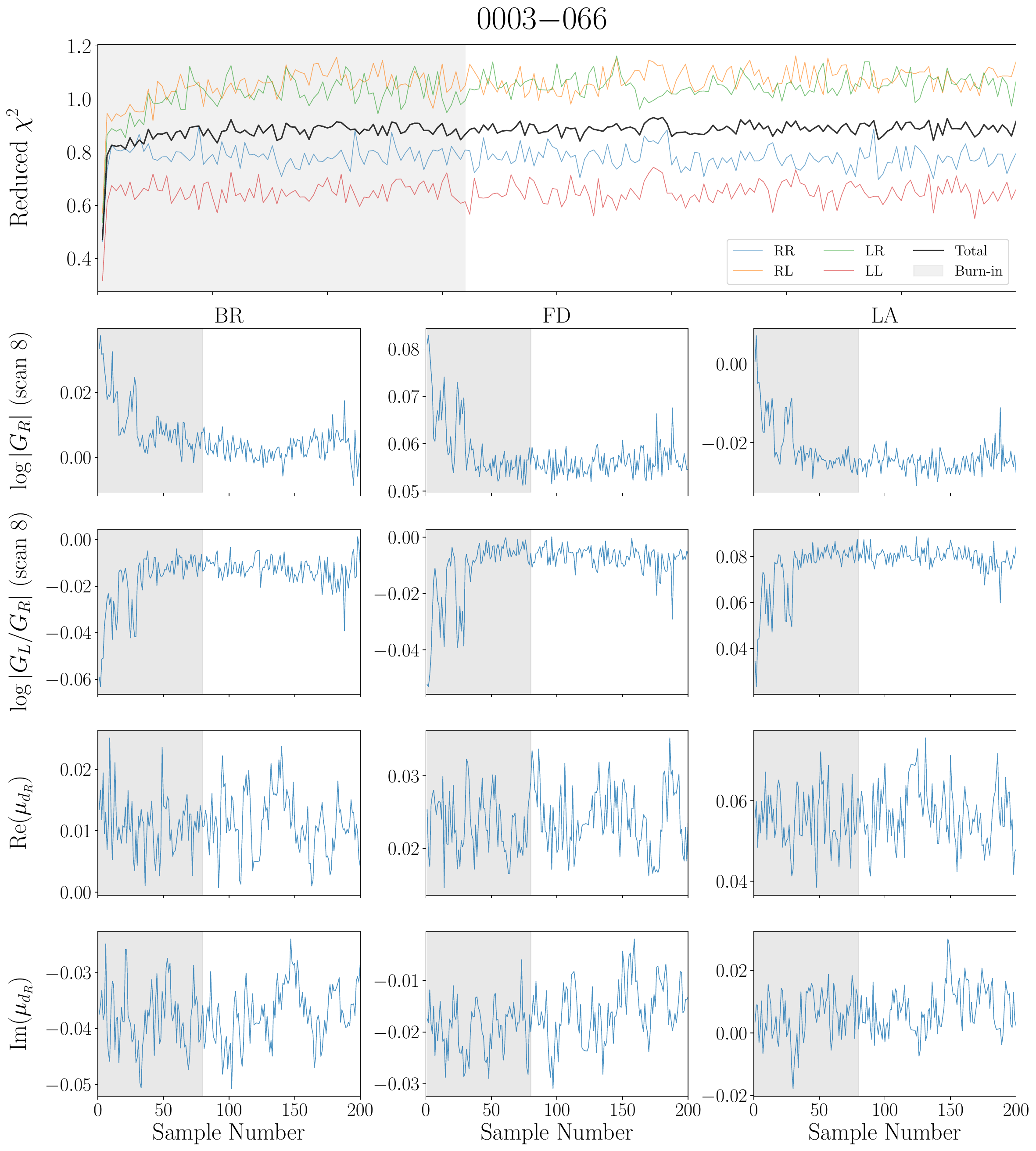}
    \caption{Chain convergence diagnostics for 0003$-$066 (representative source). \textit{Top panel:} Reduced $\chi^2$ per correlation product (RR, RL, LR, LL) and their mean (Total) as a function of sample number. The gray shaded region indicates the burn-in phase. The parallel-hand products (RR, LL) show $\chi^2_\mathrm{red} \approx 0.7$, while the cross-hand products (RL, LR) show $\chi^2_\mathrm{red} \approx 1.0$. \textit{Lower panels:} Trace plots for representative instrumental parameters at three stations (BR, FD, LA): log gain amplitude $\log|G_R|$ and gain ratio $\log|G_L/G_R|$ at scan~8, and on-axis D-term components Re($\mu_{d_R}$) and Im($\mu_{d_R}$). All parameters are stationary after burn-in, confirming convergence.}
    \label{fig:chain_convergence}
\end{figure*}

As an independent cross-check, we also determined a data-driven convergence point from the total $\chi^2$ chain. We scanned the trace from sample index $N_{\rm min}=0.05\,N$ to $N_{\rm max}=0.8\,N$, where $N$ is the total number of stored samples, and for each candidate split point $i$ computed the pooled variance
\begin{equation}\label{eq:pooled_var}
    V_{\rm pool}(i)
    = \mathrm{Var}(\chi^2_{1:i})\,i
    + \mathrm{Var}(\chi^2_{i:N})\,(N-i).
\end{equation}
The index $i^{*}$ minimizing $V_{\rm pool}$, plus an adaptive safety margin of $\max(0.1\,i^{*},0.05\,N)$, defines the $\chi^2$-based convergence point. Across the 50 reconstructions, the resulting convergence points range from 20 to 40 stored samples and lie well within the adopted 80-sample burn-in in every case. The adaptation-based cut therefore conservatively exceeds the convergence point identified from the $\chi^2$ traces. We further cross-checked the adopted burn-in by visually inspecting the $\chi^2$ traces and the chains of the instrumental parameters, including the gain amplitudes, gain phases, gain ratios, and D-terms (see Fig.~\ref{fig:chain_convergence}).

\begin{figure*}[t!]
    \centering
    \includegraphics[width=\textwidth]{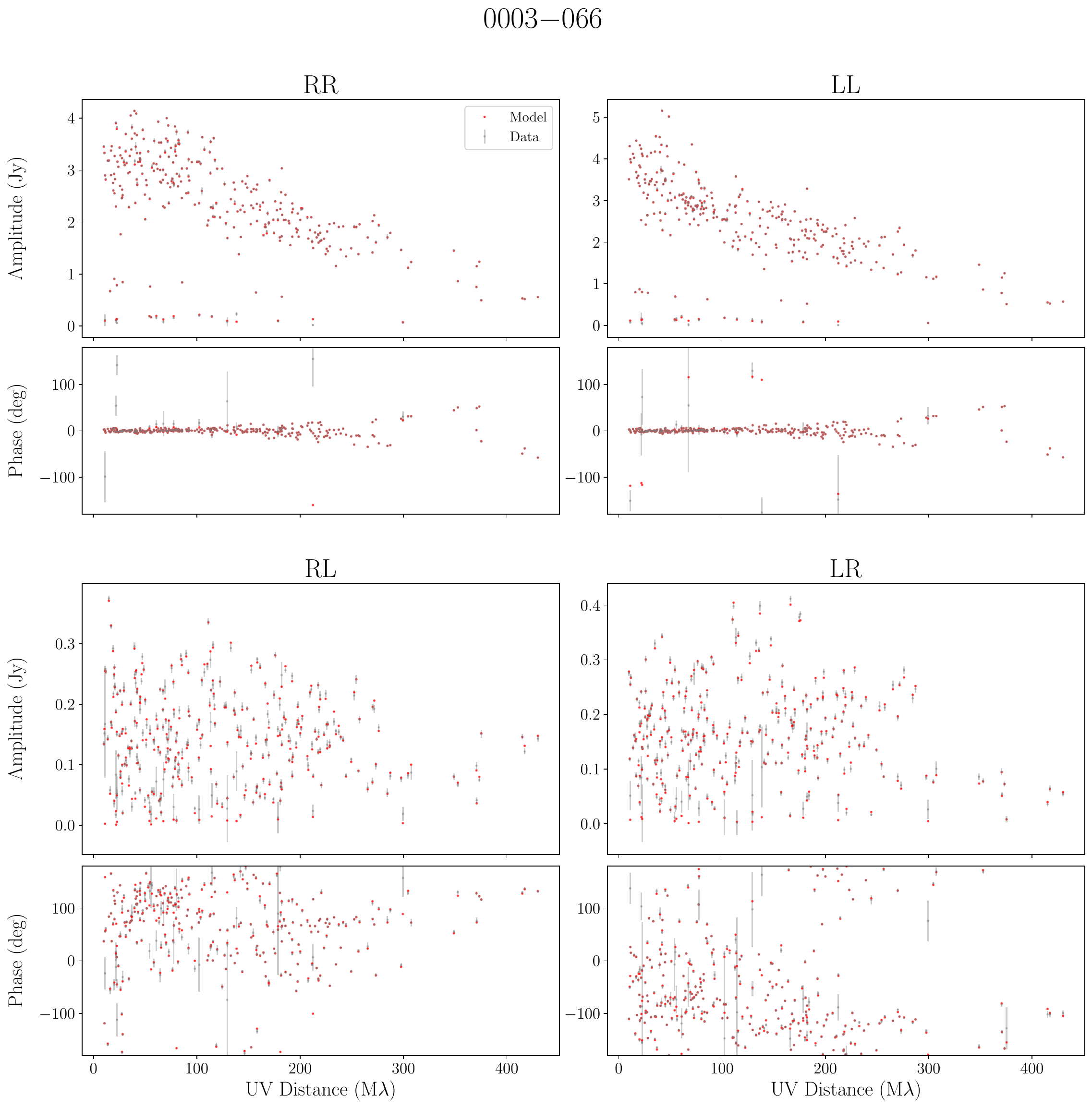}
    \caption{Visibility amplitude and phase as a function of UV distance for 0003$-$066 (posterior sample 200; the same representative source as in Figure~\ref{fig:chain_convergence}). \textit{Top row:} Parallel-hand products (RR, LL). \textit{Bottom row:} Cross-hand products (RL, LR). Each panel pair shows amplitude (upper) and phase (lower). Gray points with error bars show the data; red points show the model visibilities. The data are the pre-calibrated input visibilities, and the model includes the inferred residual antenna gains and D-terms. We kept data with very low RR and LL amplitudes to test the recovery of large antenna gain offsets (Section~\ref{sec:obs}).}
    \label{fig:vis_amp_phase}
\end{figure*}

\begin{figure*}[t!]
    \centering
    \includegraphics[width=\textwidth]{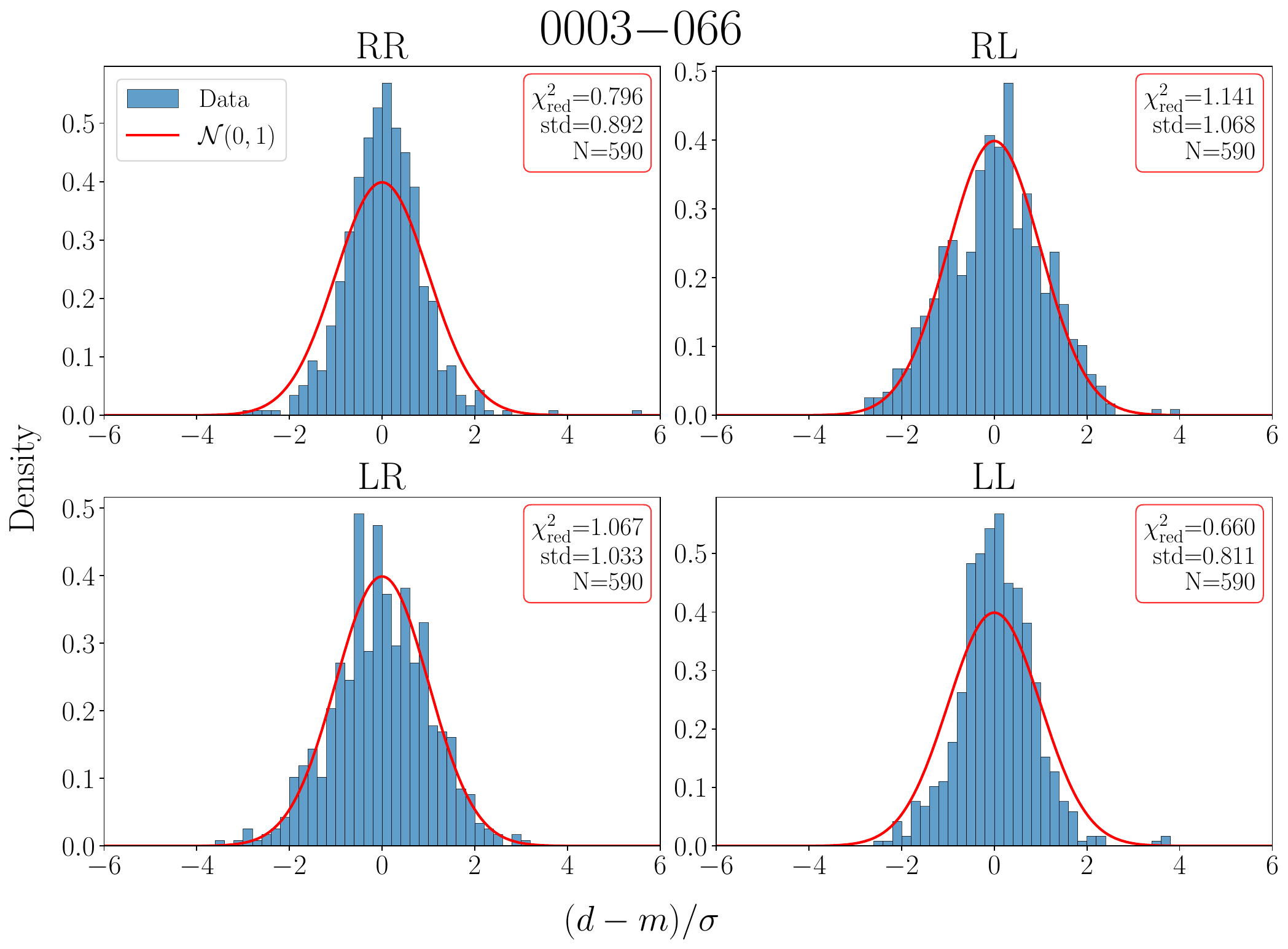}
    \caption{Normalized visibility residual histograms for 0003$-$066. Each panel shows the distribution of $(d - m)/\sigma$ (real and imaginary parts combined) for a single correlation product. The red curve shows the expected $\mathcal{N}(0,1)$ distribution. The reduced $\chi^2$, standard deviation, and number of data points are annotated. The parallel-hand products (RR, LL) show $\chi^2_\mathrm{red} < 1$, indicating slight overfitting or noise overestimation, while the cross-hand products (RL, LR) are consistent with $\chi^2_\mathrm{red} \approx 1$.}
    \label{fig:vis_residual}
\end{figure*}

\begin{figure*}[t!]
    \centering
    \includegraphics[width=0.9\textwidth]{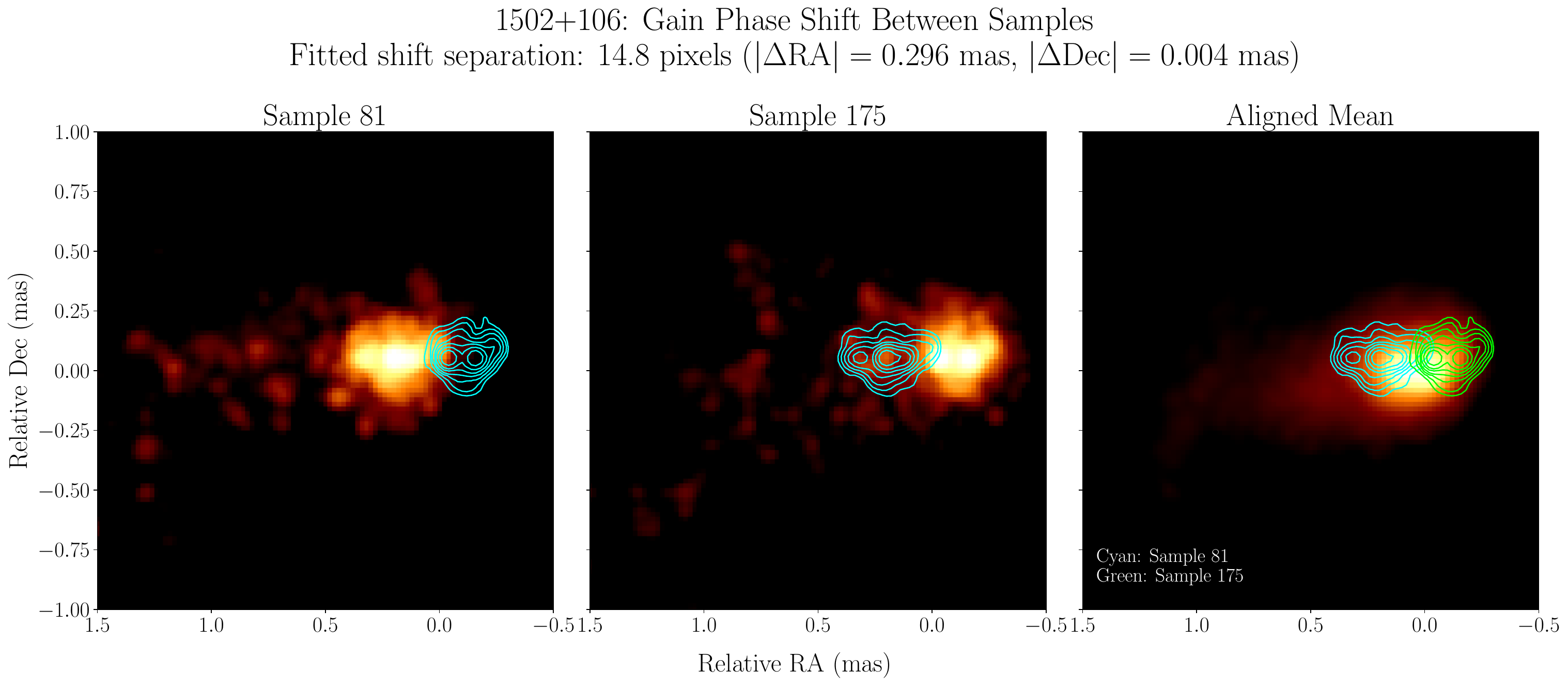}
    \caption{Illustration of the gain phase shift degeneracy for 1502$+$106, the source with the largest inter-sample position variation in our dataset. \textit{Left and center:} The two posterior samples with the largest mutual separation: their fitted image shifts differ by 14.8~pixels (0.30\,mas, almost entirely in RA). The cyan contours (factor-of-2 spaced levels starting from the peak) show the structure of the \textit{other} sample, illustrating the position offset. \textit{Right:} The gain-phase-aligned posterior mean, with contours from both samples overlaid (cyan: Sample~81, green: Sample~175). Without gain phase shift correction, averaging these shifted samples would produce a blurred mean image. This motivates our gain-phase-based image alignment procedure (Section~\ref{sec:gain_alignment}). An animated version showing the corresponding images before and after the alignment for all post-burn-in posterior samples is available as online material.}
    \label{fig:shift_comparison}
\end{figure*}

\begin{figure*}[tp]
    \centering
    \includegraphics[width=0.93\textwidth]{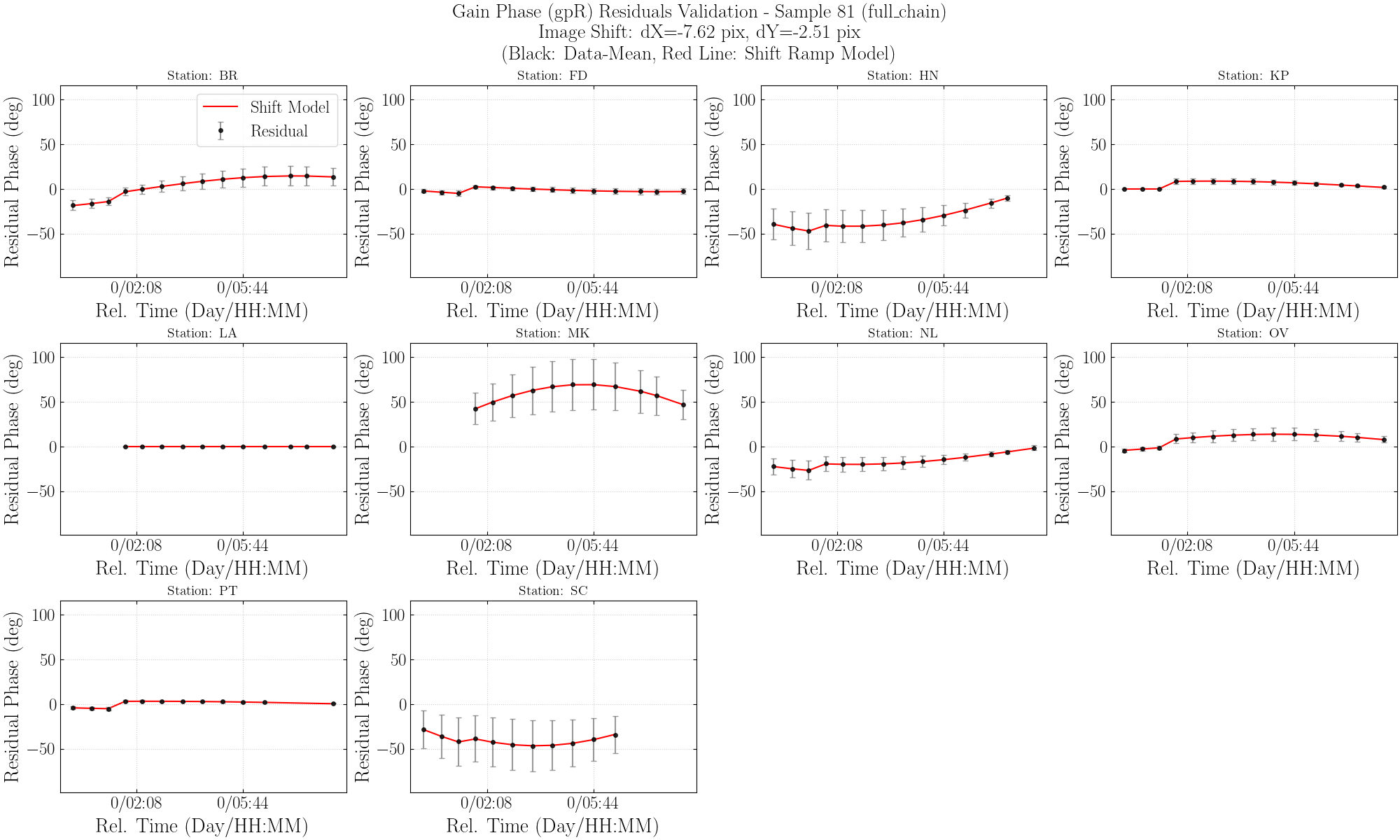}
    \caption{Validation of the gain-phase-based image alignment for 1502$+$106. Each panel shows, for a single post-burn-in posterior sample, the residual gain phase $\Delta\phi_j^{(s)}(t)$ (Eq.~\ref{eq:phase_resid}; black points) at each VLBA station as a function of time, together with the best-fit image-shift phase-ramp model of Eq.~\ref{eq:phase_ramp} (red line). For this sample the fitted offset is $(\Delta x, \Delta y) = (-7.6, -2.5)$ pixels. The residual phases closely follow the ramp expected from a pure positional offset of the reconstructed image, validating the image-shift interpretation of the sample-to-sample gain-phase fluctuations. An animated version showing the corresponding fits for all post-burn-in posterior samples is available as online material.}
    \label{fig:gainphase_fit}
\end{figure*}

\begin{figure*}[t!]
    \centering
    \includegraphics[width=\textwidth]{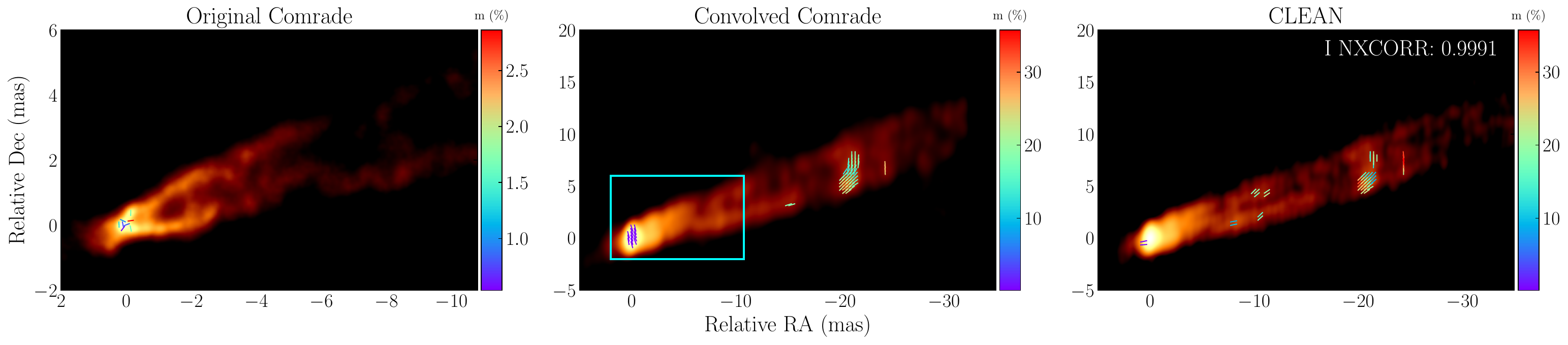}
    \caption{Polarimetric imaging results for M87. Each panel shows Stokes~$I$ on a logarithmic scale with EVPA ticks overlaid, colored by the fractional linear polarization~$m$ (\%). \textit{Left:} Original \texttt{Comrade} posterior mean at native resolution. \textit{Center:} \texttt{Comrade} image convolved with the CLEAN restoring beam; the cyan box indicates the field of view of the original image. \textit{Right:} CLEAN image. The combination of weak or undetected linear polarization in the core and inner jet and prominent polarized emission at $\sim20$\,mas from the core is consistent with previous VLBI polarimetric studies of M87 \citep{ZT2002, Walker2018, Park2019b, Park2021c, Park2026, Nikonov2023}. The Stokes~$I$ NXCORR between the convolved \texttt{Comrade} and CLEAN images is 0.9991.}
    \label{fig:highlight}
\end{figure*}

\begin{figure*}[p!]
    \centering
    \includegraphics[width=\textwidth, height=0.85\textheight, keepaspectratio]{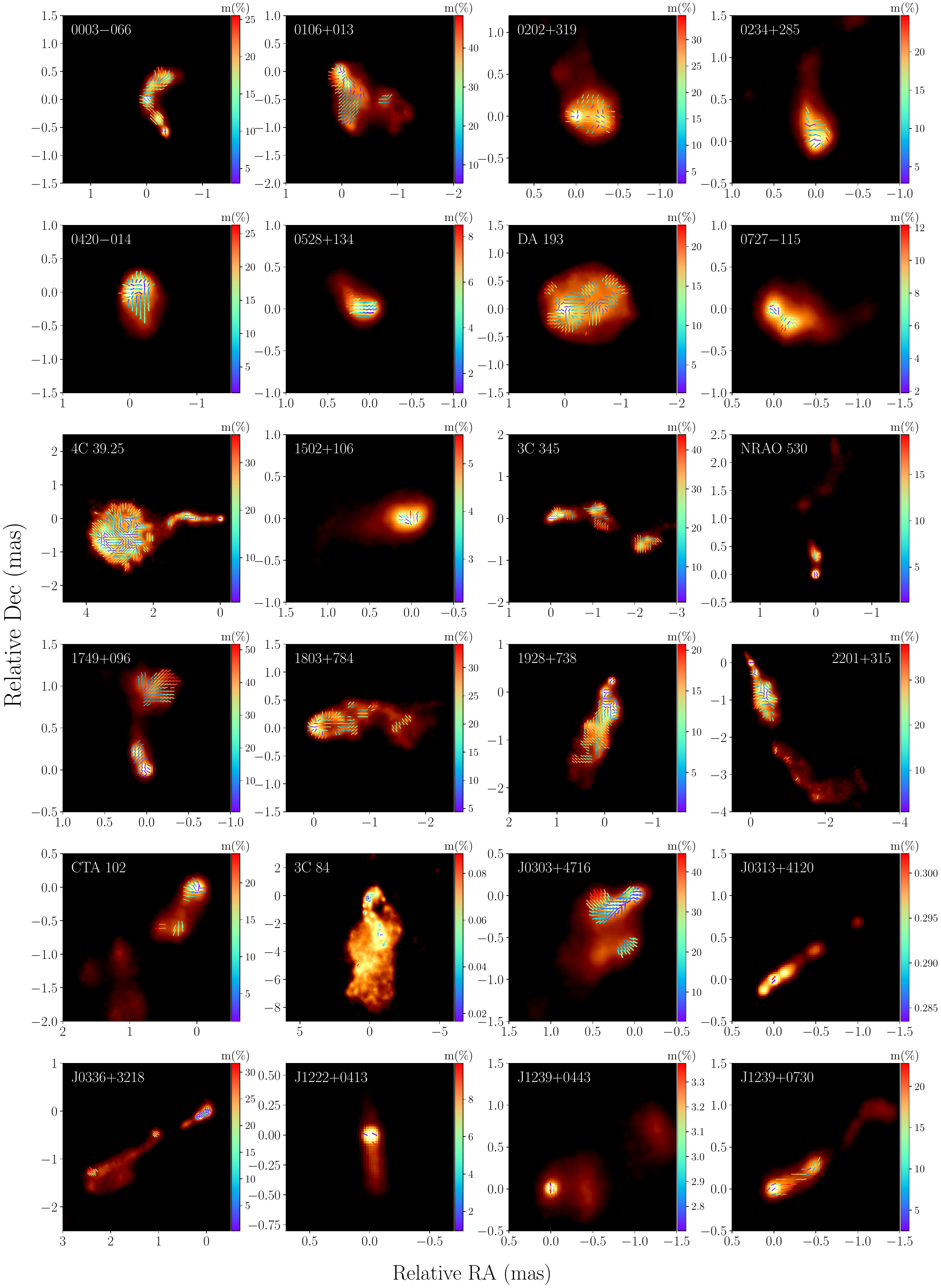}
    \caption{Polarimetric image mosaic of the 24 AGN sources (excluding M87, shown in Figure~\ref{fig:highlight}) observed in the BP266B VLBA experiment at 15\,GHz. Each panel shows the \texttt{Comrade} posterior mean Stokes~$I$ image (logarithmic scale) with EVPA ticks overlaid, colored by the fractional linear polarization~$m$ (\%). For J1239+0443 only, the lower bound of the Stokes~$I$ color scale is set to the DR-matched equivalent of $0.5\,\sigma_I$ to enhance the visibility of its low-surface-brightness jet; EVPA ticks are still plotted only where $I>\tau_I$ and $P>\tau_P$, as in the other panels. The sample spans a wide range of morphologies including compact cores, one-sided jets, and complex extended structures. The fields of view and DR-matched display thresholds emphasize the bright inner emission; Figure~\ref{fig:native_extended_emission} shows the fainter, larger-scale emission in 0003$-$066. These posterior means are representative reconstructions; the reliability of individual sub-beam features requires assessment from the posterior samples (Section~\ref{sec:uncertainty}). The display thresholds are not posterior significance cuts.}
    \label{fig:mosaic}
\end{figure*}

\begin{figure*}[t!]
    \centering
    \includegraphics[width=\textwidth]{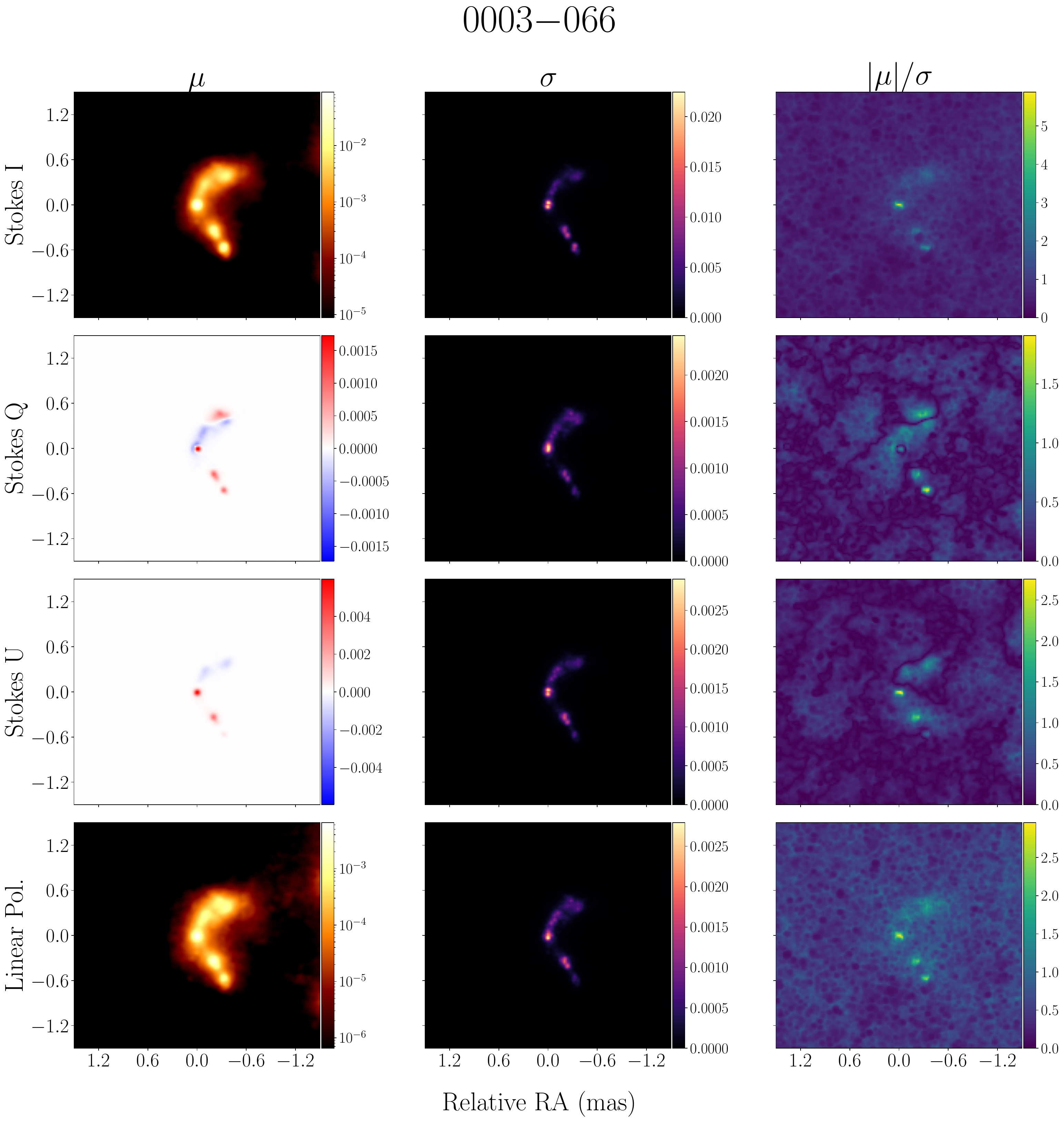}
    \caption{Posterior uncertainty maps for 0003$-$066. Rows from top to bottom: Stokes~$I$, $Q$, $U$, and linear polarization intensity~$P$. Columns from left to right: posterior mean~$\mu$, posterior standard deviation~$\sigma$, and signal-to-noise ratio $|\mu|/\sigma$. The Stokes~$I$ and $P$ mean images are shown on a logarithmic scale. The $P$ mean and standard deviation are computed directly from individual posterior samples (not from error propagation of $Q$ and $U$). The ratios describe local posterior precision for this source and are not, in general, Gaussian detection significances. Pixel correlations are not shown; uncertainties in structural measurements must be evaluated from the posterior samples (Section~\ref{sec:uncertainty}).}
    \label{fig:snr_map}
\end{figure*}

\begin{figure*}[t!]
    \centering
    \includegraphics[width=0.95\textwidth]{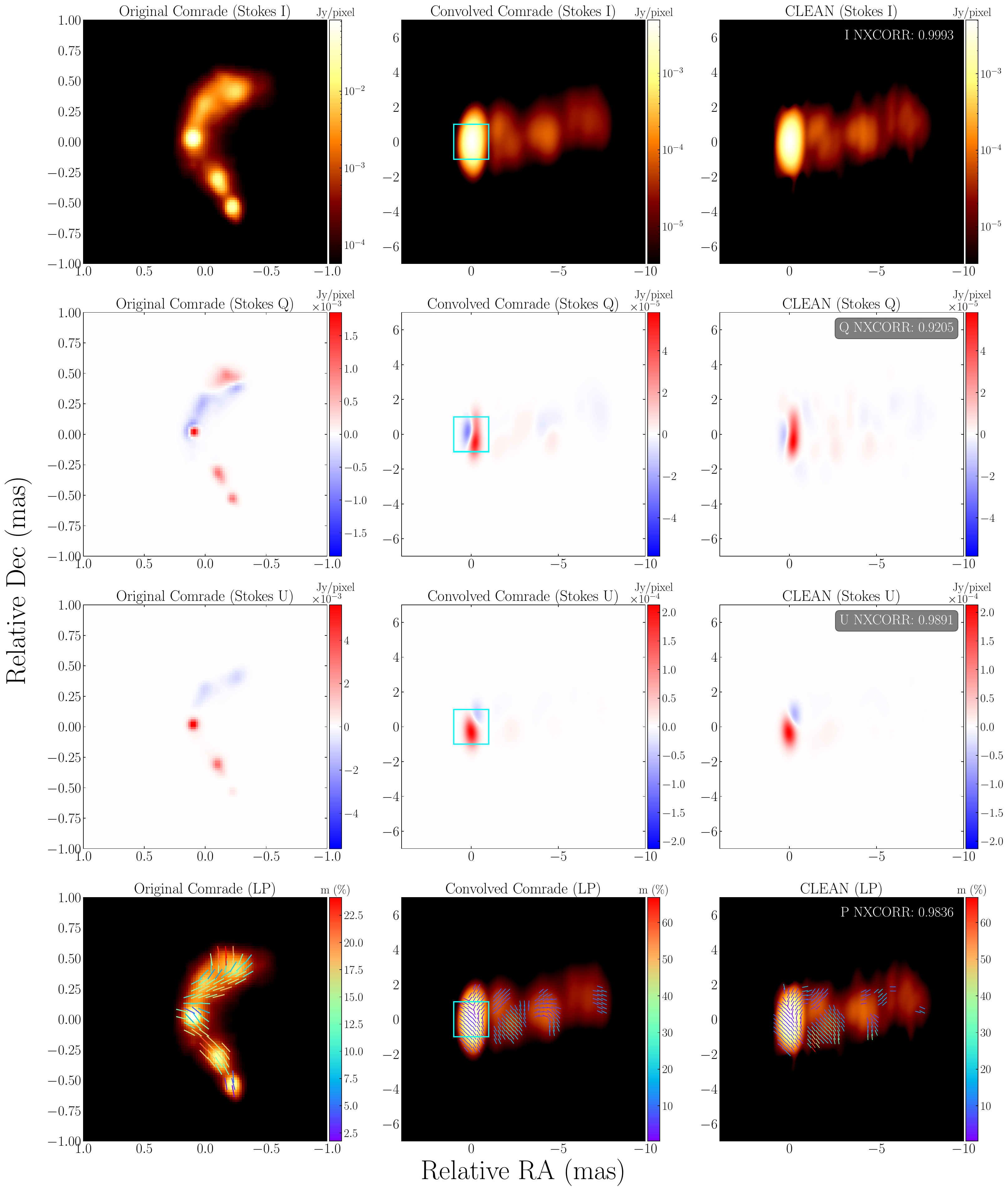}
    \caption{Total intensity and linear polarization image comparison between \texttt{Comrade} and CLEAN for 0003$-$066. From top to bottom: Stokes~$I$, $Q$, $U$, and linearly polarized intensity~$P$ (with EVPA ticks). In each row, the panels show (left to right): original \texttt{Comrade} posterior mean, \texttt{Comrade} convolved with the CLEAN restoring beam (cyan box indicates the original field of view), and the CLEAN image. The normalized cross-correlation coefficient is annotated on the CLEAN panel. The native-resolution mean is shown over the CLEAN field of view at a lower display threshold in Figure~\ref{fig:native_extended_emission}. Corresponding plots for the remaining 24 sources are provided in the online supplementary material.}
    \label{fig:comparison_4x3}
\end{figure*}

\begin{figure*}[t!]
    \centering
    \begin{minipage}[t]{0.48\textwidth}
        \centering
        \includegraphics[width=\textwidth]{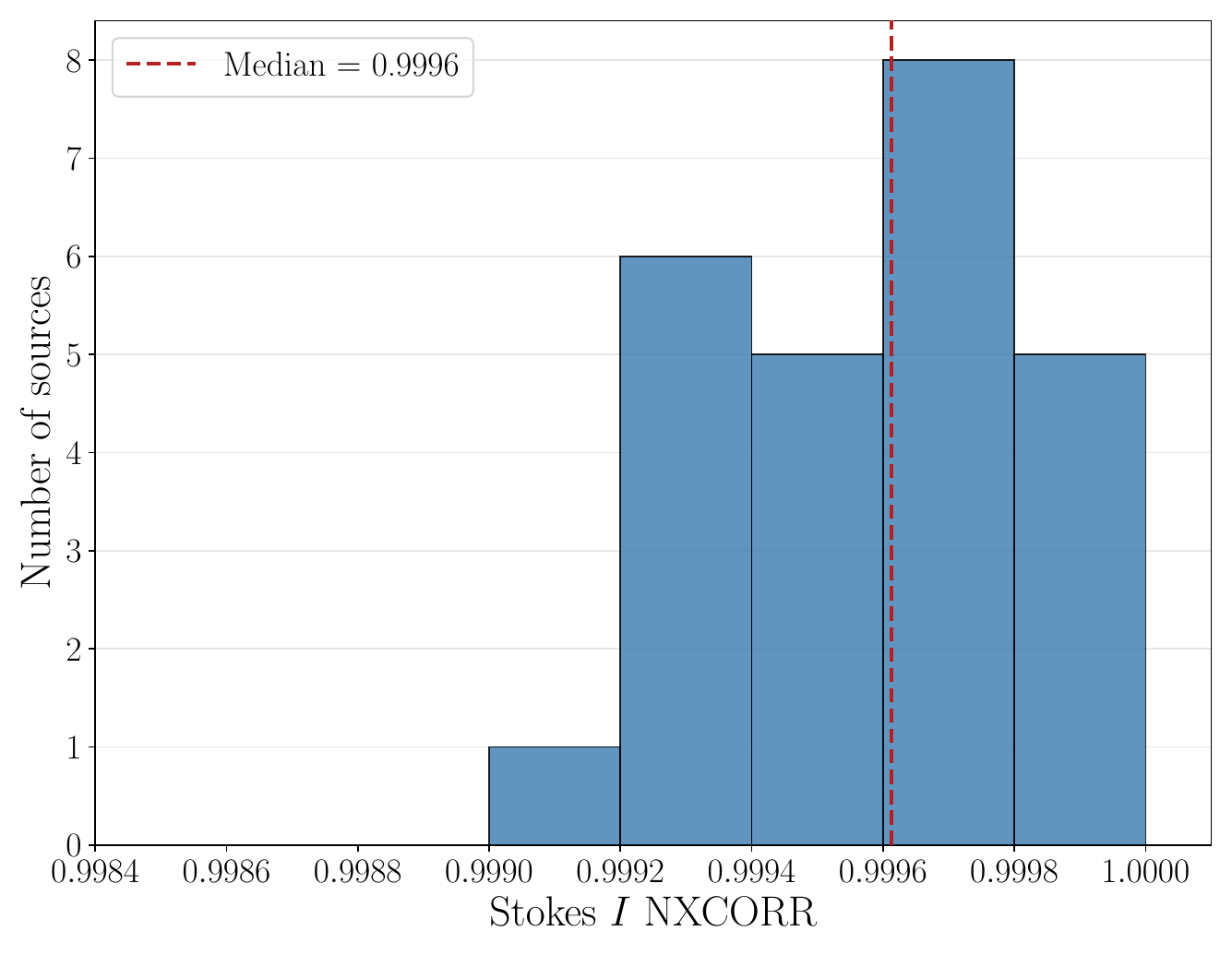}
    \end{minipage}\hfill
    \begin{minipage}[t]{0.48\textwidth}
        \centering
        \includegraphics[width=\textwidth]{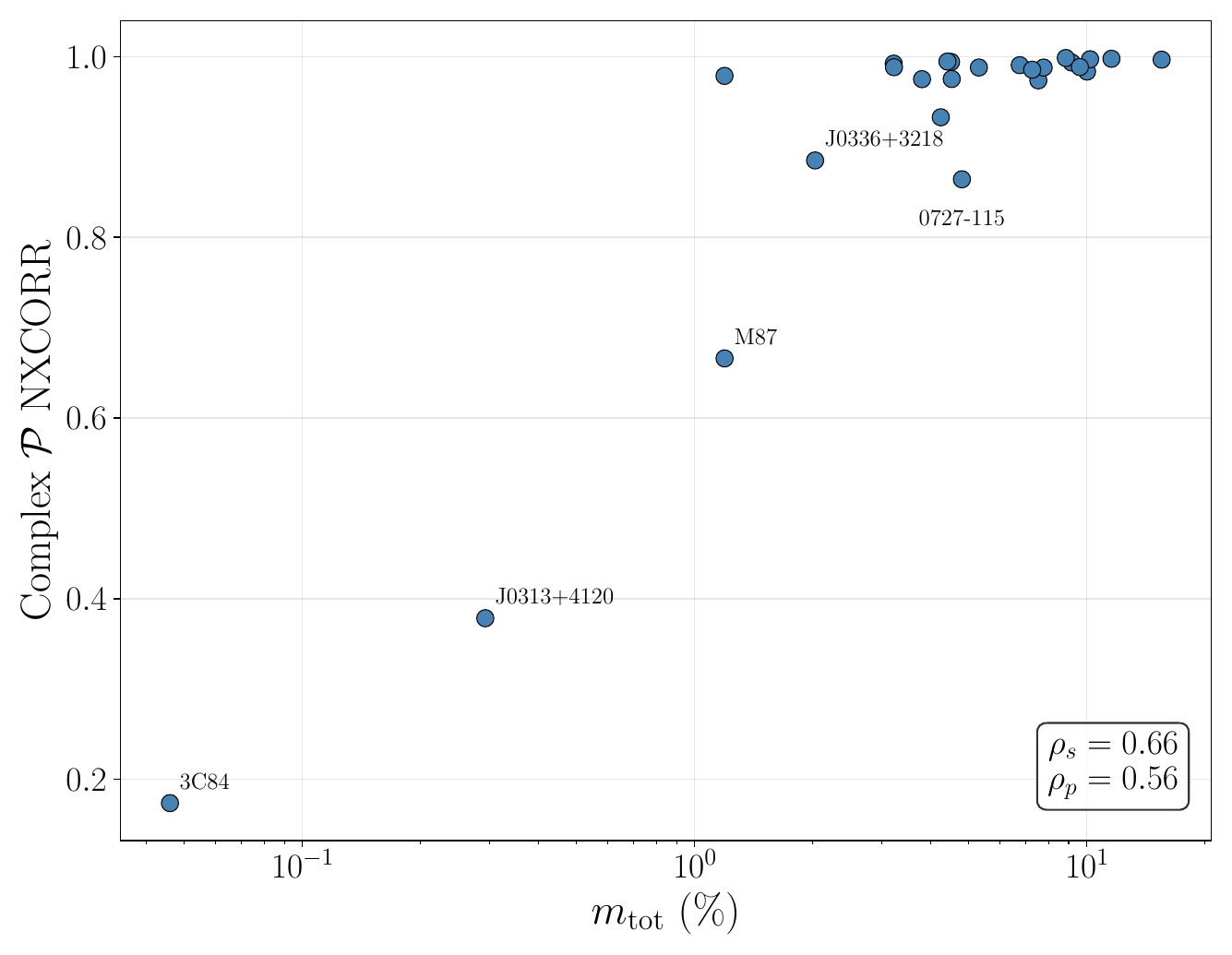}
    \end{minipage}
    \caption{\textit{Left:} Distribution of Stokes~$I$ NXCORR (Eq.~\ref{eq:nxcorr}) between the convolved \texttt{Comrade} PolExp posterior mean and CLEAN images for all 25 sources. The dashed line indicates the median value of 0.9996. All sources achieve NXCORR $> 0.999$, confirming excellent agreement in total intensity. \textit{Right:} Complex linear polarization NXCORR (Eq.~\ref{eq:complex_nxcorr}) as a function of total fractional polarization $m_{\rm tot}$ measured from the native-resolution PolExp reconstruction. Sources with higher fractional polarization show systematically better agreement ($\rho_s = 0.66$). The labeled sources fall below the general trend, with the three most extreme outliers (3C\,84, J0313$+$4120, and M87) being among the most weakly polarized sources in our sample.}
    \label{fig:nxcorr_polexp_clean}
\end{figure*}

\begin{figure*}[t!]
    \centering
    \includegraphics[width=\textwidth]{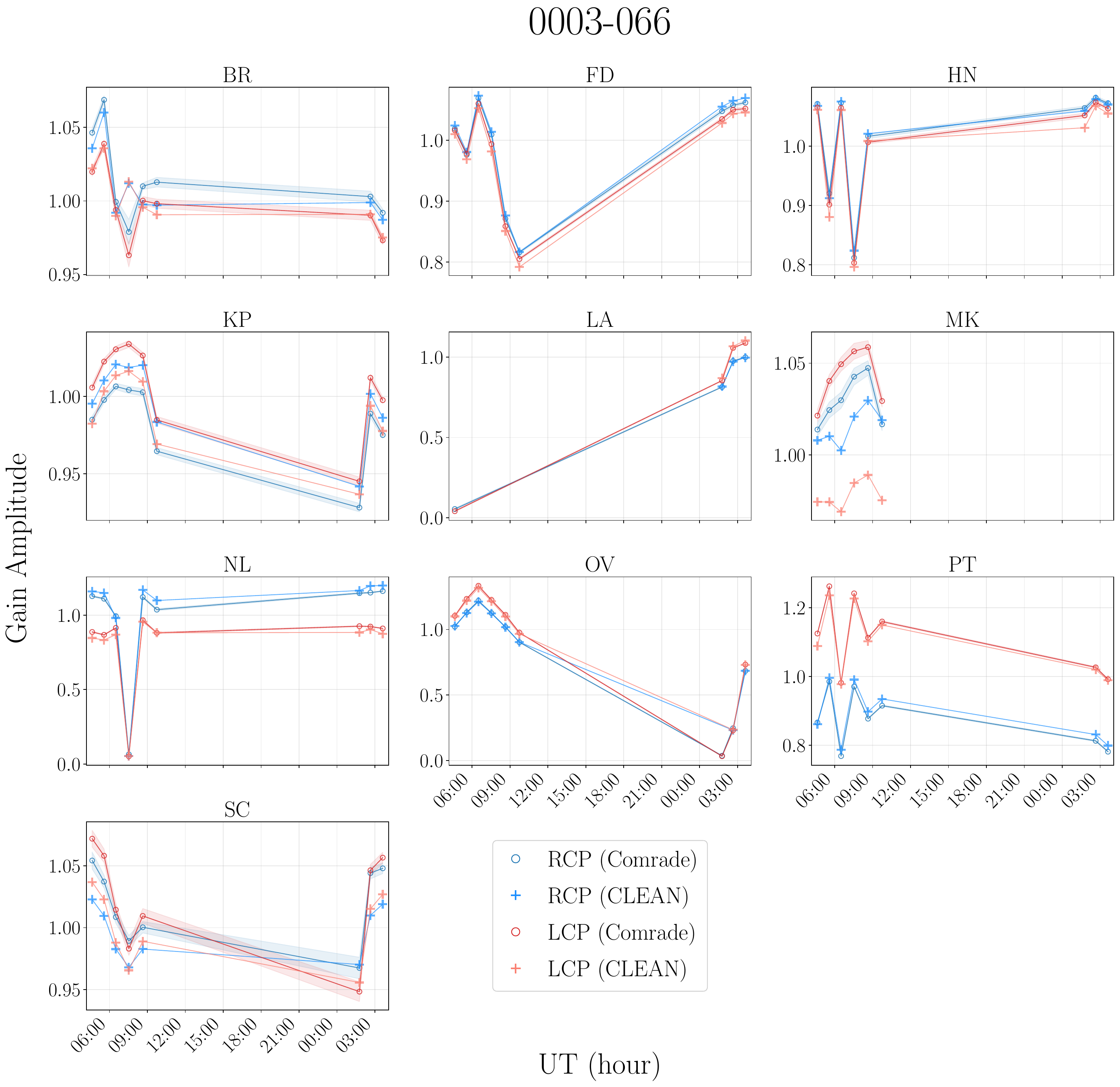}
    \caption{Gain amplitude comparison between \texttt{Comrade} and CLEAN for 0003$-$066. Each panel shows one VLBA station. \texttt{Comrade} posterior mean (open circles with $1\sigma$ bands) and CLEAN self-calibration solutions (crosses with connecting lines) are shown for both RCP (blue) and LCP (red). Corresponding plots for the remaining 24 sources are provided in the online supplementary material.}
    \label{fig:gain_persource}
\end{figure*}

\begin{figure*}[t!]
    \centering
    \includegraphics[width=\textwidth]{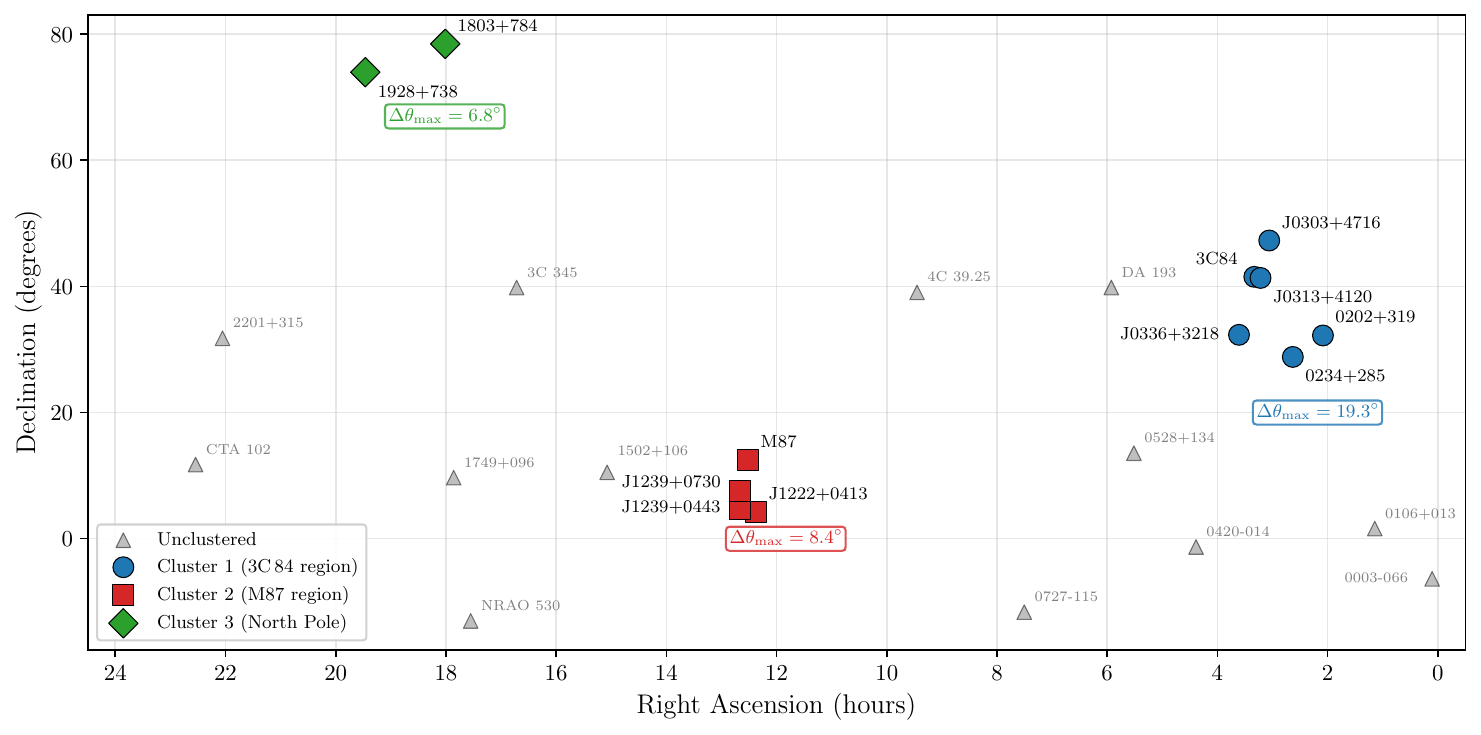}
    \caption{Sky distribution of the 25 sources of our data set in J2000 equatorial coordinates. Colored symbols indicate the three sky-region clusters used for the qualitative gain, D-term, and gain ratio comparisons (Sections~\ref{sec:gain_comparison}--\ref{sec:gain_ratio}); gray triangles mark the remaining unclustered sources. The maximum angular separation $\Delta\theta_{\max}$ within each cluster is annotated.}
    \label{fig:sky_cluster}
\end{figure*}

\begin{figure*}[t!]
    \centering
    \includegraphics[width=\textwidth]{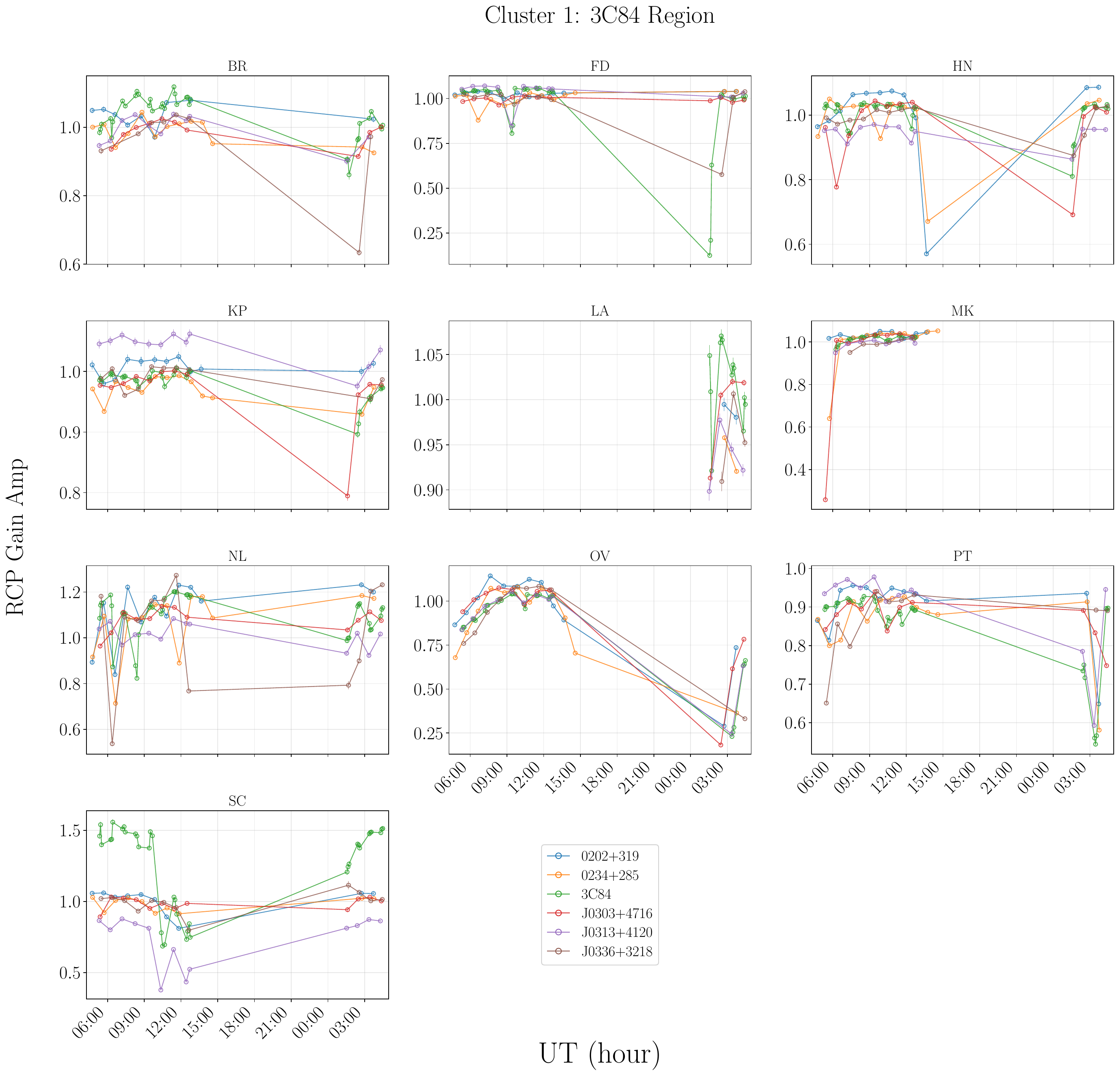}
    \caption{\texttt{Comrade} RCP gain amplitude for the Cluster~1 sources (3C\,84 region). Each subplot shows one VLBA station. Different colors represent different sources, with error bars indicating the posterior $1\sigma$ uncertainty. Although the solutions generally trace similar broad trends, SC exhibits pronounced source-dependent offsets, especially for 3C\,84 and J0313$+$4120. The corresponding CLEAN self-calibration solutions reproduce the same SC behavior (see the per-source comparisons in the online supplementary material), indicating that it is not specific to the \texttt{Comrade} reconstruction and instead likely reflects SC-specific residual amplitude-calibration or data systematics.}
    \label{fig:gain_cluster1}
\end{figure*}

\begin{figure*}[t!]
    \centering
    \includegraphics[width=\textwidth]{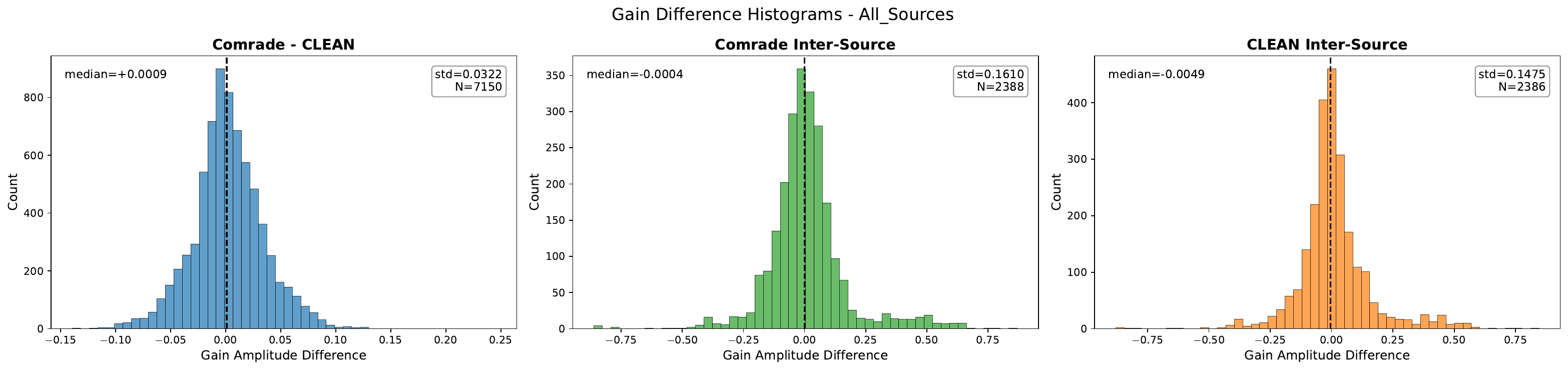}
    \caption{Gain amplitude difference histograms. \textit{Left:} Comrade$-$CLEAN gain amplitude difference for all 25 sources. \textit{Center and right:} \texttt{Comrade} and CLEAN inter-source gain spread, respectively, computed from pairwise differences between source pairs observed within 5\,min of each other and separated by $<0.5$\,h in right ascension and $<10^\circ$ on the sky (10 pairs from Clusters~1 and 2). The Comrade$-$CLEAN differences are centered near zero, confirming consistency. The inter-source spreads are comparable between the two methods; the pronounced SC offsets discussed in the text contribute to the high-difference tails of both distributions.}
    \label{fig:gain_diff_hist}
\end{figure*}

\begin{figure*}[t!]
    \centering
    \includegraphics[width=\textwidth, height=0.9\textheight, keepaspectratio]{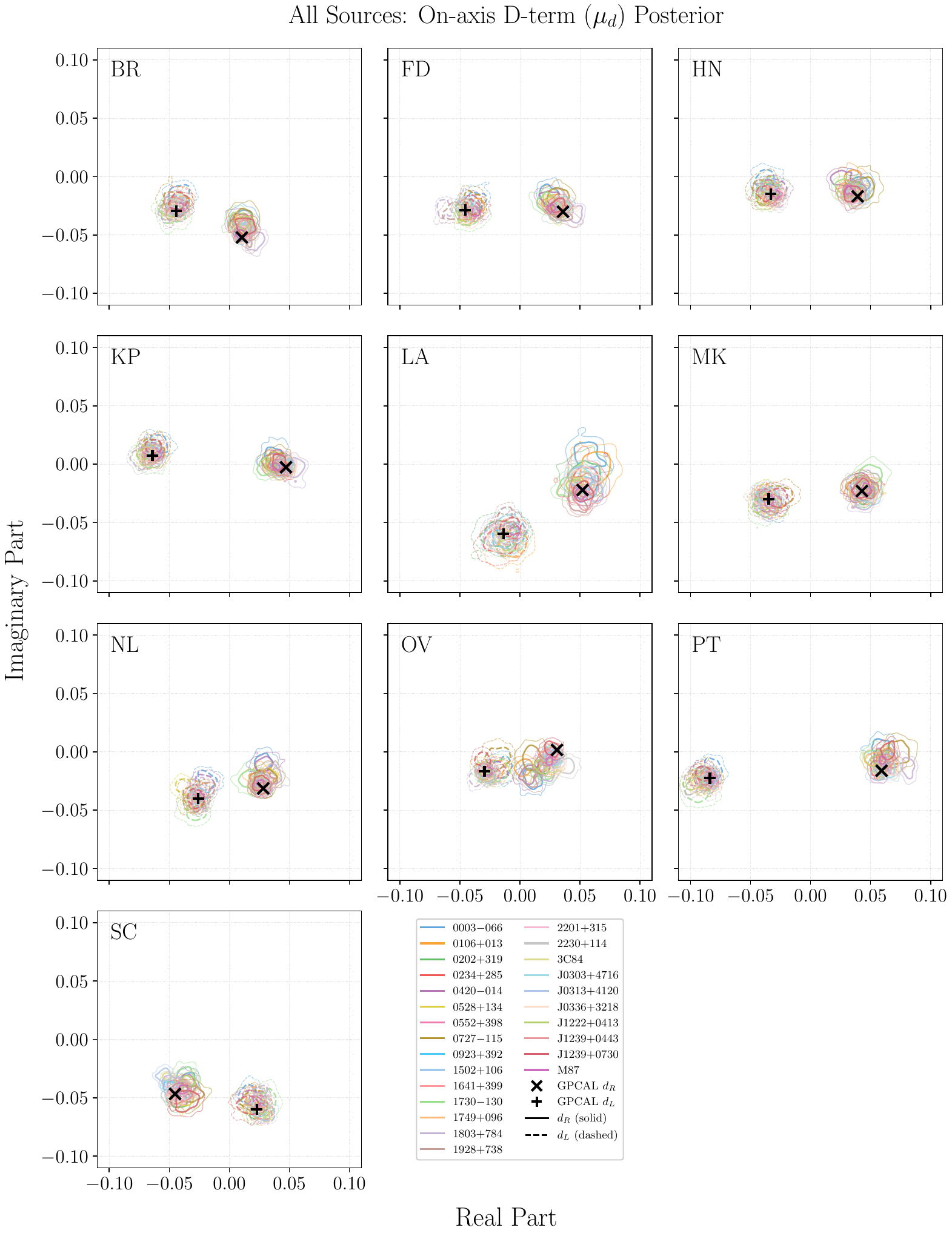}
    \caption{On-axis D-term ($\mu_d$) posterior distributions on the complex plane for all 25 sources combined. Each panel shows one VLBA antenna. Solid/dashed contours: \texttt{Comrade} posterior 68\% and 95\% credible regions for $d_R$/$d_L$ (colors indicate different sources). Black crosses/plus signs: \texttt{GPCAL} $d_R$/$d_L$ on-axis solutions. The \texttt{Comrade} posteriors are consistent with the \texttt{GPCAL} point estimates while revealing the full uncertainty structure.}
    \label{fig:dterm_posterior}
\end{figure*}

\begin{figure*}[t!]
    \centering
    \includegraphics[width=\textwidth]{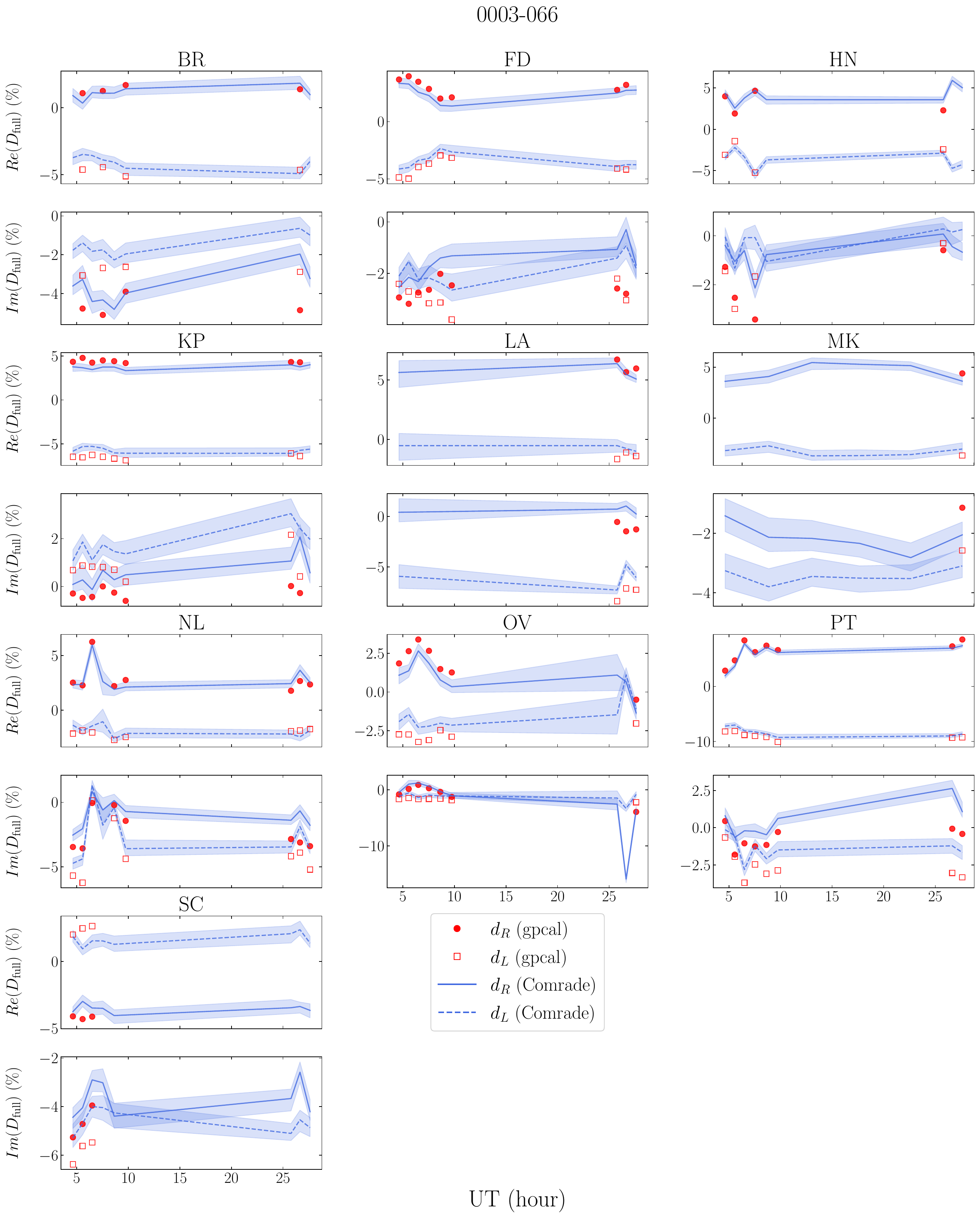}
    \caption{Full D-term (on-axis $\mu_d$ + off-axis $d(t)$) comparison between \texttt{GPCAL} and \texttt{Comrade} for 0003$-$066. Each antenna is shown in a separate panel with real (top) and imaginary (bottom) parts of the D-terms in percent. Red filled circles and open squares: \texttt{GPCAL} $d_R$ and $d_L$. Blue solid and dashed lines: \texttt{Comrade} posterior median $d_R$ and $d_L$ with $1\sigma$ uncertainty bands. Per-source D-term comparisons for the remaining 24 sources are available in the online supplementary materials.}
    \label{fig:dterm_persource}
\end{figure*}

\begin{figure*}[t!]
    \centering
    \includegraphics[width=\textwidth]{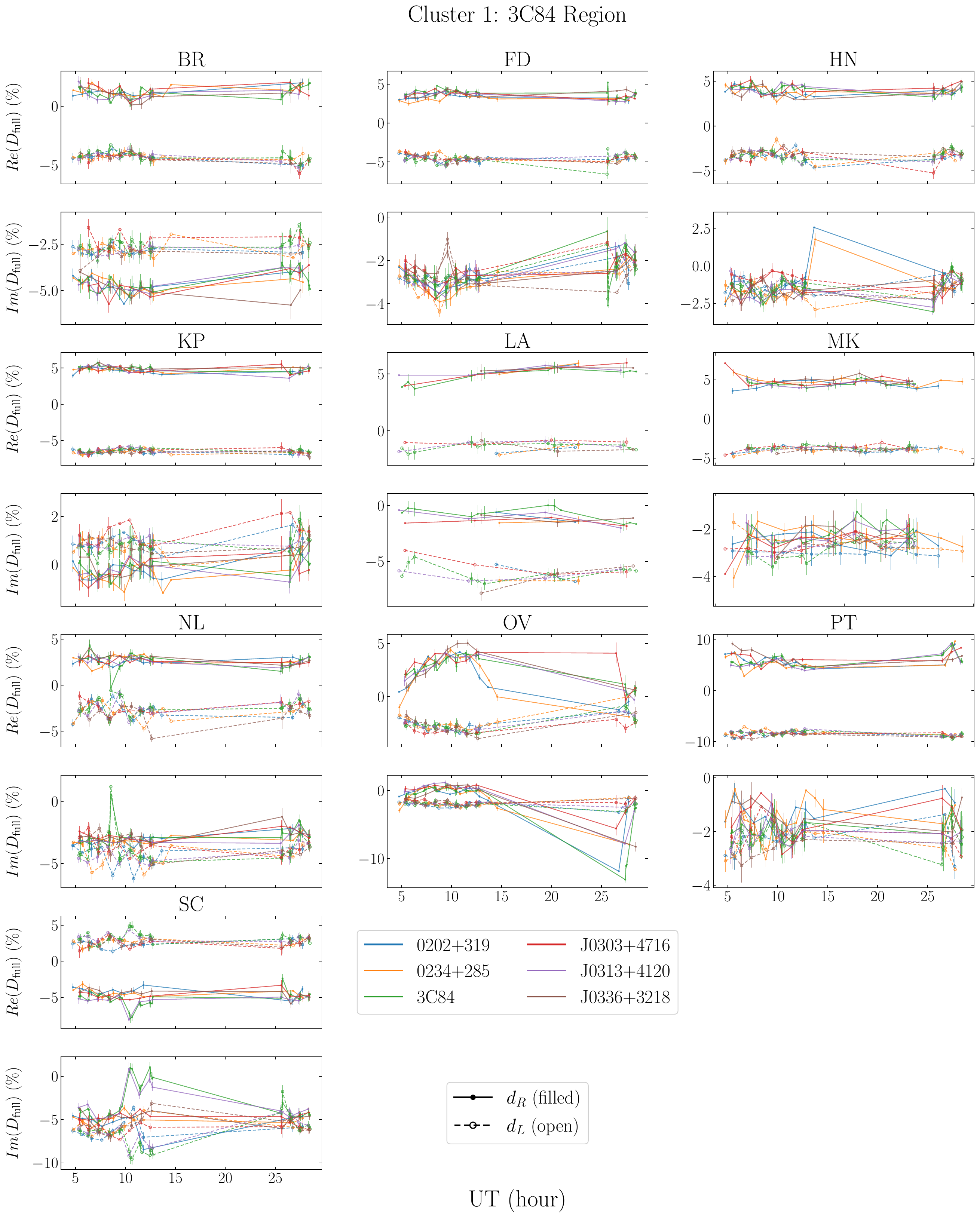}
    \caption{\texttt{Comrade}-only D-term comparison across sources within Cluster~1 (3C\,84 region). Each antenna panel shows the full D-term time series (on-axis $\mu_d$ + off-axis $d(t)$) for all sources in the cluster, with error bars indicating the posterior $1\sigma$ uncertainty. Sources within the same sky cluster show consistent D-term evolution, supporting the hierarchical time-dependent D-term model.}
    \label{fig:dterm_comrade_cluster1}
\end{figure*}

\begin{figure*}[t!]
    \centering
    \includegraphics[width=\textwidth]{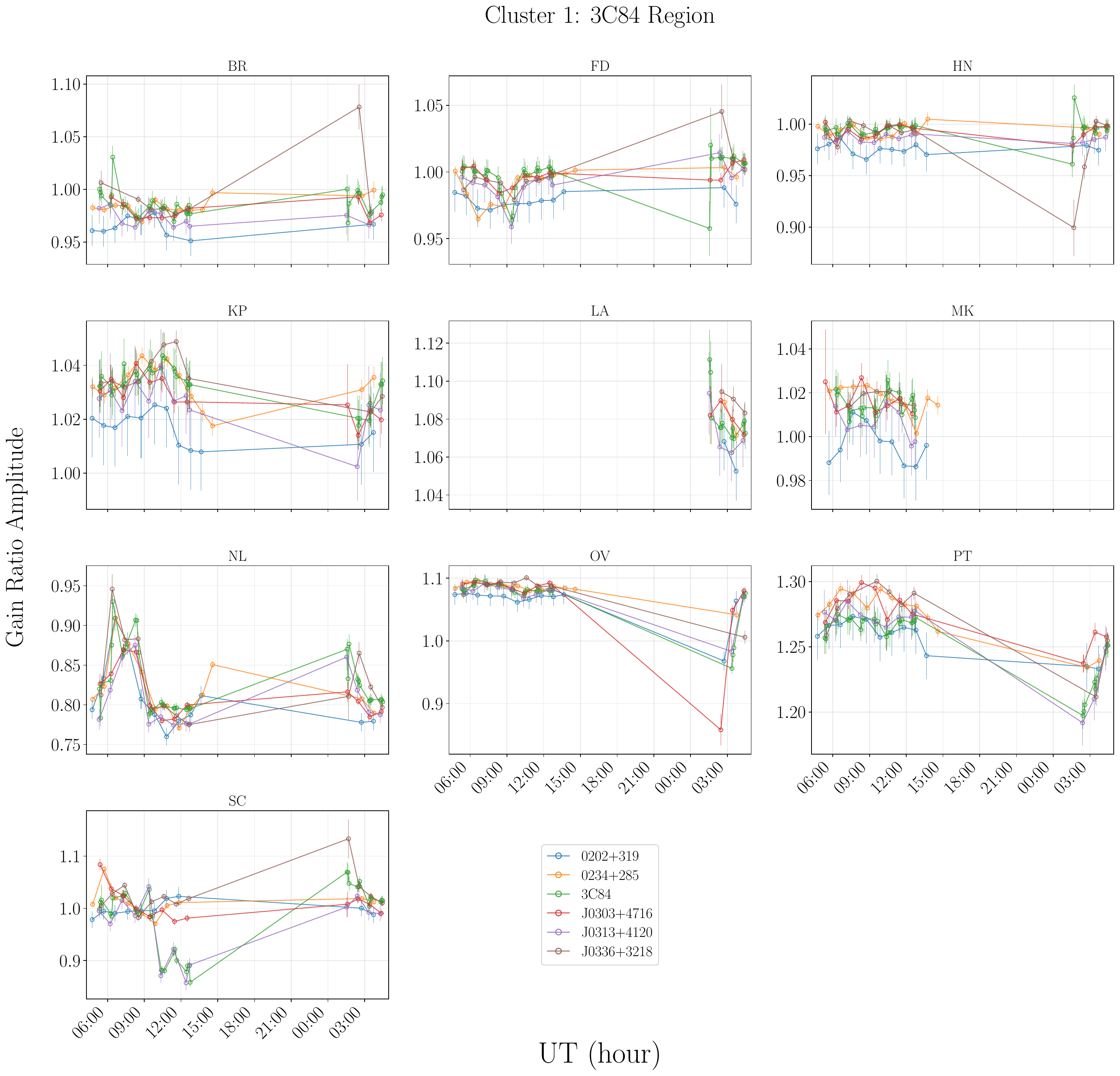}
    \caption{\texttt{Comrade} gain ratio amplitude ($|G_L/G_R|$) for the Cluster~1 sources (3C\,84 region). Each subplot shows one VLBA station. Different colors represent different sources, with error bars indicating the posterior $1\sigma$ uncertainty. Although the gain-ratio amplitudes appear broadly consistent across sources, the residual inter-source scatter remains large compared with the sub-percent calibration accuracy required to measure the typically weak Stokes~$V$ signals of AGN, reflecting the degeneracy between the R/L gain ratio and Stokes~$V$.}
    \label{fig:gain_ratio_amp_cluster1}
\end{figure*}

\begin{figure*}[t!]
    \centering
    \includegraphics[width=\textwidth]{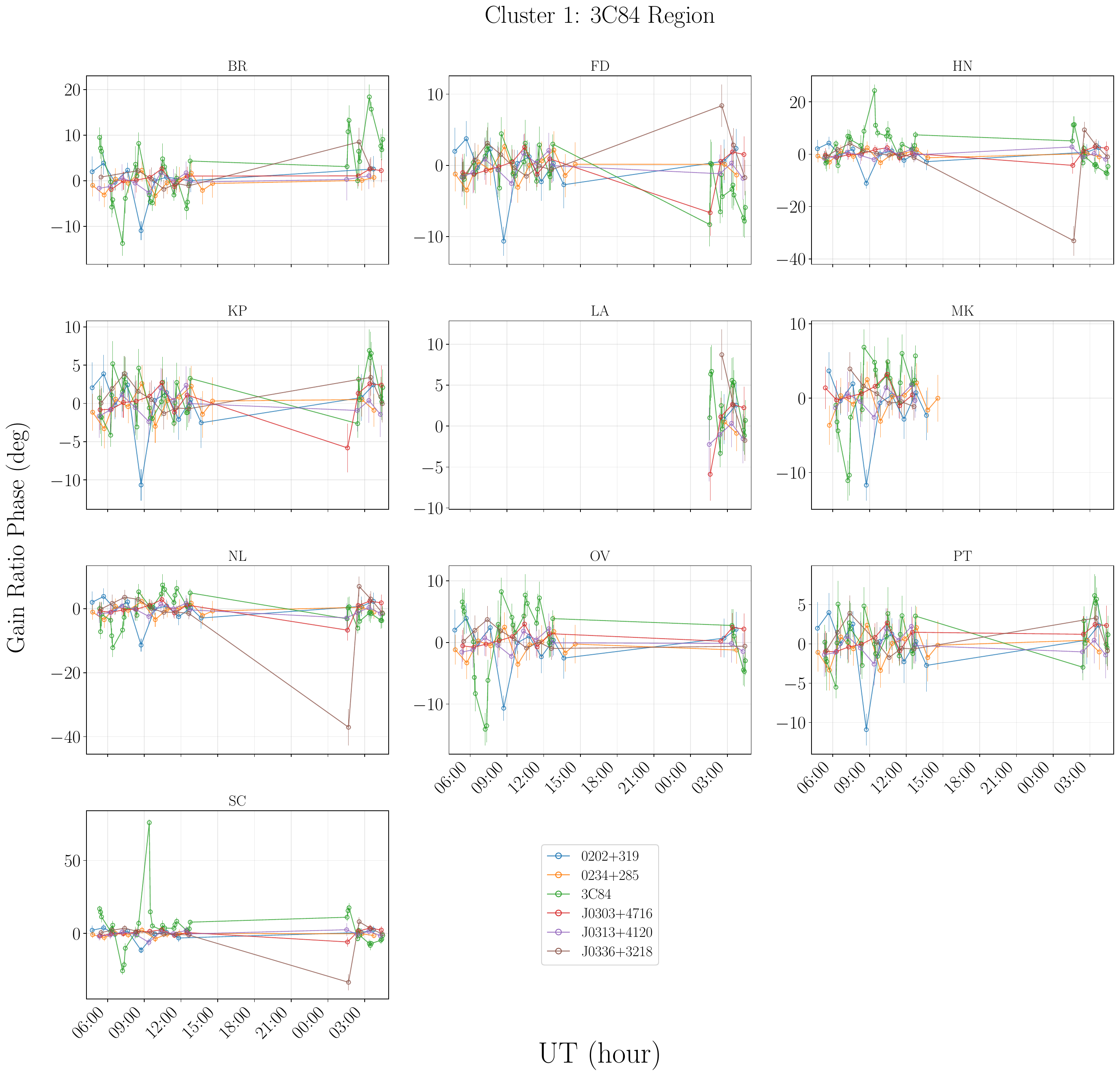}
    \caption{\texttt{Comrade} gain ratio phase ($\arg(G_L/G_R)$, in degrees) for the Cluster~1 sources (3C\,84 region). Symbols are as in Figure~\ref{fig:gain_ratio_amp_cluster1}.}
    \label{fig:gain_ratio_phase_cluster1}
\end{figure*}

\begin{figure*}[t!]
    \centering
    \includegraphics[width=\textwidth]{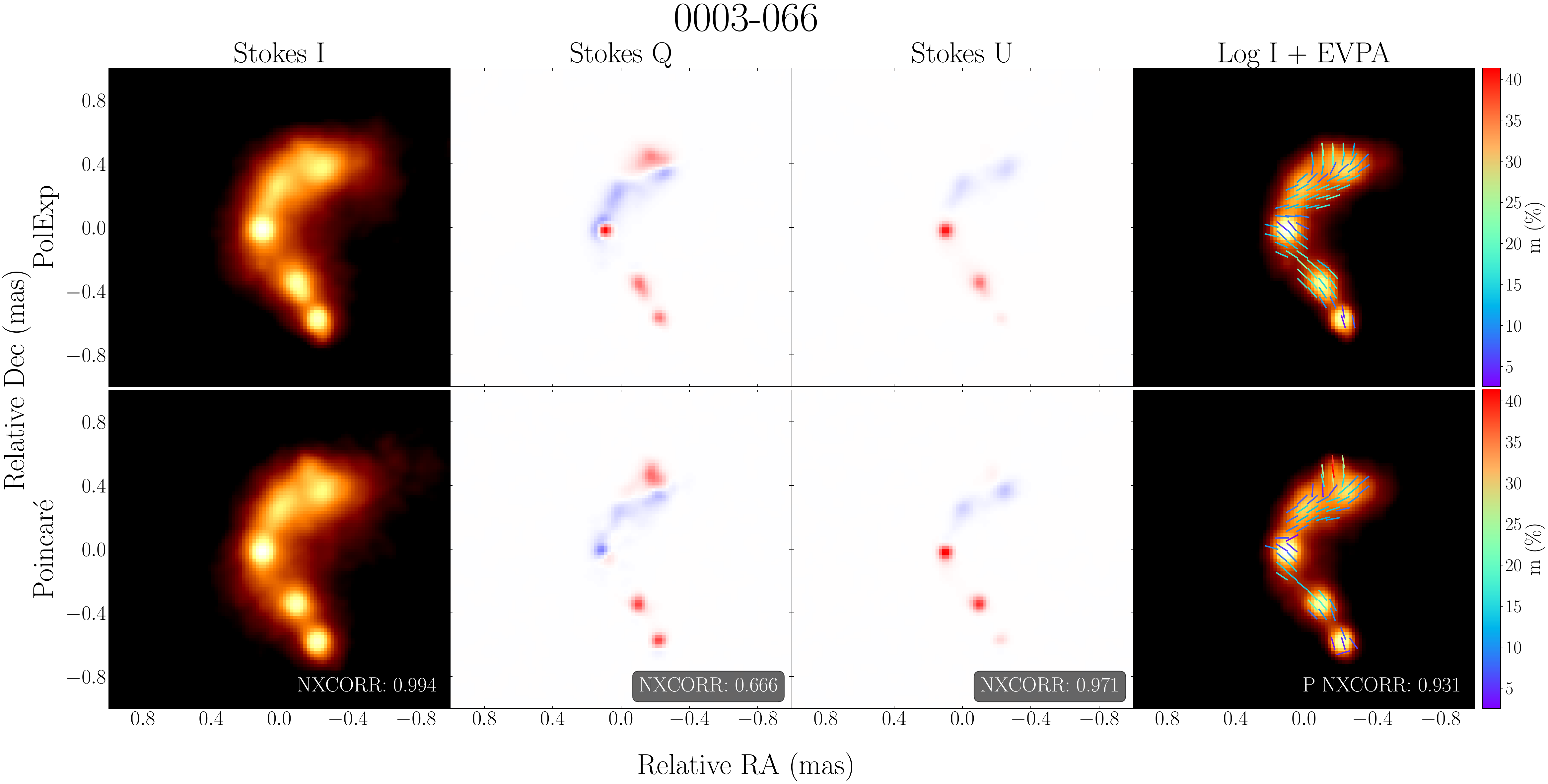}
    \caption{Direct comparison between PolExp (top row) and Poincar\'e (bottom row) polarization parameterizations for 0003$-$066 at native resolution. Columns from left to right: Stokes~$I$ (logarithmic scale), Stokes~$Q$, Stokes~$U$, and EVPA (logarithmic $I$ background with EVPA ticks colored by $m$\%). All NXCORR values are measured after a single mutual alignment of the Poincar\'e image to the PolExp frame via cross-correlation on Stokes~$I$, identical to the procedure used for the sample-wide comparison in Figure~\ref{fig:nxcorr_beam_frac}. The Stokes~$I$ maps are nearly identical (NXCORR $= 0.994$), while the $Q$ and $U$ maps show structural differences, particularly in low-SNR regions. The NXCORR values are annotated on the Poincar\'e panels.}
    \label{fig:polexp_vs_poincare}
\end{figure*}

\begin{figure*}[t!]
    \centering
    \begin{minipage}[t]{0.48\textwidth}
        \centering
        \includegraphics[width=\textwidth]{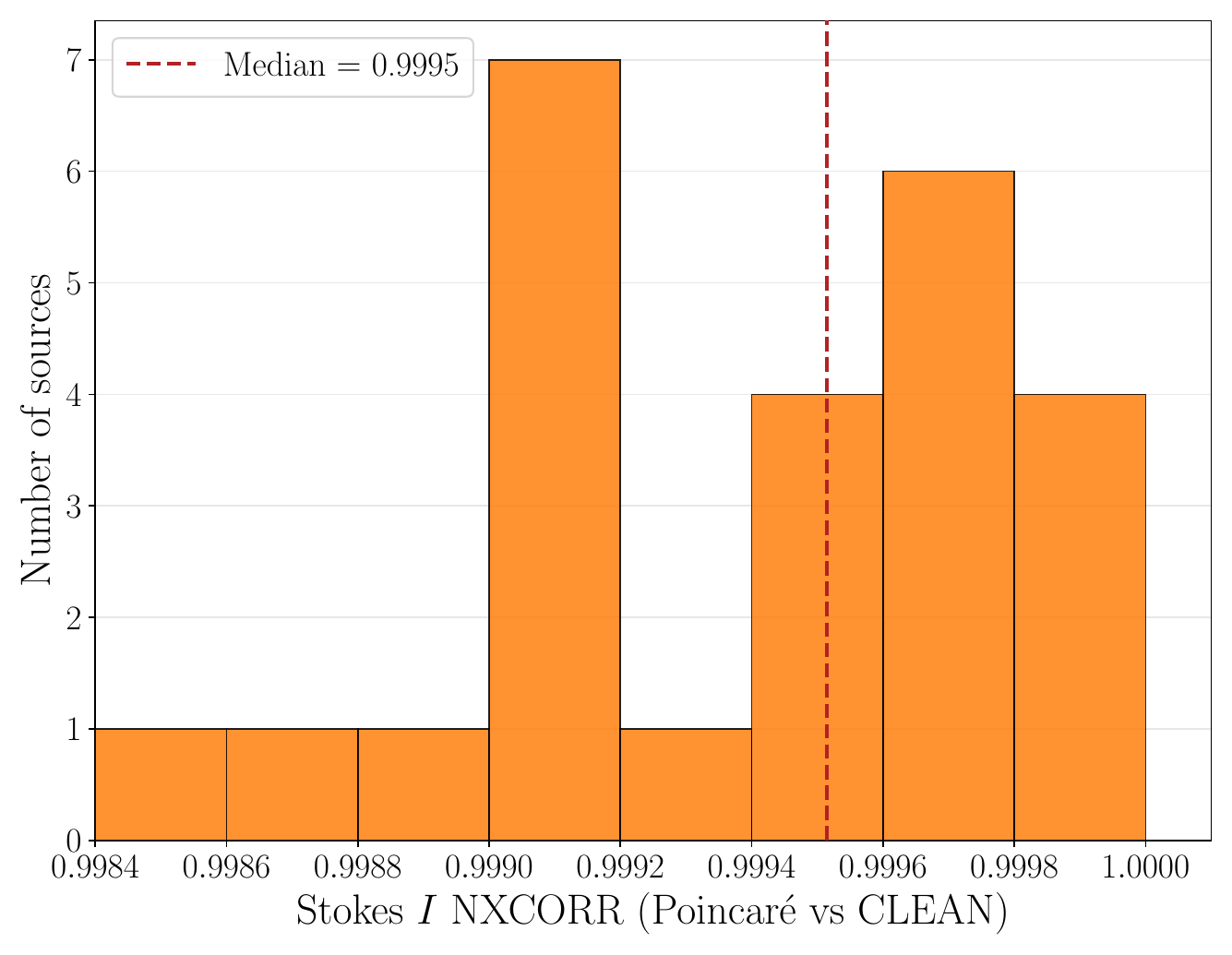}
    \end{minipage}\hfill
    \begin{minipage}[t]{0.48\textwidth}
        \centering
        \includegraphics[width=\textwidth]{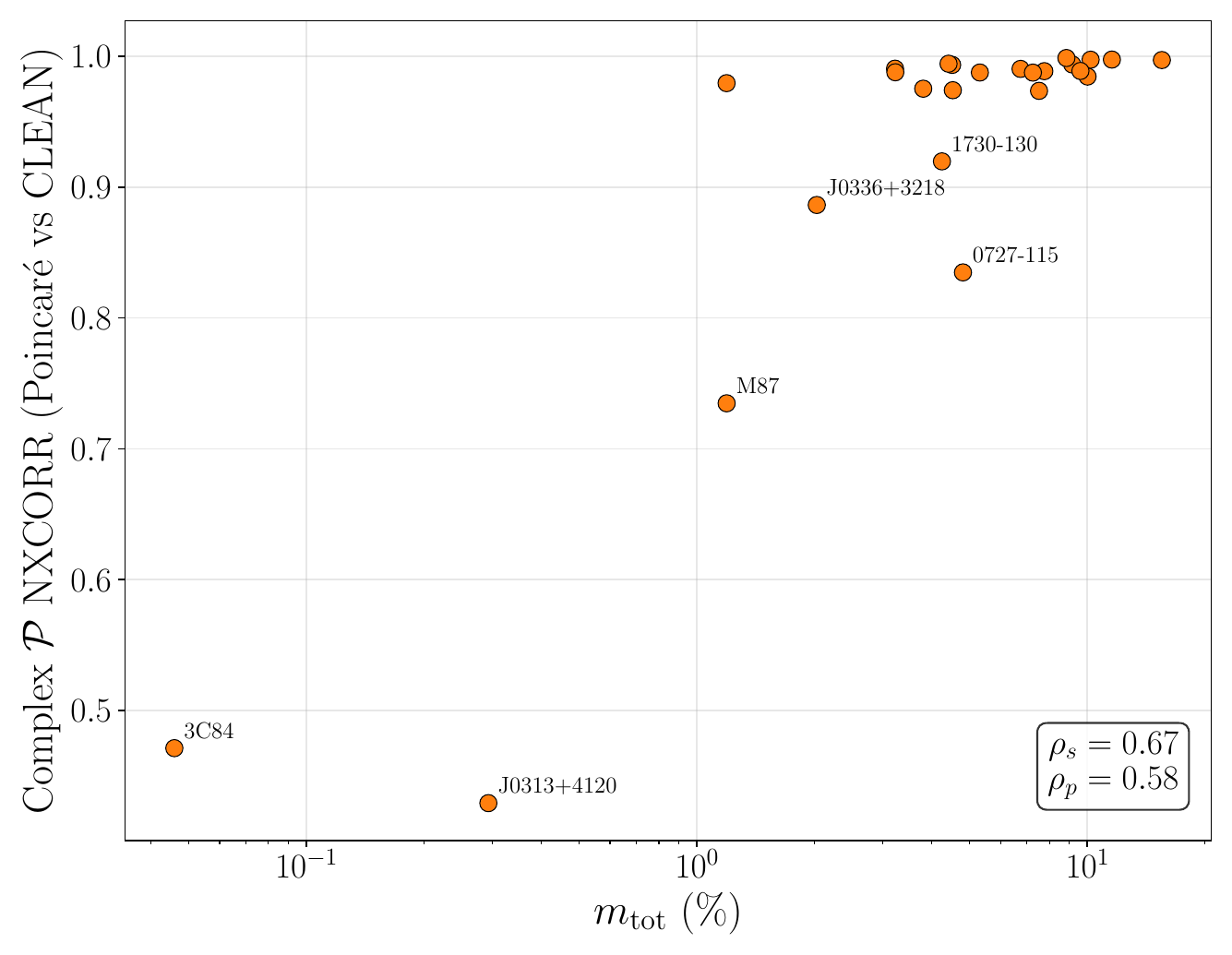}
    \end{minipage}
    \caption{\textit{Left:} Distribution of Stokes~$I$ NXCORR between the convolved Poincar\'{e} posterior mean and CLEAN images for all 25 sources (median 0.9995). \textit{Right:} Complex linear polarization NXCORR between Poincar\'{e} and CLEAN as a function of $m_{\rm tot}$ ($\rho_s = 0.67$); $m_{\rm tot}$ is measured from the native-resolution PolExp reconstruction and serves as a common x-axis across parameterizations. Compare with the PolExp results in Figure~\ref{fig:nxcorr_polexp_clean}: the two parameterizations achieve virtually identical agreement with CLEAN at beam resolution.}
    \label{fig:nxcorr_poincare_clean}
\end{figure*}

\begin{figure*}[t!]
    \centering
    \begin{minipage}[t]{0.48\textwidth}
        \centering
        \includegraphics[width=\textwidth]{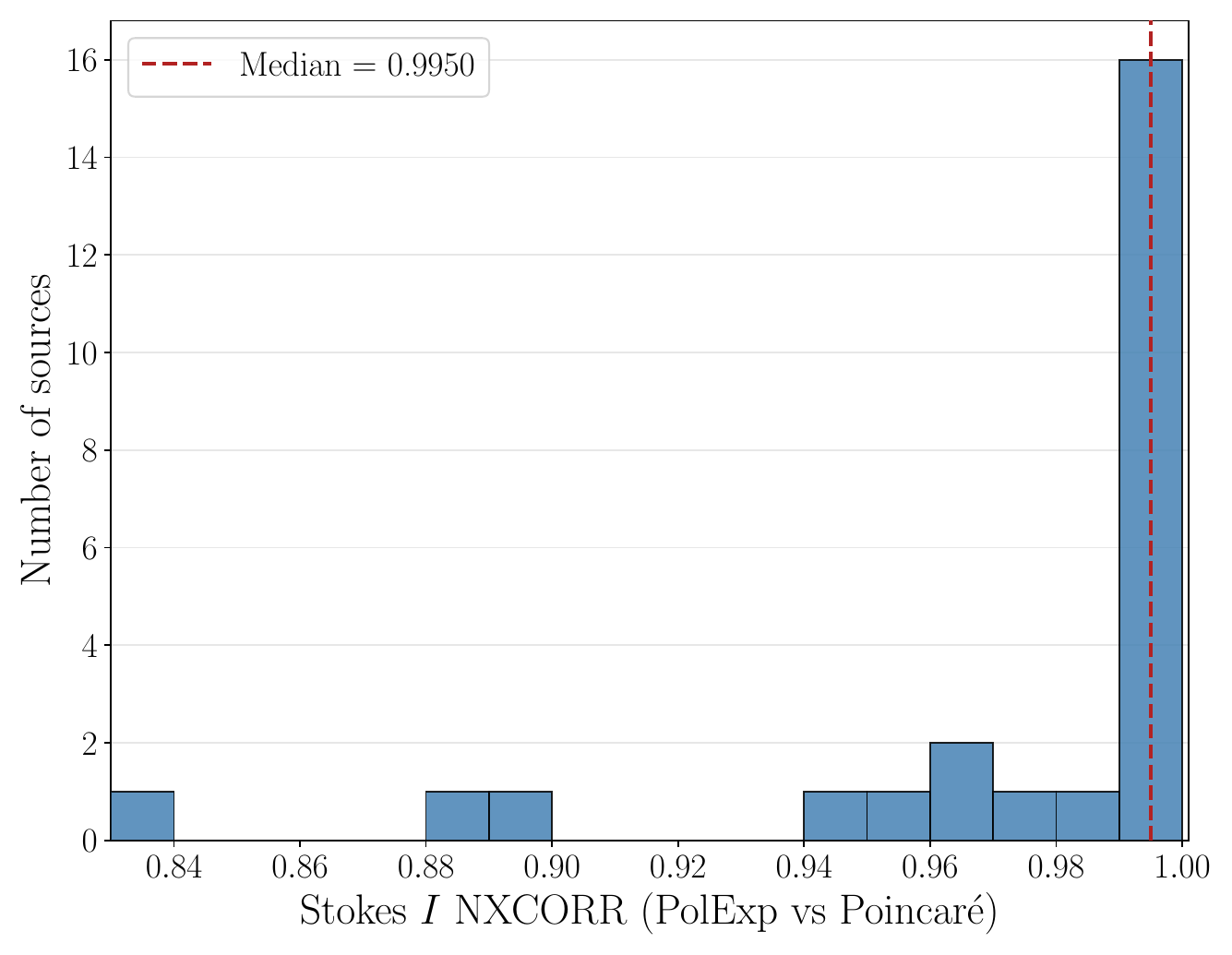}
    \end{minipage}\hfill
    \begin{minipage}[t]{0.48\textwidth}
        \centering
        \includegraphics[width=\textwidth]{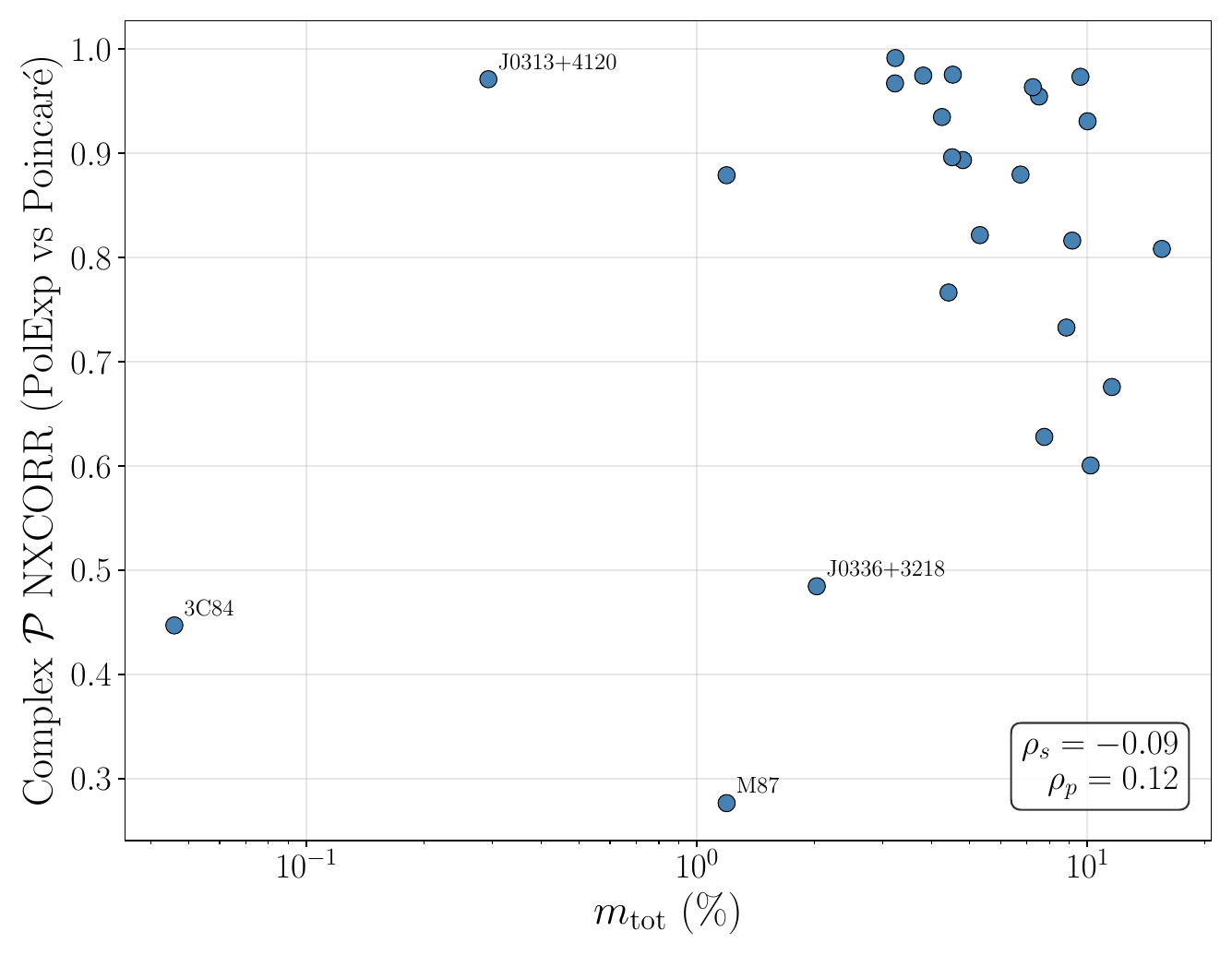}
    \end{minipage}
    \caption{\textit{Left:} Distribution of Stokes~$I$ NXCORR between the PolExp and Poincar\'{e} reconstructions for all 25 sources (median 0.995). The spread is significantly larger than the Comrade--CLEAN comparisons (Figure~\ref{fig:nxcorr_polexp_clean} and \ref{fig:nxcorr_poincare_clean}, all $> 0.998$), indicating that the inter-parameterization differences arise at super-resolution scales. \textit{Right:} Complex polarization NXCORR between PolExp and Poincar\'{e} as a function of $m_{\rm tot}$ ($\rho_s = -0.09$); $m_{\rm tot}$ is measured from the native-resolution PolExp reconstruction (common x-axis). Unlike the Comrade--CLEAN comparisons ($\rho_s = 0.66$--$0.67$), there is no significant correlation, suggesting that the differences are more plausibly associated with prior geometry than with calibration systematics.}
    \label{fig:nxcorr_pe_pc}
\end{figure*}

\begin{figure*}[t!]
    \centering
    \includegraphics[width=\textwidth]{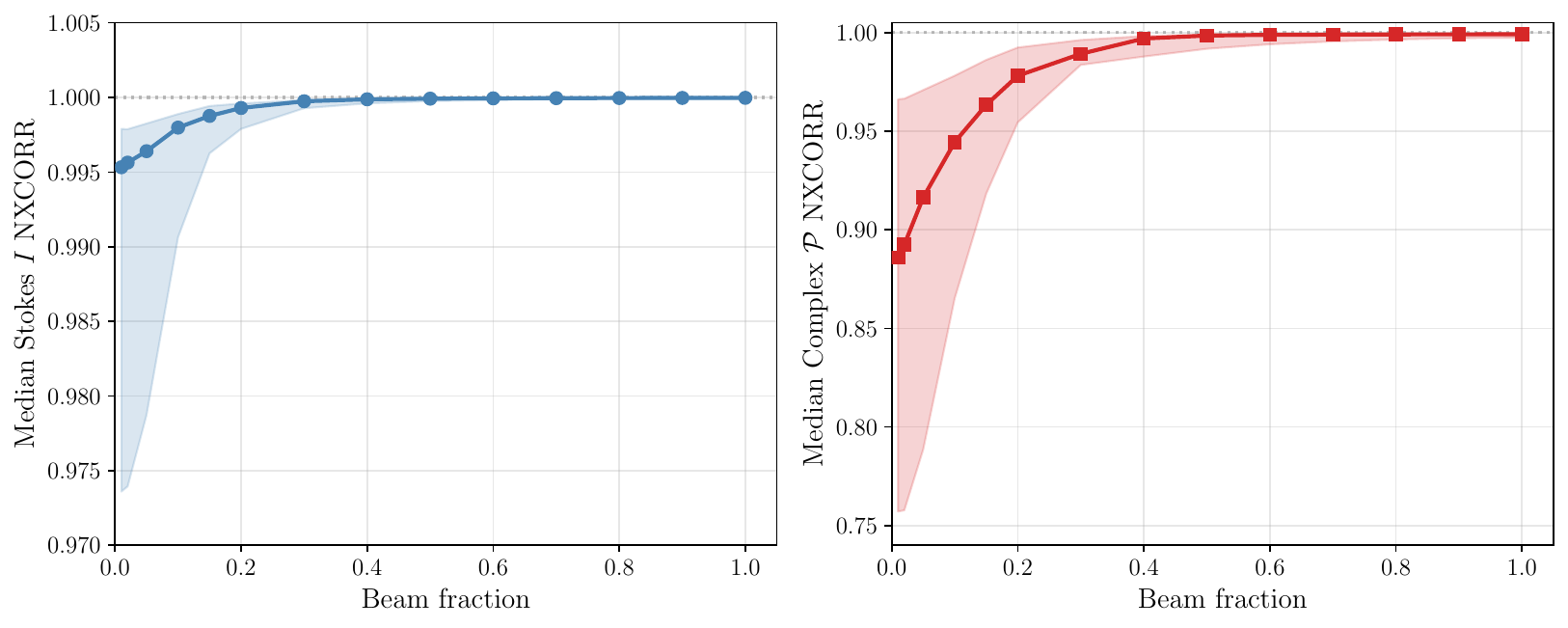}
    \caption{Median NXCORR between the PolExp and Poincar\'{e} reconstructions as a function of the convolving beam fraction $f$ (fraction of the CLEAN restoring beam). \textit{Left:} Stokes~$I$ NXCORR. \textit{Right:} Complex linear polarization NXCORR. Solid lines show the median over all 25 sources; shaded bands indicate the interquartile range (25th--75th percentile). For each source, both parameterizations were first aligned at native resolution, then convolved with an elliptical Gaussian kernel of size $f \times \theta_{\rm beam}$ applied independently to Stokes~$I$, $Q$, and $U$. The Stokes~$I$ agreement converges by $f \approx 0.15$, while the polarization agreement requires $f \approx 0.3$, indicating that polarimetric structures are more sensitive to the choice of parameterization at sub-beam scales.}
    \label{fig:nxcorr_beam_frac}
\end{figure*}

\subsubsection{Gain-phase-based image alignment}\label{sec:gain_alignment}

In Bayesian imaging with per-integration gain fitting, the absolute position of the reconstructed image is degenerate with the station-based gain phases. Although the source structure itself is well constrained by the data, \textit{individual} posterior samples may exhibit stochastic positional jitter from sample to sample, driven by the gain phase degrees of freedom. If left uncorrected, this jitter would broaden the posterior mean image and inflate the uncertainty estimates. We therefore implemented a gain-phase-based alignment procedure to register the individual posterior samples before averaging.

\paragraph{Phase residual extraction.}
For each posterior sample $s$, the RCP gain phase at station $j$ and integration $t$ is denoted $\phi_{j}^{(s)}(t)$. We first computed the circular mean phase across all posterior samples for each $(j,\,t)$ pair:
\begin{equation}\label{eq:circular_mean}
    \bar{\phi}_j(t) = \arg\!\left[\frac{1}{N_s}\sum_{s=1}^{N_s} e^{i\phi_j^{(s)}(t)}\right],
\end{equation}
where the sum runs over the post-burn-in samples. The residual phase for each sample was then defined as
\begin{equation}\label{eq:phase_resid}
    \Delta\phi_j^{(s)}(t) = \arg\!\left[e^{i(\phi_j^{(s)}(t) - \bar{\phi}_j(t))}\right],
\end{equation}
which isolates the sample-to-sample phase fluctuation from the mean gain solution.

\paragraph{Image shift model.}
A constant positional offset $(\Delta x,\, \Delta y)$ in the image plane introduces a linear phase gradient across the $(u,v)$ plane:
\begin{equation}\label{eq:phase_ramp}
    \Delta\phi_j^{\rm model}(t)
    = 2\pi\left[u_j^{\rm rel}(t)\,\Delta x
    + v_j^{\rm rel}(t)\,\Delta y\right],
\end{equation}
where the coordinates of station $j$ relative to the reference station are
\begin{equation}
    u_j^{\rm rel}(t)=u_j(t)-u_{j_{\rm ref}(t)}(t), \qquad
    v_j^{\rm rel}(t)=v_j(t)-v_{j_{\rm ref}(t)}(t).
\end{equation}
At each timestamp, the reference station was identified as the antenna whose RCP gain phase was constrained to zero by the sampler (i.e., $\phi_{j_{\rm ref}(t)}^{(s)}(t) \approx 0$ for all $s$). Because the reference station can change with time, we recomputed the relative $(u,v)$ coordinates at every timestamp before jointly fitting a single image shift to all available station--time measurements. Timestamps for which no reference station could be identified were excluded from the fit.

% A constant positional offset $(\Delta x,\, \Delta y)$ in the image plane introduces a linear phase gradient across the $(u,v)$ plane:
% \begin{equation}\label{eq:phase_ramp}
%     \Delta\phi_j^{\rm model} = 2\pi\left(u_j^{\rm rel}\,\Delta x + v_j^{\rm rel}\,\Delta y\right),
% \end{equation}
% where $(u_j^{\rm rel},\, v_j^{\rm rel})$ are the baseline coordinates of station $j$ relative to a reference station $j_{\rm ref}$:
% \begin{equation}
%     u_j^{\rm rel} = u_j - u_{j_{\rm ref}}, \qquad v_j^{\rm rel} = v_j - v_{j_{\rm ref}}.
% \end{equation}
% The reference station was chosen as the antenna whose gain phase is constrained to zero by the sampler (i.e., $\phi_{j_{\rm ref}}^{(s)} \approx 0$ for all $s$). Scans in which no reference station could be identified were excluded from the fit.

\paragraph{Shift estimation.}
For each posterior sample $s$, the image offset $(\Delta x^{(s)},\, \Delta y^{(s)})$ was estimated by minimizing the cosine loss function:
% \begin{equation}\label{eq:cosine_loss}
%     \mathcal{L}^{(s)}(\Delta x, \Delta y) = -\sum_{j \neq j_{\rm ref}} \cos\!\left[\Delta\phi_j^{(s)} - 2\pi\!\left(u_j^{\rm rel}\,\Delta x + v_j^{\rm rel}\,\Delta y\right)\right],
% \end{equation}
  \begin{equation}\label{eq:cosine_loss}
      \mathcal{L}^{(s)}(\Delta x,\Delta y)
      = -\sum_t\sum_{j\ne j_{\rm ref}(t)}
      \cos\!\left[
      \Delta\phi_j^{(s)}(t)
      -\Delta\phi_j^{\rm model}(t)
      \right].
  \end{equation}
using the Nelder--Mead simplex algorithm with initial guess $(\Delta x, \Delta y) = (0, 0)$. The cosine loss was chosen instead of a least-squares formulation to handle the $2\pi$ periodicity of phase residuals without requiring explicit unwrapping. The optimized offsets were converted from radians to pixel units via the known pixel scale.

\paragraph{Shift application.}
The estimated shifts were applied to all four Stokes channels ($I$, $Q$, $U$, $V$) of each posterior sample simultaneously using sub-pixel interpolation \citep[\texttt{scipy.ndimage.shift} with cubic spline interpolation;][]{Virtanen2020}. This ensures that the polarization structure remains self-consistent after alignment.

\subsubsection{Posterior summary images}\label{sec:posterior_summary}

Each posterior sample was evaluated on a spatial grid refined by a factor of two in each dimension relative to the model grid listed in Table~\ref{tab:sources}, yielding output images with half the pixel size at fixed field of view. This refinement does not introduce additional free parameters or change the inferred model complexity; it simply samples the same posterior sky model more finely in the image domain, since the \texttt{Comrade} sky model is represented continuously and can be evaluated on an arbitrary output grid. The refined grid provides a more faithful numerical representation of the continuous model for posterior summary products, visualization, and comparison with external images, but should not be interpreted as by itself increasing the physical resolving power of the reconstruction.

After the alignment step, a small number of post-burn-in samples with aberrant shift solutions were additionally excluded from all posterior summaries: a sample was flagged if its fitted radial shift deviated from the sample median by more than five MAD-based modified $z$-scores. This criterion acts on the shift solutions alone and is independent of the reconstructed image morphology, so it cannot preferentially remove samples that explore different but statistically valid source structures. Across the 50 reconstructions, samples were removed in only six cases (21 of the 6{,}000 post-burn-in samples in total, at most 11 for a single reconstruction); in four of the six cases the flagged samples lie immediately after the adopted burn-in boundary. After burn-in removal, gain-phase alignment, and this outlier rejection, the representative Stokes images were computed as the unweighted arithmetic mean over the remaining $N_s$ posterior samples on this refined grid:
\begin{equation}\label{eq:mean_image}
    \bar{S}_{ij} = \frac{1}{N_s}\sum_{s=1}^{N_s} S_{ij}^{(s)}, \qquad S \in \{I,\, Q,\, U,\, V\},
\end{equation}
with the per-pixel posterior standard deviation computed analogously. The linearly polarized intensity, fractional linear polarization, and electric vector position angle (EVPA) were derived from the mean Stokes maps as
\begin{align}
    P_{ij} &= \sqrt{\bar{Q}_{ij}^2 + \bar{U}_{ij}^2}, \label{eq:linpol}\\
    m_{ij} &= P_{ij} \,/\, \bar{I}_{ij}, \label{eq:pfrac}\\
    \chi_{ij} &= \tfrac{1}{2}\arctan\!\left(\bar{U}_{ij},\, \bar{Q}_{ij}\right). \label{eq:evpa}
\end{align}
These derived maps ($P$, $m$, $\chi$) serve as representative display products; for quantitative inference, uncertainties should be propagated directly from the individual posterior samples rather than from the mean Stokes maps (see Section~\ref{sec:uncertainty}).

\subsection{Comparison with the conventional pipeline}\label{sec:clean_comparison}

To validate the \texttt{Comrade} reconstructions, we compared the posterior mean images against CLEAN images produced through standard self-calibration and imaging with \texttt{Difmap} \citep{Shepherd1997}. For a fair comparison, the native-grid \texttt{Comrade} images were zero-padded, regridded to the CLEAN field of view and pixel grid, and then convolved with the CLEAN restoring beam using the \texttt{eht-imaging} library \citep{Chael2016, Chael2018}. The optimal positional registration between the two images was determined using the \texttt{find\_shift} function in \texttt{eht-imaging} \citep{Chael2018}, which maximizes the normalized cross-correlation on a $2\times$-padded grid via FFT-based correlation of the Stokes~$I$ maps. The resulting shift was then applied identically to the Stokes~$Q$ and $U$ maps to maintain polarimetric consistency. The \texttt{Comrade}--CLEAN NXCORR values reported in this paper were computed after this alignment step.

We quantified the agreement using the normalized cross-correlation coefficient \citep[following][]{EHTC2019d}:
\begin{equation}\label{eq:nxcorr}
    \rho_{\rm NX} = \frac{\sum_{ij}\left(a_{1,ij} - \langle a_1\rangle\right)\left(a_{2,ij} - \langle a_2\rangle\right)}{\sqrt{\sum_{ij}\left(a_{1,ij} - \langle a_1\rangle\right)^2 \;\sum_{ij}\left(a_{2,ij} - \langle a_2\rangle\right)^2}},
\end{equation}
applied independently to each Stokes parameter. For the complex linear polarization $\mathcal{P} = Q + iU$, we computed the complex cross-correlation without mean subtraction:
\begin{equation}\label{eq:complex_nxcorr}
    \rho_{\mathcal{P}} = \frac{\mathrm{Re}\!\left[\sum_{ij}\mathcal{P}_{1,ij}\,\mathcal{P}_{2,ij}^{*}\right]}{\sqrt{\sum_{ij}|\mathcal{P}_{1,ij}|^2 \;\sum_{ij}|\mathcal{P}_{2,ij}|^2}},
\end{equation}
following the convention of \citet{EHTC2021a}. For the \texttt{Comrade}--CLEAN comparisons, the $2\times$-padded correlation grid was used only to determine the positional shift. After applying this shift, the final Stokes~$I$, $Q$, $U$, and complex-polarization NXCORR values were all evaluated on the same unpadded CLEAN image grid. This registration is specific to each \texttt{Comrade}--CLEAN comparison; the direct PolExp--Poincar\'{e} comparison instead uses a separate mutual alignment of the two native-resolution reconstructions, as described in Section~\ref{sec:polexp_vs_poincare}.

Throughout this paper, unless explicitly stated otherwise, all \texttt{Comrade} and CLEAN images are displayed above a common source-specific intensity threshold determined as follows. The Stokes~$I$ noise level~$\sigma_I$ is estimated from the median absolute deviation (MAD) of the off-source edge pixels (50-pixel margin) in the CLEAN image, converted to an equivalent Gaussian standard deviation via $\sigma_I = \mathrm{MAD}/0.6745$. The CLEAN-level threshold is then $\tau_{I,\mathrm{CLEAN}} = 3\,\sigma_I$. For a consistent visual comparison, we adopt a dynamic-range (DR) matching approach. This matches the displayed intensity range of each reconstruction relative to its peak; it does not impose a posterior significance criterion (Section~\ref{sec:uncertainty}). The convolved \texttt{Comrade} and CLEAN images share the threshold $\tau_{I,\mathrm{CLEAN}}$ directly, while the native-resolution \texttt{Comrade} threshold is scaled by the peak intensity ratio between the original \texttt{Comrade} image and the CLEAN image, $\tau_I = I_{\rm peak}^{\rm orig}\,\tau_{I,\mathrm{CLEAN}}\,/\,I_{\rm peak}^{\rm CLEAN}$, so that the native-resolution \texttt{Comrade} image and the CLEAN image share the same dynamic range. The polarization threshold is determined independently from the Stokes~$Q$ and $U$ noise levels: $\sigma_P = \sqrt{\sigma_Q^2 + \sigma_U^2}$, where $\sigma_Q$ and $\sigma_U$ are each estimated via MAD on the corresponding CLEAN images. The CLEAN-level polarization threshold is $\tau_{P,\mathrm{CLEAN}} = 3\,\sigma_P$, and the native-resolution threshold is obtained by the same DR-matching procedure, $\tau_P = P_{\rm peak}^{\rm orig}\,\tau_{P,\mathrm{CLEAN}}\,/\,P_{\rm peak}^{\rm CLEAN}$. These display thresholds use the CLEAN-image peak as the DR-matching reference. For J1239+0443 in Figure~\ref{fig:mosaic}, only the lower bound of the native-resolution Stokes~$I$ background color scale is lowered to the DR-matched equivalent of $0.5\,\sigma_I$ to reveal the low-surface-brightness jet; EVPA ticks are still plotted only where $I>\tau_I$ and $P>\tau_P$, as in the other panels.

% ============================================================
\section{Results and Discussion}\label{sec:results}

Unless stated otherwise, all results presented in this section are based on the PolExp polarization parameterization, which we adopted as the default for the 25-source production campaign owing to its stability in the weakly polarized regime, where the Poincar\'{e}-sphere representation can become prior-sensitive due to the ill-defined polarization direction at low polarization fractions. The results from both parameterizations are compared and their differences are discussed in Section~\ref{sec:polexp_vs_poincare}.

\subsection{Chain convergence and data fit quality}\label{sec:convergence}

Figure~\ref{fig:chain_convergence} shows representative convergence diagnostics for 0003$-$066. The per-product $\chi^2$ traces stabilize well within the adopted 80-sample burn-in interval (gray shaded region). Defining the plotted normalized statistic as $\chi^2/(2N_{\rm vis})$, the parallel-hand products yield values of approximately 0.7 (RR: 0.78; LL: 0.65), while the cross-hand products yield values close to unity (RL: 1.08; LR: 1.05). The slightly sub-unity values for the parallel hands may indicate mildly overestimated visibility uncertainties. The representative instrumental-parameter traces shown in the lower panels---gain amplitudes, gain ratios, and D-term components---are stationary after burn-in and show no systematic post-burn-in trends.

Figure~\ref{fig:vis_amp_phase} presents the model and data visibilities for 0003$-$066 as a function of projected baseline length. The parallel-hand products (RR, LL) are primarily sensitive to the Stokes~$I$ emission structure and the complex antenna gains $G_R$, $G_L$; their amplitudes trace the expected decline with baseline length characteristic of resolved source structure. The cross-hand products (RL, LR), on the other hand, are sensitive to the linearly polarized emission as well as the instrumental polarization leakage (D-terms) and gains. Their amplitudes are typically $\sim$1--10\% of the parallel-hand level, reflecting both the fractional polarization of the source and the instrumental polarization leakage. Overall, the \texttt{Comrade} model (red) follows the amplitude and phase distributions of the data (gray) across all four products.

The very low RR and LL amplitudes in Figure~\ref{fig:vis_amp_phase}, including those at short baselines, occur predominantly on baselines involving NL and OV during scans with strongly suppressed antenna gains (Figure~\ref{fig:gain_persource}). We did not flag these data for the reasons described in Section~\ref{sec:obs}. Because the plotted model includes the inferred gains and D-terms, a successful fit reproduces the low amplitudes in the input data.

Figure~\ref{fig:vis_residual} shows the normalized residual distributions $(d - m)/\sigma$ for the same source, where $d$, $m$, and $\sigma$ are the data, model, and noise uncertainty for each visibility, respectively. For all four products, the residuals are consistent with the expected $\mathcal{N}(0,1)$ distribution.

\subsection{Gain phase alignment}\label{sec:alignment_results}

As described in Section~\ref{sec:gain_alignment}, the absolute position of the reconstructed image is degenerate with the station-based gain phases. Figure~\ref{fig:shift_comparison} illustrates this effect for 1502$+$106, the source with the largest inter-sample positional variation among the reconstructions. The two posterior samples with the greatest mutual separation differ by 14.8 pixels (0.30\,mas, almost entirely in RA), corresponding to $\sim$25\% of the synthesized beam at 15\,GHz. Without the gain-phase-based alignment, averaging these shifted samples would produce a blurred posterior mean image with artificially reduced peak flux density and inflated uncertainty. Across our full sample of 25 sources, the mean inter-sample shift ranges from 0.1\,pixels to 3.4\,pixels, with a sample-averaged value of 1.3\,pixels. Figure~\ref{fig:gainphase_fit} validates the underlying shift model: for a representative post-burn-in posterior sample of 1502$+$106, the per-station residual gain phases (Eq.~\ref{eq:phase_resid}) closely follow the fitted phase-ramp model of Eq.~\ref{eq:phase_ramp}, confirming that the sample-to-sample gain-phase fluctuations are dominated by a coherent image shift rather than by station-specific noise. An animation showing the corresponding fits for all post-burn-in posterior samples is available as online material.

\subsection{Imaging results}\label{sec:imaging_results}

Figure~\ref{fig:highlight} highlights the M87 reconstruction. The native-resolution posterior mean shows an edge-brightened jet structure emerging from a compact core, as well as edge-brightened counterjet emission. These features are qualitatively consistent with previous higher-frequency, higher-resolution VLBA observations at 43\,GHz \citep{Walker2018} and VLBI observations at 86\,GHz \citep{Hada2016, Lu2023}. The linear polarization is largely undetected in the core and inner jet regions, but prominent polarized emission appears at $\sim$20\,mas from the core. This is also consistent with earlier cm-VLBI polarimetric studies that reported weak polarization near the core and significant linear polarization emission in the $\sim$20\,mas region \citep{ZT2002, Park2019b, Park2021c, Nikonov2023, Park2026}. When the \texttt{Comrade} image is convolved with the CLEAN restoring beam, it becomes nearly identical to the CLEAN Stokes~$I$ image (NXCORR $= 0.9991$), and the polarimetric features --- weak core/inner-jet polarization and prominent linear polarization at $\sim$20\,mas --- are also closely reproduced. This agreement establishes consistency at the CLEAN resolution. The native-resolution posterior mean contains finer structure that is smoothed out by beam convolution; the confidence in a particular feature must be evaluated from the posterior samples as described in Section~\ref{sec:uncertainty}.

Figure~\ref{fig:mosaic} shows the linear polarization images for the remaining 24 sources, illustrating the diversity of morphologies in our sample: compact cores (e.g., 0528$+$134, J1239$+$0443), bent jets (e.g., 0003$-$066), and complex extended structures (e.g., 3C\,84, 1928$+$738). The \texttt{Comrade} pipeline successfully reconstructs a wide range of polarimetric structures, from compact linearly polarized sources (e.g., 0528$+$134, J1239$+$0443) to weakly polarized sources (e.g., 3C\,84) and sources with complex, spatially extended polarization patterns (e.g., 4C\,39.25, 1928$+$738). Fainter extended emission can be hidden by the adopted fields of view and DR-matched display thresholds (Appendix~\ref{sec:appendix_extended_emission}).

\subsection{Posterior uncertainty quantification}\label{sec:uncertainty}

A key advantage of Bayesian imaging is that the posterior samples provide uncertainty estimates not only for the pixel brightnesses but also for any quantity measured from the images. Figure~\ref{fig:snr_map} presents the posterior mean ($\mu$), standard deviation ($\sigma$), and signal-to-noise ratio ($|\mu|/\sigma$) maps for 0003$-$066 across Stokes~$I$, $Q$, $U$, and linear polarization $P$. The standard deviation of $P$ was computed directly from the individual posterior samples (i.e., $P^{(s)} = \sqrt{Q^{(s)2} + U^{(s)2}}$ for each sample $s$), without assuming independence between $Q$ and $U$.

The per-pixel SNR reaches $\sim$5 for Stokes~$I$ and $\sim$2--3 for Stokes~$Q$ and $U$ in the brightest regions. These ratios describe the local brightness relative to its posterior uncertainty for this source and are not automatically Gaussian detection significances, particularly for the positive and asymmetric intensity distributions of Stokes~$I$. We note that neighboring pixel brightnesses vary together because they share the same visibility constraints and image prior, and these correlations are not shown in the per-pixel maps. A low per-pixel SNR alone therefore does not determine how well a quantity such as a jet width or a separation between components is constrained; conversely, a well-constrained integrated flux does not establish that the internal structure of the emission is accurately known.

\paragraph{Interpreting the posterior mean.}
The mean image represents the expected brightness at each position and is the best single estimate of the source in the least-squares sense, i.e., its squared difference from the true brightness, averaged over the posterior, is smaller than that of any other image. However, the mean image does not necessarily resemble a typical posterior sample in every morphological detail, and fine structures that vary substantially among the aligned samples remain uncertain even when they appear sharp in the mean image. Drawing more posterior samples reduces only the numerical error in estimating the mean, not the posterior uncertainty itself, since the samples do not constitute additional observations. Under the adopted model, the root-mean-square (rms) error of the mean image is the posterior standard deviation, whereas that of a single sample is larger by a factor of $\sqrt{2}$ (Appendix~\ref{sec:appendix_mean_error}).

\paragraph{Resolution and model dependence.}
A component smaller than the nominal resolution can still affect the measured visibilities, and with sufficient SNR and an appropriate source model its size can be constrained below that scale \citep[e.g.,][]{Lobanov2005}. This does not imply that spatial frequencies beyond the longest baseline are measured or that an arbitrary image is uniquely determined at those scales; the inferred fine structure depends on both the data and the adopted image and calibration priors \citep{Tiede2026}. The output pixel size ($0.02$\,mas for 0003$-$066, compared with a CLEAN beam of $\sim$1.1\,$\times$\,0.6\,mas) reflects how finely the model is represented rather than an independently established resolving power (Section~\ref{sec:posterior_summary}).

\paragraph{Measuring a structural quantity.}
For quantitative structural measurements, the relevant uncertainty is the variation of the measured quantity among the posterior images. For example, a transverse jet width can be evaluated in three steps:
\begin{enumerate}
    \item The measurement is first defined by specifying the location and orientation of the transverse cut, the region used to extract the intensity profile, and a consistent definition of width (e.g., the full width at half maximum, when appropriate for the profile).
    \item The width is then measured in each aligned posterior image with the same prescription, yielding a set of widths, one from each sample, rather than a single width measured from the mean image.
    \item Finally, the distribution of these widths is examined. Its spread describes the uncertainty in the width under the adopted model, and a median with a stated posterior credible interval can summarize a single-peaked distribution. If several distinct widths are allowed, or if the feature is absent or its width is undefined in some samples, this ambiguity and the fraction of such samples should be reported rather than silently discarded.
\end{enumerate}
Because each measurement uses a complete posterior image, this procedure retains the effects of correlated changes in pixel brightness. The same approach applies to other structural quantities, such as the contrast between the jet edges and center or the separation of two components; for such nonlinear measurements, the value measured from the mean image generally differs from the mean of the values measured from the samples. If smoothing or binning is appropriate, the same operation should be applied to every sample before the measurement, and any improvement in precision should be assessed together with the loss of spatial detail. Similar sample-based measurements have been used in previous Bayesian VLBI imaging studies, for example, to estimate ring properties such as the diameter and width of M87* from EHT data \citep{EHTC2024_M87shadow, EHTC2025_M87pol, Tiede2026} and of the ring in M87 from \texttt{resolve} reconstructions of GMVA$+$ALMA data at 86\,GHz \citep{Kim2025_M87_86GHz}.

In Appendix~\ref{sec:appendix_transverse}, we illustrate this procedure using the M87 images in Figure~\ref{fig:highlight}: we extracted transverse profiles from each aligned sample and measured the separation and widths of two fitted ridges. The comparison between a thin slice and a finite longitudinal average shows that these quantities have different uncertainties and respond differently to binning.

In Appendix~\ref{sec:appendix_multiscale}, we compare the posterior mean-to-standard-deviation ratio of the M87 images after smoothing each sample with Gaussian kernels of different sizes, up to the CLEAN beam. The ratio increases over much of the jet with smoothing, showing that the brightness averaged over a larger area is better constrained than that of individual pixels in the native-resolution images, at the cost of spatial detail.

\subsection{Comparison with CLEAN images}\label{sec:clean_image_comparison}

Figure~\ref{fig:comparison_4x3} presents the Stokes $I$, $Q$, $U$ and linear-polarization image comparison for 0003$-$066, demonstrating consistent results between \texttt{Comrade} and CLEAN across the Stokes parameters. The Stokes~$I$ NXCORR (Eq.~\ref{eq:nxcorr}) is 0.9993, and the linear polarization complex NXCORR (Eq.~\ref{eq:complex_nxcorr}) is 0.9836. The native-resolution \texttt{Comrade} map reveals a continuous bent jet with polarization structure that is smoothed out in the CLEAN reconstruction. Full $4\times3$ comparison plots for the remaining 24 sources are provided in the online supplementary materials.

Figure~\ref{fig:nxcorr_polexp_clean} shows the distribution of Stokes~$I$ NXCORR values across all 25 sources. The values are tightly clustered above 0.999, with a median of 0.9996, demonstrating that the convolved \texttt{Comrade} posterior mean images are nearly identical to the CLEAN reconstructions in total intensity for all sources in our sample.

In contrast to the uniformly high Stokes~$I$ agreement (Figure~\ref{fig:nxcorr_polexp_clean}, left), the complex linear polarization NXCORR (Eq.~\ref{eq:complex_nxcorr}) shows considerably more scatter, ranging from $\sim$0.17 (3C\,84) to $>$0.99, with several sources falling well below the Stokes~$I$ floor. To investigate the origin of this spread, Figure~\ref{fig:nxcorr_polexp_clean} (right) shows the complex polarization NXCORR as a function of the total fractional polarization $m_{\rm tot}$ for all 25 sources. Here $m_{\rm tot} \equiv \sum_i P_i \,/\, \sum_i I_i$, where $P_i = \sqrt{Q_i^2 + U_i^2}$ and the sums run over pixels in the native-resolution PolExp posterior mean satisfying $I_i > \tau_I$ and $P_i > \tau_P$, with the CLEAN-DR-matched thresholds $\tau_I$ and $\tau_P$ defined in Section~\ref{sec:clean_comparison}. We note that $m_{\rm tot}$ is a summed polarized-intensity fraction, not the vector-integrated fractional polarization $|\sum_i(Q_i + iU_i)| / \sum_i I_i$; the former avoids vector cancellation between pixels with different EVPA orientations and is therefore a more representative measure of the total resolved polarized flux. The two quantities are positively correlated (Spearman $\rho_s = 0.66$, $p < 0.001$; Pearson $\rho_p = 0.56$, $p = 0.004$), indicating that the polarimetric agreement between \texttt{Comrade} and CLEAN degrades systematically for weakly polarized sources. This is expected because the dominant systematic uncertainties (residual gain and D-term errors) scale with the source flux density, so weakly polarized sources have a lower signal-to-systematic-noise ratio regardless of the thermal noise level. The three clear outliers --- 3C\,84 ($m_{\rm tot} \approx 0.05\%$), J0313$+$4120 ($m_{\rm tot} = 0.29\%$), and M87 ($m_{\rm tot} = 1.19\%$) --- are among the most weakly polarized sources in our sample.

\subsection{Validation of the \texttt{Comrade} instrumental solutions}\label{sec:validation}

A central goal of this paper is to show that the Bayesian framework is not only producing plausible images, but also recovering credible instrumental solutions. We therefore compared the posterior gain and D-term products from \texttt{Comrade} against the corresponding products from an independent CLEAN$+$\texttt{GPCAL} reduction of the same 15\,GHz BP266B data set. The comparison is intentionally stringent: each source is imaged independently with \texttt{Comrade}, whereas the conventional calibration is obtained through the standard iterative workflow. Agreement between the two pipelines therefore supports both the image-domain and calibration-domain reliability of the Bayesian method.

\subsubsection{Gain amplitude comparison}\label{sec:gain_comparison}

Figure~\ref{fig:gain_persource} shows the per-source gain amplitude comparison between \texttt{Comrade} and CLEAN for 0003$-$066. At each station, the \texttt{Comrade} posterior mean gain amplitudes (circles with $1\sigma$ error bars) closely track the CLEAN self-calibration solutions (crosses) for both RCP and LCP, confirming that the Bayesian method recovers gain solutions consistent with the conventional pipeline. Per-source gain comparisons for the remaining 24 sources are provided in the online supplementary materials.

We further examined the consistency of the gain amplitudes recovered from independent \texttt{Comrade} imaging runs on different target sources. In general, antenna gains are both time- and direction-dependent \citep[e.g.,][]{Smirnov2011b}. Therefore, gain solutions from different sources can be meaningfully compared only when the sources are observed at similar times and at nearby sky positions. If these conditions are met and the instrumental model is correctly specified, the recovered gains should be mutually consistent, since they reflect the same underlying instrumental state.

For this comparison, we grouped the sources into three broad sky-region clusters based on proximity and overlapping observing scans:
\begin{itemize}
    \item \textit{Cluster~1} (3C\,84 region): 3C\,84, J0313$+$4120, J0303$+$4716, J0336$+$3218, 0234$+$285, 0202$+$319 (6 sources, $\Delta\theta_{\max} = 19.3^\circ$);
    \item \textit{Cluster~2} (M87 region): M87, J1222$+$0413, J1239$+$0443, J1239$+$0730 (4 sources, $\Delta\theta_{\max} = 8.4^\circ$);
    \item \textit{Cluster~3} (North Pole): 1803$+$784, 1928$+$738 (2 sources, $\Delta\theta_{\max} = 6.8^\circ$).
\end{itemize}
The remaining 13 sources are widely distributed across the sky and are therefore not suitable for this direction-dependent gain comparison. Figure~\ref{fig:sky_cluster} shows the sky distribution of all 25 sources color-coded by cluster membership. These clusters are intentionally defined broadly to allow a qualitative visual comparison of the gain, D-term, and gain ratio evolution across sources in the same sky region (Sections~\ref{sec:gain_comparison}--\ref{sec:gain_ratio}). For the quantitative inter-source gain consistency analysis (Section~\ref{sec:gain_histograms}), we apply stricter pairwise selection criteria that account for both angular separation and right ascension difference.

Figures~\ref{fig:gain_cluster1}, \ref{fig:gain_cluster2}, and \ref{fig:gain_cluster3} show the \texttt{Comrade} gain amplitudes (RCP and LCP) as a function of time for all 10 stations in Clusters~1--3. These plots include all sources in each broad sky-region cluster to provide a qualitative overview of the gain behavior. For each source, the posterior mean gain amplitude is plotted with error bars indicating the $1\sigma$ posterior uncertainty. Since antenna gains are both direction- and time-dependent, sources within the same cluster are observed at similar but not identical pointings and times, and their gain solutions are therefore not expected to be perfectly identical. At several stations the sources trace broadly similar temporal evolution; a clear example is OV in Cluster~1 (Figure~\ref{fig:gain_cluster1}), where all six sources show closely matching trends. A notable exception is SC. In particular, 3C\,84 and J0313$+$4120, which are separated by only $1.3^\circ$ and were observed within a few minutes of one another, show pronounced source-dependent gain offsets. The corresponding per-source comparisons in the online supplementary materials show that the independent CLEAN self-calibration solutions recover the same anomalous SC levels and temporal behavior for both sources. This agreement makes a \texttt{Comrade}-specific reconstruction artifact unlikely and instead points to an effective SC-specific residual amplitude correction shared by the two reductions. In Cluster~2 (Figure~\ref{fig:gain_cluster2}), the four sources share similar right ascensions ($\Delta\alpha < 0.3$\,h), so the elevation-dependent gain component is nearly identical at a given time; these sources accordingly show more consistent gain trends across stations than the more widely spaced Cluster~1. In Cluster~3 (Figure~\ref{fig:gain_cluster3}), 1803$+$784 and 1928$+$738, separated by $6.8^\circ$ on the sky, show distinct gain evolution patterns. Their right ascensions differ by $\sim1.5$\,h, so at a given UT the antennas observe them at different hour angles and elevations; the gain curves of 1928$+$738 appear shifted by approximately this amount relative to 1803$+$784, consistent with the elevation-dependent response of the VLBA antennas.

The corresponding plots for Clusters~2 and 3 are presented in Appendix~\ref{sec:appendix_clusters} (Figures~\ref{fig:gain_cluster2} and \ref{fig:gain_cluster3}).

\subsubsection{Quantitative assessment of gain consistency}\label{sec:gain_histograms}

Figure~\ref{fig:gain_diff_hist} quantifies the gain amplitude consistency. The left panel shows the Comrade$-$CLEAN gain amplitude difference for all 25 sources at each station and time stamp. The distribution is centered near zero with small scatter, confirming that the two methods recover consistent gain amplitudes regardless of source structure and brightness.

The center and right panels show the inter-source gain spread for \texttt{Comrade} and CLEAN, respectively. While the cluster plots in Figures~\ref{fig:gain_cluster1}, \ref{fig:gain_cluster2}, and \ref{fig:gain_cluster3} include all sources within each broad sky region for qualitative inspection, this quantitative comparison requires stricter pairwise selection criteria to ensure that the compared gains genuinely reflect the same instrumental state. Specifically, we include only source pairs that satisfy three conditions: (1) observed within 5\,minutes of each other, (2) separated by less than 0.5\,h in right ascension, and (3) separated by less than $10^\circ$ on the sky. The RA restriction is particularly important: at a given UT, sources with different right ascensions are observed at different hour angles and therefore at different elevations, so the elevation-dependent component of the antenna gain differs systematically even if the observations are simultaneous. A separation of $\Delta\alpha < 0.5$\,h ensures that the hour-angle difference --- and hence the elevation mismatch --- remains small enough for a meaningful gain comparison. The angular separation criterion further excludes widely separated pairs for which direction-dependent gain differences are expected to be significant. The 5-minute time window restricts the comparison to near-simultaneous scans, minimizing the effect of temporal gain drifts. These criteria select pairs primarily from the compact cores of Clusters~1 and 2 (e.g., 3C\,84 $-$ J0313$+$4120 at $1.3^\circ$, M87 $-$ J1239$+$0730 at $5.3^\circ$), yielding $\sim$2400 matched data points from 10 source pairs. The \texttt{Comrade} and CLEAN inter-source spreads are comparable in width, demonstrating that the Bayesian method does not introduce additional scatter relative to the standard calibration approach.

\subsubsection{D-term comparison}\label{sec:dterm_comparison}

We first compared the on-axis (track-averaged) D-term solutions. In the \texttt{Comrade} hierarchical model, this corresponds to the hyperparameter $\mu_d$, while in \texttt{GPCAL} it is the time-independent on-axis leakage term solved per antenna. Figure~\ref{fig:dterm_posterior} shows the $\mu_d$ posterior distributions on the complex plane for all 25 sources at each VLBA antenna. Since each source is imaged in a separate run, the $\mu_d$ posteriors from different sources provide separate estimates of the same instrumental quantity. The fact that these independently obtained posteriors overlap tightly at each antenna demonstrates that the hierarchical model reliably recovers the on-axis D-terms regardless of the source structure. Furthermore, the \texttt{Comrade} posteriors are consistent with the \texttt{GPCAL} point estimates while revealing the full posterior uncertainty structure, including correlations between the real and imaginary parts.

Figure~\ref{fig:dterm_persource} presents the full D-term (on-axis $\mu_d$ + time-dependent off-axis $d(t)$) time series comparison between \texttt{GPCAL} and \texttt{Comrade} for 0003$-$066. The \texttt{GPCAL} solutions trace the same temporal trends as the \texttt{Comrade} posterior median, and generally fall within the 68\% credible intervals. Notably, the time-dependent leakage variations are not negligible even at 15\,GHz: the off-axis D-term components vary by up to a few percent over the track, which is comparable to or larger than the fractional polarization of many sources in our sample. This underscores the importance of incorporating a time-dependent leakage model in full polarimetric imaging, even at the relatively low frequencies of the VLBA. Per-source comparisons for the remaining 24 sources are available in the online supplementary materials.

Figures~\ref{fig:dterm_comrade_cluster1}, \ref{fig:dterm_comrade_cluster2}, and \ref{fig:dterm_comrade_cluster3} show the \texttt{Comrade}-only D-term time series for sources within each cluster. The time-dependent leakage is known to arise primarily from direction-dependent effects: the polarization response of each antenna is inherently direction-dependent \citep{Smirnov2011b}, and the time-varying antenna pointing offsets cause the effective off-axis leakage to change over the course of an observation \citep{Park2023a}. Because sources at similar sky positions share similar direction-dependent polarization responses and experience similar pointing offsets at similar times, they should exhibit similar time-dependent leakage patterns. This is precisely what we observe: within each cluster, the D-term time series from independently imaged sources trace consistent temporal evolution. This result is in agreement with the \texttt{GPCAL}-based analysis presented in \citet{Park2023a}, where the same direction-dependent behavior was demonstrated using the conventional calibration pipeline. Our \texttt{Comrade} results thus provide independent confirmation of this effect from the Bayesian forward-modeling framework.

The corresponding plots for Clusters~2 and 3 are presented in Appendix~\ref{sec:appendix_clusters} (Figures~\ref{fig:dterm_comrade_cluster2} and \ref{fig:dterm_comrade_cluster3}).

\subsubsection{Gain ratio comparison}\label{sec:gain_ratio}

The R/L gain ratio ($|G_L/G_R|$) is directly coupled to the measurement of circular polarization (Stokes~$V$): since $V = (RR - LL)/2$, even small gain ratio errors propagate directly into the measured Stokes~$V$. At centimeter wavelengths, AGN circular polarization is typically $\lesssim$0.3--1\% of the total intensity \citep{Wardle1998, HomanWardle1999}, with the MOJAVE 15\,GHz survey finding strong detections ($\geq$0.3\%) in $\sim$15\% of sources at levels of 0.3--0.5\% \citep{HomanLister2006}. Accurate gain ratio calibration at the sub-percent level is therefore essential. In conventional CLEAN-based pipelines, recovering the circular polarization signal requires dedicated post-calibration procedures such as the gain-transfer technique \citep{HomanWardle1999, Homan2001}, which separates the instrumental R/L gain ratio from the source circular polarization by smoothing over multiple sources. In contrast, the \texttt{Comrade} forward model is based on the full RIME and jointly solves for Stokes~$V$ and the R/L gain ratio without assuming $V = 0$ at any stage.

The R/L gain ratio is a complex quantity, and we examine its amplitude $|G_L/G_R|$ and phase $\arg(G_L/G_R)$ separately. Because CLEAN derives the gain ratio under the assumption $V = 0$---absorbing any true circular polarization into the gain solutions---it does not provide a meaningful reference for the intrinsic R/L gain ratio. We therefore do not compare against CLEAN here, and instead assess the \texttt{Comrade} solutions through their internal consistency across sources observed at similar sky positions and times, as we did for the gain amplitudes and D-terms. Figures~\ref{fig:gain_ratio_amp_cluster1} and \ref{fig:gain_ratio_phase_cluster1} show the \texttt{Comrade} gain ratio amplitude and phase for the Cluster~1 sources. Although the gain-ratio amplitudes appear broadly consistent across sources---and visually more consistent than the individual gain amplitudes---the residual inter-source scatter remains large compared with the sub-percent calibration accuracy required to recover the typically weak Stokes~$V$ signals of AGN. This visual consistency alone therefore does not establish reliable circular-polarization calibration. This limitation reflects the degeneracy between the R/L gain ratio and Stokes~$V$: small Stokes~$V$ signals can be absorbed into the gain ratio or vice versa, leading to source-dependent gain-ratio solutions. We note that this degeneracy is analytic in origin: an array-wide rescaling of the station R/L gain-ratio amplitudes trades off against the source Stokes~$I$ and $V$ in a manner analogous to the EVPA degeneracy of the cross-hand phases (P.~Tiede et al., in preparation). Breaking this degeneracy across a heterogeneous sample will benefit from a dedicated modeling framework that incorporates the key insights of the gain-transfer technique \citep{HomanWardle1999, Homan2001}: namely, that the instrumental R/L gain ratio should vary smoothly over the course of an experiment, and that the ensemble-averaged Stokes~$V$ across multiple sources should be close to zero. We plan to develop such a framework in a forthcoming study. The corresponding amplitude and phase plots for Clusters~2 and 3 are presented in Appendix~\ref{sec:appendix_clusters} (Figures~\ref{fig:gain_ratio_amp_cluster2}--\ref{fig:gain_ratio_phase_cluster3}).

\subsection{PolExp vs Poincar\'{e} parameterization}\label{sec:polexp_vs_poincare}

The two polarization parameterizations explored in this work---PolExp and Poincar\'{e}---both enforce the physical constraints $I > 0$ and $I \geq \sqrt{Q^2+U^2+V^2}$ by construction, but they do so with fundamentally different prior geometries. To our knowledge, this is the first systematic comparison of the two parameterizations applied to the same VLBI data set within the same inference framework.

In the PolExp representation \citep{Arras2025, Kim2025}, the brightness matrix is parameterized as the matrix exponential of a $2 \times 2$ Hermitian matrix constructed from four latent fields $(s, q, u, v)$ --- corresponding to $(a, b, c, d)$ in Eq.~(\ref{eq:polexp_matrix}) --- yielding $I = e^s \cosh p$, $Q = (q/p)\,e^s \sinh p$, $U = (u/p)\,e^s \sinh p$, $V = (v/p)\,e^s \sinh p$, where $p = \sqrt{q^2 + u^2 + v^2}$. This construction guarantees Hermitian positive-definiteness of the coherency matrix, ensuring $I > 0$ and $I \geq \sqrt{Q^2 + U^2 + V^2}$ by construction \citep{Arras2025}.

The Poincar\'{e} representation, as implemented in \texttt{Comrade} \citep{Tiede2022, EHTC2025_M87pol}, separates the total intensity from the polarization fraction $p$ and polarization direction on the Poincar\'{e} sphere. This separation makes it straightforward to design physically motivated priors on the polarization fraction and direction independently. However, when $p \to 0$, the polarization direction becomes poorly constrained by the data, making the reconstruction more prior-sensitive in weakly polarized regions.

In practice, we find that the two parameterizations agree closely in Stokes~$I$ across all 25 sources, while noticeable differences appear in Stokes~$Q$ and $U$, particularly in low-S/N polarized regions. This is consistent with the expectation that bright total-intensity structure is strongly data-driven, whereas the polarization morphology is more sensitive to the choice of prior geometry. Figure~\ref{fig:polexp_vs_poincare} illustrates this behavior for 0003$-$066: the Stokes~$I$ maps are nearly identical (NXCORR $= 0.994$), while the $Q$ and $U$ maps show structural differences, particularly in low-SNR regions.

To disentangle the effect of parameterization from calibration systematics, we compared both parameterizations independently against the CLEAN images. At CLEAN resolution (i.e., after convolving the \texttt{Comrade} images with the CLEAN restoring beam), both PolExp and Poincar\'{e} achieve excellent agreement with CLEAN in Stokes~$I$: the median NXCORR is 0.9996 for PolExp (Figure~\ref{fig:nxcorr_polexp_clean}, left) and 0.9995 for Poincar\'{e} (Figure~\ref{fig:nxcorr_poincare_clean}, left), with all sources exceeding 0.998 in both cases. The complex polarization NXCORR against CLEAN also shows a nearly identical pattern: both parameterizations exhibit a strong correlation with $m_{\rm tot}$ ($\rho_s = 0.66$ for PolExp, $\rho_s = 0.67$ for Poincar\'{e}; Figures~\ref{fig:nxcorr_polexp_clean} and \ref{fig:nxcorr_poincare_clean}, right panels), with the same three outliers (3C\,84, J0313$+$4120, M87) appearing in both cases. This demonstrates that, when evaluated at the synthesized beam scale, the two parameterizations produce essentially indistinguishable results, and the agreement with CLEAN is governed by the same calibration systematics regardless of the polarization prior.

In contrast, the direct PolExp--Poincar\'{e} comparison at native resolution reveals significantly larger differences. Figure~\ref{fig:nxcorr_pe_pc} (left) shows the distribution of Stokes~$I$ NXCORR between the two parameterizations: the median drops to 0.995, noticeably below the $> 0.998$ achieved by either method against CLEAN, and the three most structurally complex jets in the sample --- 2201$+$315, 1641$+$399, and 0923$+$392 --- fall to 0.84--0.90 while all remaining sources stay above 0.94 (21 of the 25 above 0.95). For these three sources the disagreement is confined entirely to sub-beam scales: convolving the two reconstructions with the CLEAN restoring beam raises their mutual NXCORR to $\geq 0.998$. This additional spread is consistent with the stronger coupling between total intensity and polarization in the PolExp parameterization, together with differences in prior geometry between the two models, which manifest most strongly in the prior-dominated intra-beam structure of bright, complex sources.

The complex polarization NXCORR between PolExp and Poincar\'{e} (Figure~\ref{fig:nxcorr_pe_pc}, right) shows no significant correlation with $m_{\rm tot}$ ($\rho_s = -0.09$), in stark contrast to the strong trend seen in the Comrade--CLEAN comparisons ($\rho_s = 0.66$--$0.67$). This suggests that the inter-parameterization differences do not follow the same weak-polarization trend that characterizes the Comrade--CLEAN comparison, and are therefore more plausibly associated with differences in prior geometry than with the calibration systematics identified above. The outliers at low NXCORR --- M87 ($\rho_{\mathcal{P}} = 0.28$), 3C\,84 ($\rho_{\mathcal{P}} = 0.45$), and J0336$+$3218 ($\rho_{\mathcal{P}} = 0.48$) --- are weakly polarized sources where the reconstruction is prior-dominated.

To quantify the angular scale at which the two parameterizations converge, we performed the following experiment. For each source, we first aligned the Poincar\'{e} image directly to the PolExp image at native resolution by maximizing the Stokes~$I$ NXCORR over integer shifts within $\pm15$ pixels, followed by a Fourier-domain subpixel refinement in steps of 0.25~pixel. This mutual PolExp--Poincar\'{e} registration is distinct from the \texttt{Comrade}--CLEAN alignment described in Section~\ref{sec:clean_comparison}. We then convolved both the PolExp and Poincar\'{e} Stokes~$I$, $Q$, and $U$ images with a Gaussian kernel of size $f \times \theta_{\rm beam}$, where $f$ is a beam fraction ranging from 0.01 to 1.0 and $\theta_{\rm beam}$ is the CLEAN restoring beam. The NXCORR between the two parameterizations was then computed at each kernel size without any additional alignment. Figure~\ref{fig:nxcorr_beam_frac} shows the median NXCORR across all 25 sources as a function of beam fraction. The Stokes~$I$ NXCORR converges rapidly, reaching a median of $> 0.998$ by $f \approx 0.15$, indicating that the total-intensity reconstruction is substantially less sensitive to the choice of polarization parameterization than the polarized structure, even on sub-beam angular scales. In contrast, the complex polarization NXCORR converges more slowly, reaching a median of $\approx 0.99$ at $f \approx 0.3$ and exceeding 0.99 by $f \approx 0.4$. The interquartile range is also substantially wider for polarization, reflecting the source-to-source variation in how strongly the prior geometry affects the reconstruction.

Taken together, these results demonstrate that at the synthesized beam scale, the Bayesian reconstructions are robust to the choice of polarization parameterization: both PolExp and Poincar\'{e} recover the same data-driven structure and agree equally well with the conventional CLEAN pipeline. The dominant differences between the two methods emerge at super-resolution scales, particularly in weakly polarized regions where the prior geometry has the strongest influence. This implies that while the overall polarization morphology is reliably constrained by the data, fine-scale polarimetric structures at sub-beam resolution should be interpreted with caution, as they may depend on the specific choice of parameterization.

In this study, we adopted PolExp as the default representation, primarily because it remains regular at $p \to 0$ in its Cartesian latent variables, avoiding the geometric degeneracy of the Poincar\'{e} sphere at low polarization fractions. This is particularly important for our sample, which includes several weakly polarized sources (e.g., 3C\,84, M87, J0313$+$4120) where the Poincar\'{e} reconstruction can become strongly prior-sensitive. The Poincar\'{e} results serve as a valuable systematic cross-check for quantifying the parameterization-dependent uncertainty, and the full per-source comparison is available in the online supplementary material.

% ============================================================
\section{Summary and conclusions}\label{sec:conclusions}

We have benchmarked a hierarchical Bayesian framework for full-Stokes VLBA imaging and polarimetric calibration using \texttt{Comrade} \citep{Tiede2022}, applied to 25 AGN observed with the VLBA at 15\,GHz on 2024 October 4--5. Our main conclusions are as follows:

\begin{enumerate}
    \item All 25 sources were successfully reconstructed in full polarization. The native-resolution posterior means show fine-scale jet morphology --- including edge-brightened structures in M87 consistent with higher-frequency and higher-resolution VLBI observations --- and spatially resolved polarization features across a diverse sample of compact cores, bent jets, and extended structures. When convolved to the CLEAN beam, the Bayesian images closely reproduce the conventional reconstructions: all sources achieve Stokes~$I$ NXCORR $> 0.999$ (median 0.9996).

    \item We developed a gain-phase-based posterior sample alignment method to address the position shift degeneracy inherent in VLBI self-calibration. Individual posterior samples can be offset from one another due to the gain phase degree of freedom; without correction, averaging these shifted samples would blur the posterior mean image. Our method fits the per-sample gain phase offsets relative to a reference and applies the corresponding image shifts before averaging, yielding a sharp and physically meaningful posterior mean for all Stokes parameters.

    \item The complex polarization NXCORR between \texttt{Comrade} and CLEAN correlates strongly with the total fractional polarization ($\rho_s = 0.66$), indicating that the polarimetric agreement degrades systematically for weakly polarized sources where systematic calibration uncertainties dominate. This trend is consistent across both the PolExp and Poincar\'{e} parameterizations ($\rho_s = 0.67$ for Poincar\'{e}), confirming that it reflects calibration systematics rather than a parameterization-specific artifact.

    \item The per-pixel posterior uncertainty maps provide spatially resolved estimates of the reconstruction reliability, showing the local precision of the reconstructed brightness --- an advantage over traditional imaging pipelines that do not natively produce pixel-level uncertainty estimates. However, neighboring pixels are strongly correlated due to shared visibility constraints and the image prior, so per-pixel uncertainties should not be interpreted as independent measurements. For quantitative analyses, uncertainties should be estimated by measuring the quantity of interest in each posterior sample (Section~\ref{sec:uncertainty}).

    \item The hierarchical gain model recovers gain amplitudes consistent with the independent CLEAN pipeline. Sources within the same sky region generally show consistent gain evolution, while sources separated in right ascension (e.g., the Cluster~3 sources 1803$+$784 and 1928$+$738, $\Delta\alpha \approx 1.5$\,h) show systematically shifted gain curves consistent with direction-dependent effects \citep{Smirnov2011b}. A quantitative inter-source comparison using strict pairwise criteria ($\Delta\alpha < 0.5$\,h, angular separation $< 10^\circ$, $\Delta t < 5$\,min) yields comparable scatter for \texttt{Comrade} and CLEAN, confirming that the Bayesian method does not introduce additional systematic errors.

    \item The on-axis D-term posteriors (i.e., the track-averaged constant leakage component $\mu_d$) from independently imaged sources overlap tightly at each antenna and are consistent with the independent \texttt{GPCAL} estimates, demonstrating reliable recovery of the instrumental leakage. The full time-dependent D-term solutions (on-axis $\mu_d$ + off-axis $d(t)$) also trace the same temporal trends as the \texttt{GPCAL} solutions, which generally fall within the \texttt{Comrade} 68\% credible intervals. The time-dependent leakage variations are not negligible even at 15\,GHz: the polarization response of each antenna is inherently direction-dependent \citep{Smirnov2011b}, and time-varying pointing offsets cause the effective off-axis leakage to change over the course of an observation \citep{Park2023a}. Sources observed at similar sky positions in our sample show consistent D-term temporal evolution across clusters, providing independent confirmation of these effects from the Bayesian framework.

    \item The full RIME forward model jointly solves for Stokes~$V$ and the R/L gain ratio without assuming $V = 0$ a priori. Although the gain-ratio amplitudes show greater inter-source consistency than the individual gain amplitudes, this consistency is insufficient to establish the sub-percent accuracy required for typical Stokes~$V$ signals. The underlying degeneracy is analytic in origin, and robust circular-polarization inference will require a Bayesian multi-source extension of the gain-transfer technique \citep{HomanWardle1999, Homan2001}, in which the instrumental R/L gain ratio varies smoothly over the experiment and the ensemble-averaged Stokes~$V$ is constrained to be close to zero.

    \item The PolExp and Poincar\'{e} polarization parameterizations produce virtually identical results at the synthesized beam scale: both achieve the same NXCORR against CLEAN (median Stokes~$I$ NXCORR $> 0.999$) and exhibit the same correlation between complex polarization NXCORR and $m_{\rm tot}$. The inter-parameterization differences become apparent at native (super-)resolution scales. Unlike the Comrade--CLEAN comparison, where the polarimetric agreement degrades systematically for weakly polarized sources ($\rho_s = 0.66$), the PolExp--Poincar\'{e} polarization NXCORR shows no significant correlation with $m_{\rm tot}$ ($\rho_s = -0.09$), suggesting that the inter-parameterization differences are more plausibly associated with differences in prior geometry than with calibration systematics. A beam-fraction analysis, in which both parameterizations are convolved with a Gaussian kernel of varying size $f \times \theta_{\rm beam}$, shows that the median Stokes~$I$ NXCORR exceeds 0.998 by $f \approx 0.15$, while the median complex polarization NXCORR requires $f \approx 0.3$ to reach 0.99, indicating that polarimetric structures are more sensitive to the choice of parameterization at sub-beam scales. PolExp was adopted as the default for this study due to its regularity at low polarization fractions.
\end{enumerate}

The advantages of Bayesian imaging for VLBI --- principled uncertainty quantification, joint calibration and imaging, and super-resolution capability --- have been established by previous studies, but primarily in the context of compact sources with small fields of view (e.g., EHT observations of M87 and Sgr\,A*), through synthetic data tests, or on a limited number of specific data sets. Whether these methods can be reliably applied to general AGN jet imaging, which involves diverse source morphologies, a wide range of fractional polarization levels, and sometimes large fields of view, has remained an open question. In this work, we have addressed this by applying the Bayesian framework to 25 sources spanning a broad range of jet structures and polarimetric properties. The close agreement with the conventional CLEAN$+$\texttt{GPCAL} pipeline in both imaging and calibration products, and in particular the successful recovery of direction- and time-dependent instrumental effects, are results that cannot be obtained from idealized synthetic data tests alone. Together, they provide a robust demonstration that Bayesian polarimetric imaging is practical and reliable for general-purpose VLBI studies.

\begin{acknowledgements}
Jongho Park thanks Kazunori Akiyama for his guidance and support in getting started with \texttt{Comrade} imaging. This work was supported by the National Research Foundation of Korea (NRF) grant funded by the Korea government (MSIT; RS-2024-00449206; RS-2025-02214038). This research has been supported by the POSCO Science Fellowship of POSCO TJ Park Foundation. This research was supported by Global-Learning \& Academic research institution for Master's \textperiodcentered~PhD students, and Postdocs(G-LAMP) Program of the National Research Foundation of Korea(NRF) grant funded by the Ministry of Education(RS-2025-25442355). This work was supported by the BK21 FOUR program through the National Research Foundation of Korea (NRF) under the Ministry of Education (Kyung Hee University, Human Education Team for the Next Generation of Space Exploration). This research was supported by the Korea Astronomy and Space Science Institute under the R\&D program (Project No. 2026-2-844-01/RS-2026-25587698) supervised by the Korea AeroSpace Administration. The VLBA is an instrument of the National Radio Astronomy Observatory. The National Radio Astronomy Observatory is a facility of the National Science Foundation operated by Associated Universities, Inc. This research has made use of data from the MOJAVE database that is maintained by the MOJAVE team \citep{Lister2018}. This work made use of the \texttt{Comrade} \citep{Tiede2022}, \texttt{eht-imaging} \citep{Chael2016, Chael2018}, \texttt{Difmap} \citep{Shepherd1997}, AIPS \citep{Greisen2003}, and \texttt{GPCAL} \citep{Park2021, Park2023a} software packages. During the preparation of this manuscript, J.P. used OpenAI ChatGPT and Anthropic Claude for proofreading and grammatical correction. All AI-assisted changes were critically reviewed and edited by the authors, who take full responsibility for the content of the manuscript.
\end{acknowledgements}

% ============================================================
\begin{appendix}
\nolinenumbers

\clearpage
\twocolumn[{
\section{Larger-scale emission in the native-resolution images}\label{sec:appendix_extended_emission}
\begin{minipage}{\textwidth}
\centering
\includegraphics[width=\textwidth]{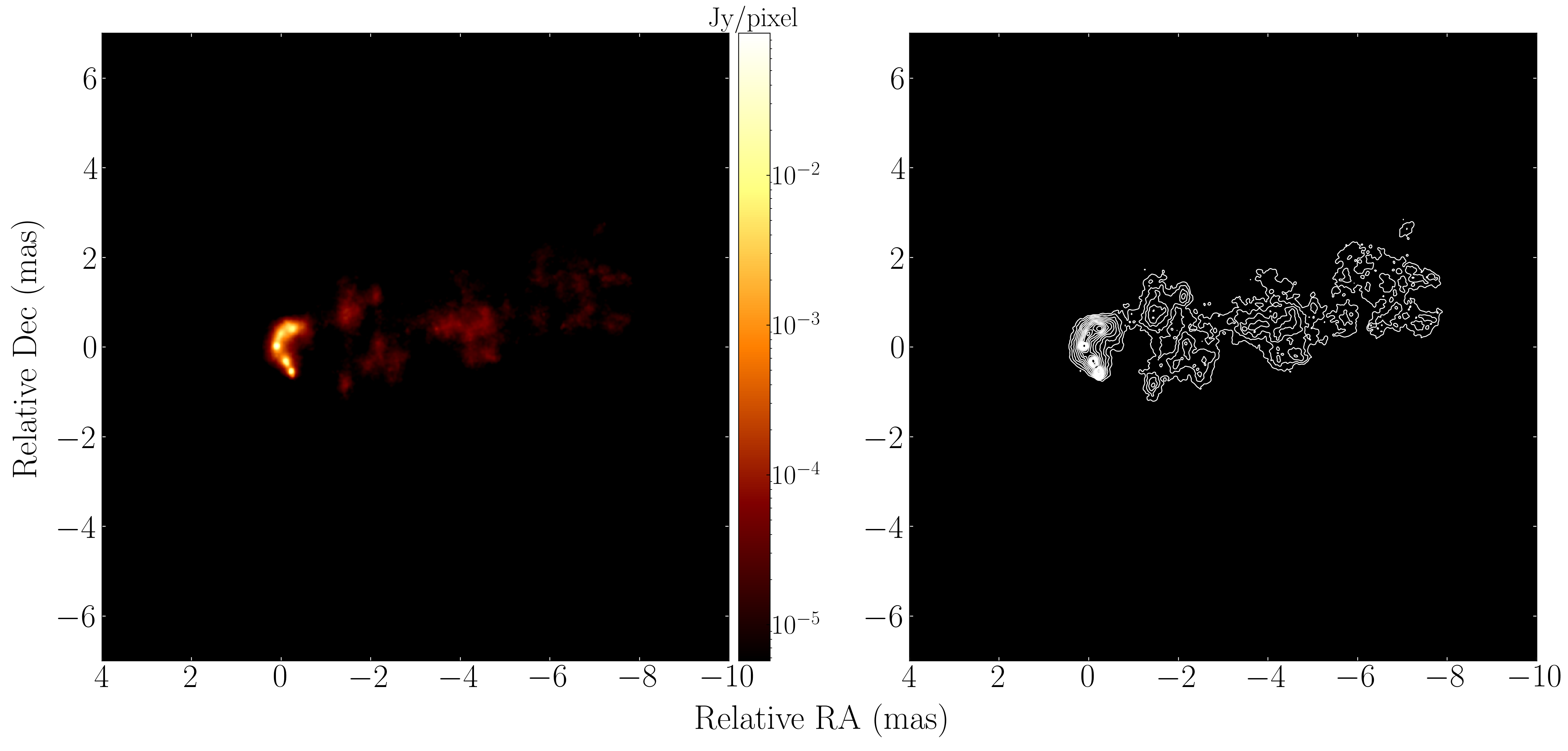}
\captionof{figure}{Native-resolution Stokes~$I$ posterior mean for 0003$-$066, without beam convolution, shown over the same field of view as the CLEAN image in Figure~\ref{fig:comparison_4x3}. \textit{Left:} Logarithmic intensity image. \textit{Right:} Contours at $I_{\min}\,2^n$ for $n=0,\ldots,13$. Both panels use a display threshold of $I_{\min}=5.73\times10^{-6}$\,Jy/pixel, ten times lower than the DR-matched threshold used for the native-resolution intensity image in Figure~\ref{fig:comparison_4x3}.}
\label{fig:native_extended_emission}
\end{minipage}
\vspace{1em}
}]

The native-resolution images in Figures~\ref{fig:mosaic}--\ref{fig:comparison_4x3} are displayed over small fields of view to emphasize the bright inner structure. For 0003$-$066, the \texttt{Comrade} posterior mean also contains the larger-scale emission seen in the CLEAN image. Its lower surface brightness places much of this emission below the DR-matched intensity threshold used in the main figures (Section~\ref{sec:clean_comparison}). Figure~\ref{fig:native_extended_emission} shows the original Stokes~$I$ posterior mean without beam convolution, using the CLEAN field of view and an intensity threshold ten times lower. Diffuse emission extends in the same direction as the jet in the CLEAN image.

\newpage

The faint emission has an irregular morphology and less clearly defined boundaries than the bright inner jet. At lower SNR, the detailed source structure is harder to constrain \citep[e.g.,][]{Lobanov2005}. Convolution with the larger CLEAN restoring beam smooths these variations into a more regular, apparently well-defined jet (Figure~\ref{fig:comparison_4x3}). However, this smooth appearance does not by itself establish that the intrinsic jet width or transverse profile is well constrained. Measurements based on such low-surface-brightness regions, such as jet collimation profiles, therefore require image uncertainty estimates, which can be obtained from the posterior samples as described in Section~\ref{sec:uncertainty}.

\clearpage

\section{Expected error of the posterior mean}\label{sec:appendix_mean_error}

Here we show that the posterior mean image is a more accurate estimate of the source than an individual posterior sample, and that the posterior standard deviation describes the uncertainty of the mean image itself. Let $\theta$ be the true brightness at a given position, and let $\mu$ and $\sigma^2$ be its posterior mean and variance given the data $D$. If the adopted model is correct, our knowledge of $\theta$ given the data is described by its posterior distribution, so that $\theta$ can be treated as a random draw from this distribution \citep[e.g.,][]{Cook2006}. For any estimate $a$, the expected squared error is
\begin{equation}
\begin{split}
\mathrm{E}[(a-\theta)^2\,|\,D] &= (a-\mu)^2 + 2(a-\mu)\,\mathrm{E}[\mu-\theta\,|\,D] \\
&\quad + \mathrm{E}[(\mu-\theta)^2\,|\,D] \\
&= (a-\mu)^2 + \sigma^2 ,
\end{split}
\label{eq:mean_error}
\end{equation}
because $\mathrm{E}[\theta\,|\,D]=\mu$. The posterior mean ($a=\mu$) therefore minimizes the expected squared error, and its root-mean-square (rms) error is $\sigma$. This is a standard result of Bayesian decision theory \citep[e.g.,][]{Berger1985, Gelman2013}.

A posterior sample $X$ follows the same distribution as $\theta$ but is drawn independently of the true brightness. Since $X$ and $\theta$ have the same mean, the expected squared error of $X$ is
\begin{equation}
\begin{split}
\mathrm{E}[(X-\theta)^2\,|\,D] &= \mathrm{Var}(X\,|\,D) + \mathrm{Var}(\theta\,|\,D) \\
&\quad - 2\,\mathrm{Cov}(X,\theta\,|\,D) = 2\sigma^2 .
\end{split}
\label{eq:sample_error}
\end{equation}
The rms error of a single sample is thus $\sqrt{2}\,\sigma$, larger by a factor of $\sqrt{2}$ than that of the mean. The posterior standard deviation is therefore not an overestimate of the uncertainty of the mean image; under the adopted model, it is the rms error of the mean image.

These relations do not assume a Gaussian posterior. They also apply to a quantity measured from the images, such as a jet width or a component separation, when $\mu$ and $\sigma$ are the mean and standard deviation of that quantity across the posterior samples.

\clearpage

\twocolumn[{
\section{Transverse structure of the M87 jet measured from posterior samples}\label{sec:appendix_transverse}
\begin{minipage}{\textwidth}
\centering
\includegraphics[width=\textwidth]{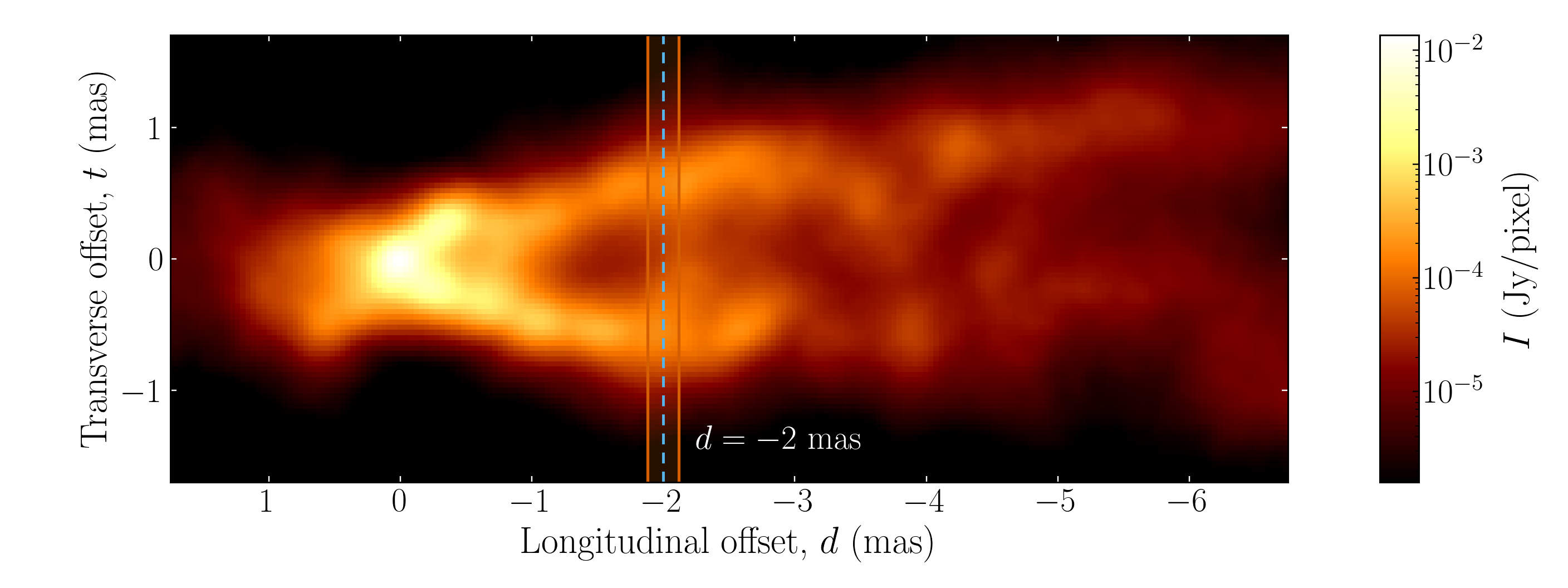}\par
\includegraphics[width=\textwidth]{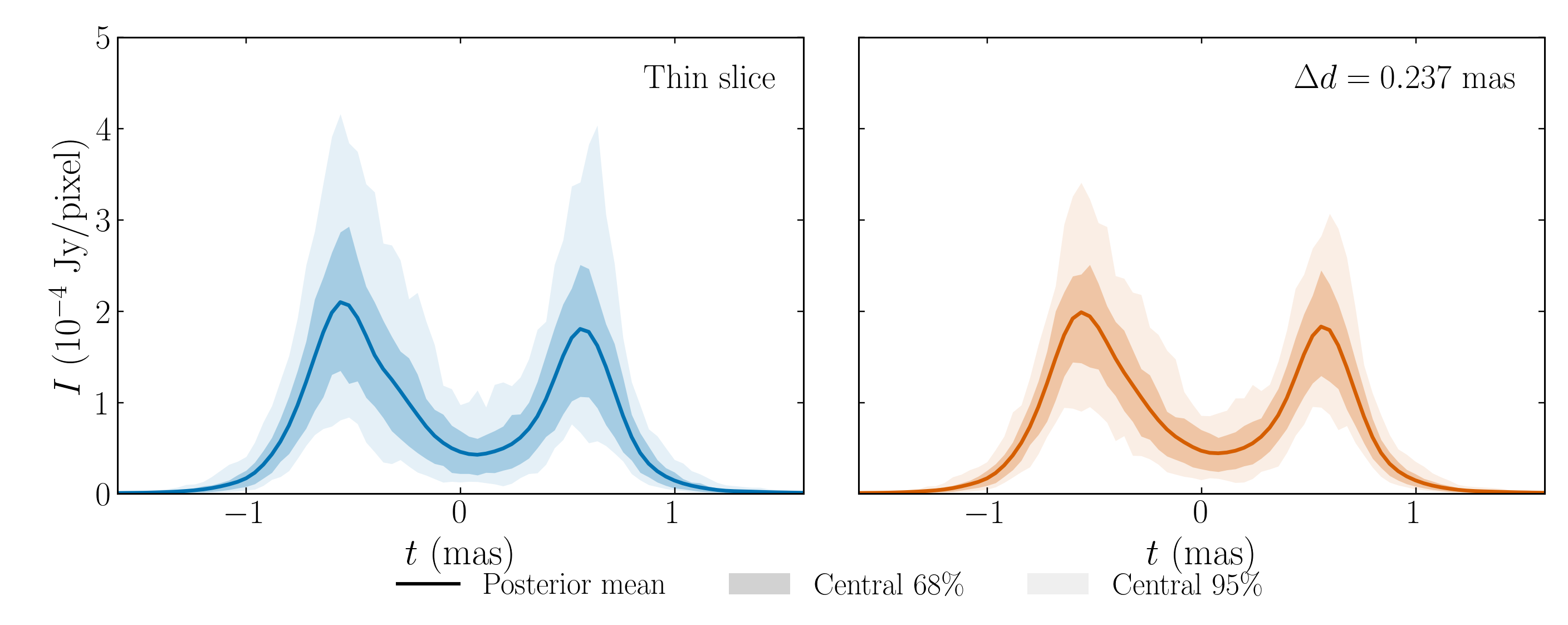}
\captionof{figure}{Transverse profiles from the native-resolution M87 Stokes~$I$ samples used for Figure~\ref{fig:highlight}, without beam convolution. \textit{Top:} Posterior mean rotated clockwise by $18^\circ$, with a common origin at its peak. The dashed line marks $d=-2$\,mas; the orange lines enclose a longitudinal bin of width $\Delta d=0.237$\,mas, one third of the CLEAN beam projected along the jet. \textit{Bottom:} Mean profiles for the thin slice (left) and longitudinal average (right). Dark and light shading show the pointwise central 68\% and 95\% intervals across all 120 samples.}
\label{fig:m87_transverse_profiles}
\end{minipage}
\vspace{1em}
}]

The transverse structure of the M87 jet, including its width and limb brightening, has been measured in many VLBI studies to determine the jet collimation profile \citep[e.g.,][]{AsadaNakamura2012, NakamuraAsada2013, Hada2016, Nakamura2018, Walker2018, Lu2023}. Here, we applied the procedure described in Section~\ref{sec:uncertainty} to the 120 post-burn-in, gain-phase-aligned PolExp samples used for the M87 images in Figure~\ref{fig:highlight} (Sections~\ref{sec:burnin}--\ref{sec:posterior_summary}); none of the M87 samples were excluded by the outlier rejection described in Section~\ref{sec:posterior_summary}. Figure~\ref{fig:m87_transverse_profiles} shows the posterior mean rotated clockwise by $18^\circ$ \citep[following][]{Park2019a, Park2026}, so that the jet extends toward negative longitudinal offset $d$, with $t$ denoting the offset perpendicular to the jet. All samples share a common origin at the peak of the posterior mean, and no additional shifts were applied to align the peaks of individual samples. No convolution with the CLEAN beam was applied.

At $d=-2$\,mas, we extracted the transverse intensity profile in two ways: along a single line at this position (thin slice) and averaged over a short segment of the jet centered on it (longitudinal average). For the longitudinal average, we used a top-hat bin of width $\Delta d=b_\parallel/3=0.237$\,mas without any smoothing in the transverse direction, where $b_\parallel=0.712$\,mas is the FWHM of the CLEAN beam projected onto the jet axis. The same extraction was applied to every sample. Because the average is computed within each sample before the intervals are derived, the intervals of the averaged profile include the correlations between neighboring pixels.

\begin{figure*}[t]
\centering
\includegraphics[width=\textwidth]{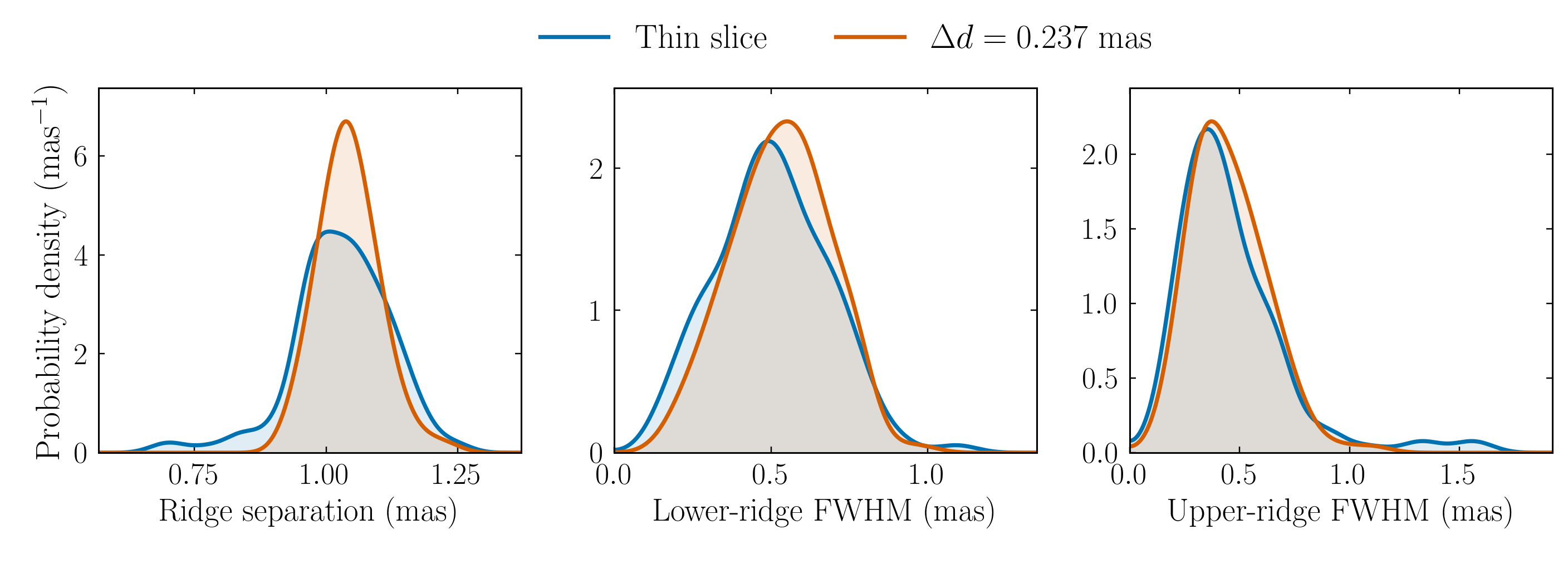}
\caption{Distributions of the fitted ridge separation (left), lower-ridge FWHM (middle), and upper-ridge FWHM (right) for the thin slice (blue) and longitudinal average (orange) in Figure~\ref{fig:m87_transverse_profiles}. The curves show Gaussian kernel density estimates of all 120 measurements for each case, including the broad tails, computed with a common bandwidth within each panel and reflected at zero because the quantities are positive. The ridge separation is the difference between the two fitted Gaussian centers, and each ridge width is the FWHM of the corresponding Gaussian component without beam deconvolution. These distributions represent the spread of the same measurement applied to each sample, rather than the formal uncertainties of a single fit to the mean profile.}
\label{fig:m87_transverse_parameters}
\end{figure*}

For each extracted profile, we fitted the sum of two positive Gaussians with a fixed zero baseline by unweighted least squares over $-1.6\lesssim t\lesssim1.6$\,mas. The two components define the lower and upper ridges in $t$, and we measured their separation from the distance between the two Gaussian centers and their widths from the FWHMs of the components. Because the same fit is applied to the full profile of each sample, correlated variations in pixel brightness propagate into the measured separation and widths, and their uncertainties are obtained from the distribution of these measurements across samples.

Figure~\ref{fig:m87_transverse_parameters} shows that the ridge separation has a smaller fractional spread than either ridge width. Longitudinal averaging reduces the spread of all three quantities, although by different amounts; for example, the standard deviation of the separation decreases from $0.093$ to $0.056$\,mas. This gain in precision comes at the cost of spatial information: averaging over the bin smooths out any fine-scale structure along the jet within the bin, so that the averaged profile describes the jet over a finite length rather than at a single position. The choice of the bin size therefore represents a trade-off between precision and spatial detail, which should be evaluated for each measurement.

This example demonstrates that confidence must be assessed for the particular quantity and extraction region of interest. Since the Gaussian components provide only an approximate description of the profiles, the derived widths also depend on this definition, as well as on the extraction region, imaging model, and priors.

\clearpage

\twocolumn[{
\section{Image uncertainty at different spatial scales}\label{sec:appendix_multiscale}
\begin{minipage}{\textwidth}
\centering
\includegraphics[width=0.62\textwidth]{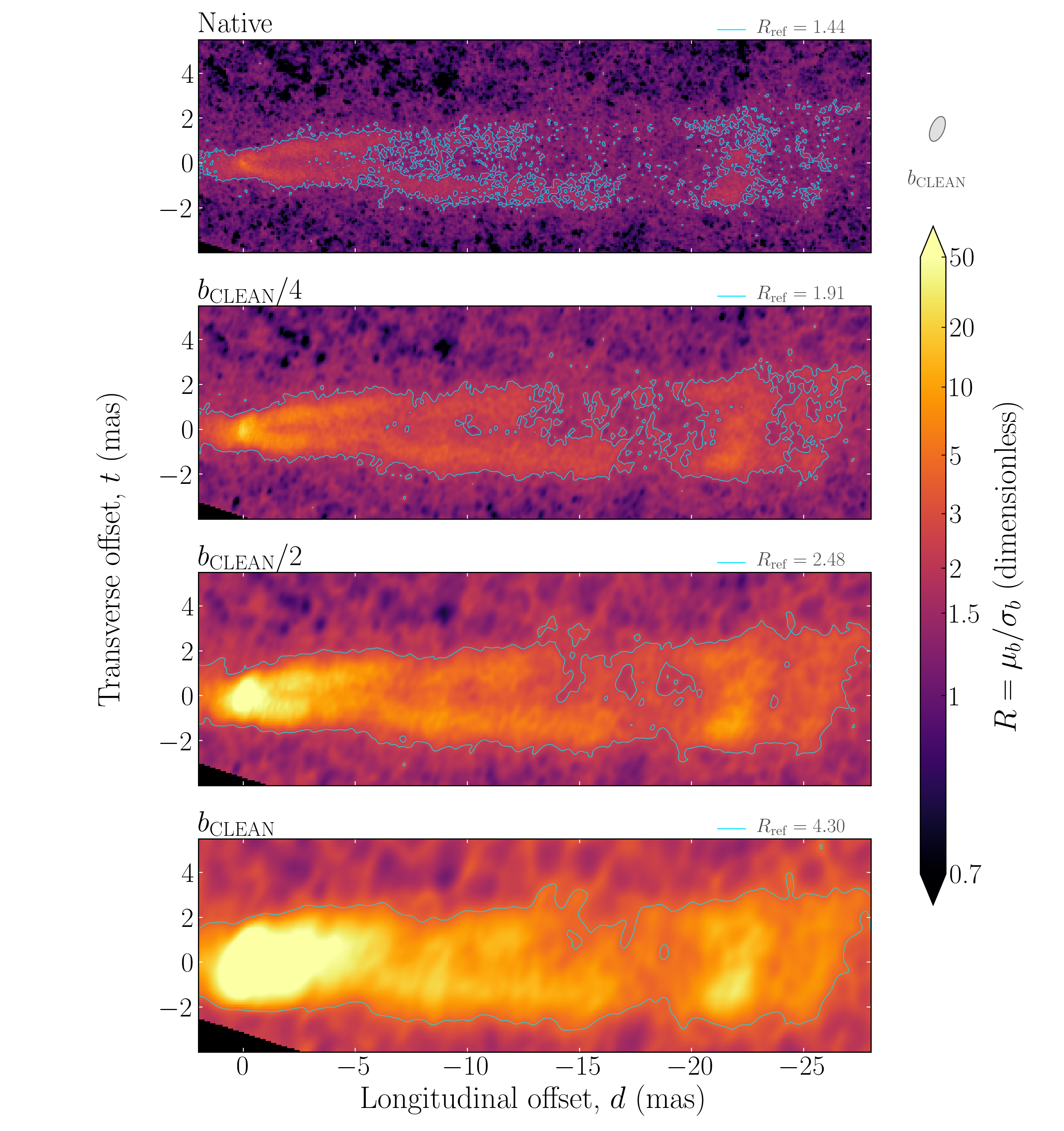}
\captionof{figure}{Posterior mean-to-standard-deviation ratio, $R=\mu_b/\sigma_b$, of the M87 Stokes~$I$ samples used for Figure~\ref{fig:highlight}, after convolving each sample with Gaussian kernels of different sizes. From top to bottom, the panels show the native-resolution samples without convolution and kernels equal to the CLEAN restoring beam ($1.175\times0.594$\,mas at a position angle of $-4.79^\circ$) with its FWHMs scaled by 1/4, 1/2, and 1. As in Figure~\ref{fig:m87_transverse_profiles}, the images are rotated clockwise by $18^\circ$, and $d$ and $t$ denote the longitudinal and transverse offsets from the common origin. All panels use the same nonlinear color scale, which saturates at $R=50$ so that values near $R_{\rm ref}$ remain distinguishable; the maximum values of $R$, reached near the origin, are 6.4, 27, 158, and 259 from top to bottom. Since $\mu_b$ and $\sigma_b$ are both in Jy/pixel, $R$ is dimensionless. Cyan contours mark $R_{\rm ref}$, the 99.9th percentile of $R$ in off-source regions at $-26<d<-1$\,mas and either $t<-3$\,mas or $t>4.5$\,mas within the evaluated field ($-5\lesssim t\lesssim6$\,mas), derived separately for each kernel size. These levels serve as empirical references rather than calibrated detection thresholds. The gray ellipse shows the CLEAN beam on the same angular scale as the images.}
\label{fig:m87_multiscale_snr}
\end{minipage}
\vspace{1em}
}]

Section~\ref{sec:uncertainty} and Appendix~\ref{sec:appendix_transverse} describe the uncertainties of quantities measured from each posterior sample. Here, we examine how the uncertainty of the image itself depends on the spatial scale over which the brightness is averaged. We used the same 120 gain-phase-aligned PolExp samples of M87 as in Appendix~\ref{sec:appendix_transverse}. For each kernel size, we convolved every sample with a Gaussian kernel equal to the CLEAN beam with its FWHMs scaled by 1/4, 1/2, or 1, and calculated the posterior mean, $\mu_b$, the standard deviation, $\sigma_b$, and their ratio, $R=\mu_b/\sigma_b$, at each pixel across the 120 smoothed samples. Because the convolution is applied within each sample before $\mu_b$ and $\sigma_b$ are derived, $\sigma_b$ includes the correlations between neighboring pixels.

Figure~\ref{fig:m87_multiscale_snr} shows that $R$ increases over much of the jet as the kernel size increases, while the fine-scale structure becomes less distinct. The brightness averaged over a larger area is therefore better constrained than the brightness of individual pixels in the native-resolution image. This gain in precision does not come from additional information in the data: smoothing changes the quantity being summarized, from the brightness at each pixel to the brightness averaged over the kernel. This is analogous to CLEAN imaging, in which the CLEAN components are convolved with the restoring beam, so that the brightness and SNR of the image refer to the scale of the beam.

The choice of the kernel size therefore represents a trade-off between precision and spatial detail, which depends on the scientific question and on the scales the data can constrain. Gaussian smoothing averages the brightness both along and across the jet, whereas the longitudinal average in Appendix~\ref{sec:appendix_transverse} averages only along it. A high $R$ for the smoothed brightness does not by itself establish that an intrinsic jet width or ridge separation is well constrained; such quantities should be measured from each sample, as described in Section~\ref{sec:uncertainty}.

\clearpage

\twocolumn[{
\section{Cluster comparison plots for Clusters~2 and 3}\label{sec:appendix_clusters}

This appendix presents the gain amplitude, D-term, and gain ratio cluster comparison plots for Clusters~2 (M87 region) and 3 (North Pole), complementing the Cluster~1 plots shown in the main text (Sections~\ref{sec:gain_comparison}--\ref{sec:gain_ratio}).
\par\vspace{1em}
\begin{minipage}{\textwidth}
\centering
    \includegraphics[width=\textwidth]{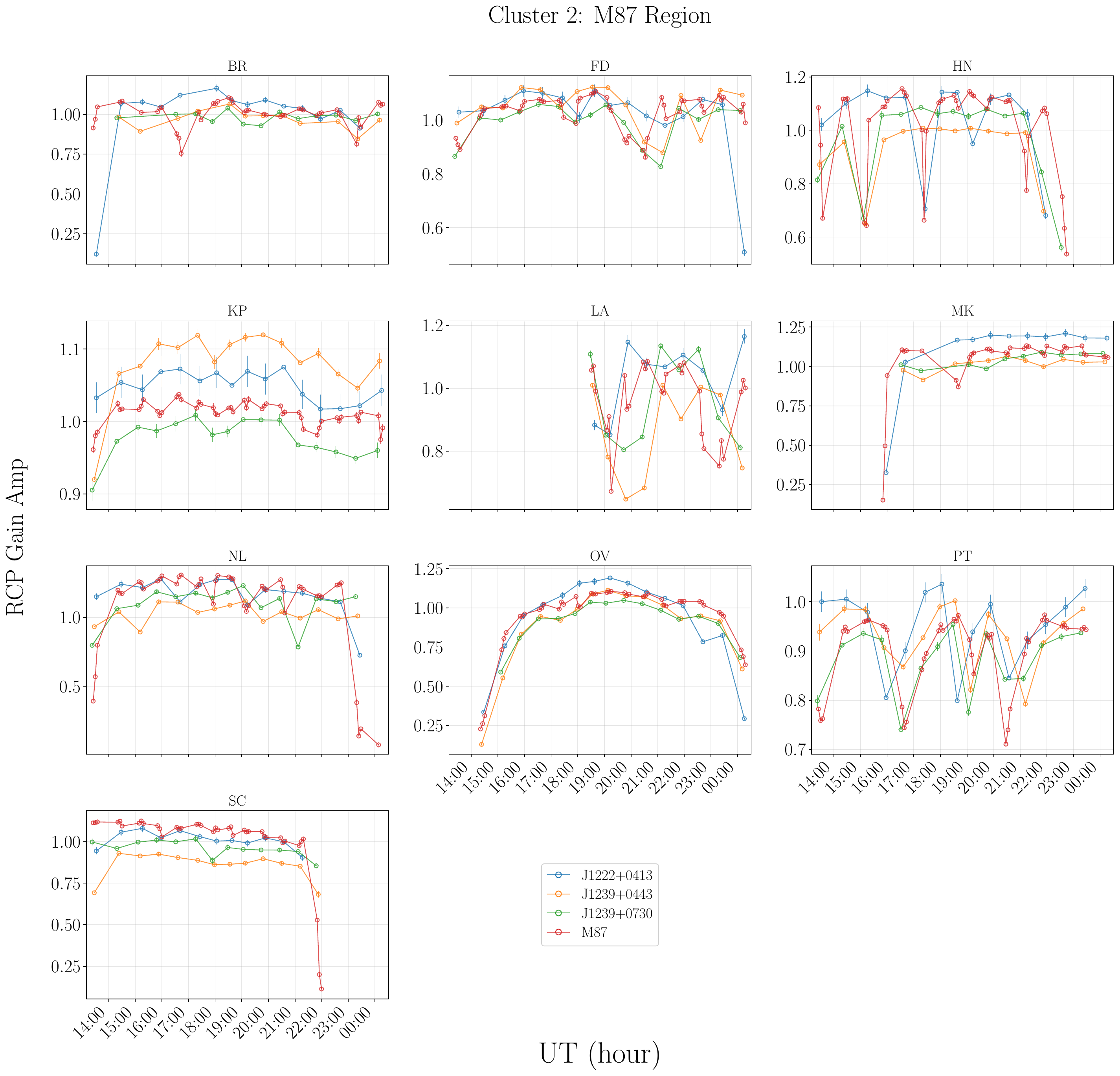}
    \captionof{figure}{Same format as Figure~\ref{fig:gain_cluster1}, but for the Cluster~2 sources (M87 region).}
    \label{fig:gain_cluster2}
\end{minipage}
\vspace{1em}
}]

\begin{figure*}[t!]
    \centering
    \includegraphics[width=\textwidth]{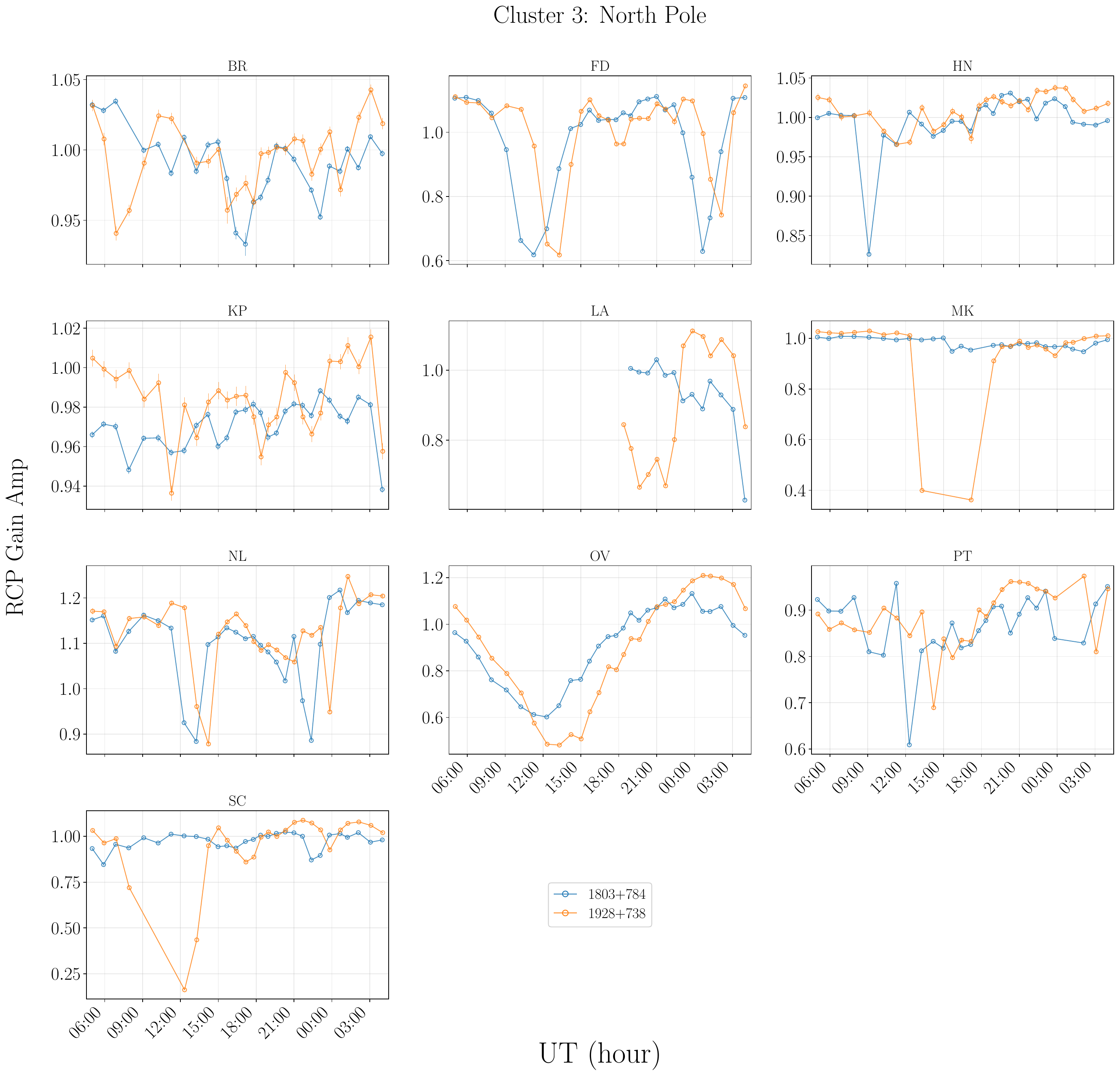}
    \caption{Same format as Figure~\ref{fig:gain_cluster1}, but for the Cluster~3 sources (North Pole region). Note the distinct gain amplitude evolution for 1803$+$784 and 1928$+$738, separated by $6.8^\circ$ on the sky, demonstrating direction-dependent gain variations.}
    \label{fig:gain_cluster3}
\end{figure*}

\begin{figure*}[t!]
    \centering
    \includegraphics[width=\textwidth]{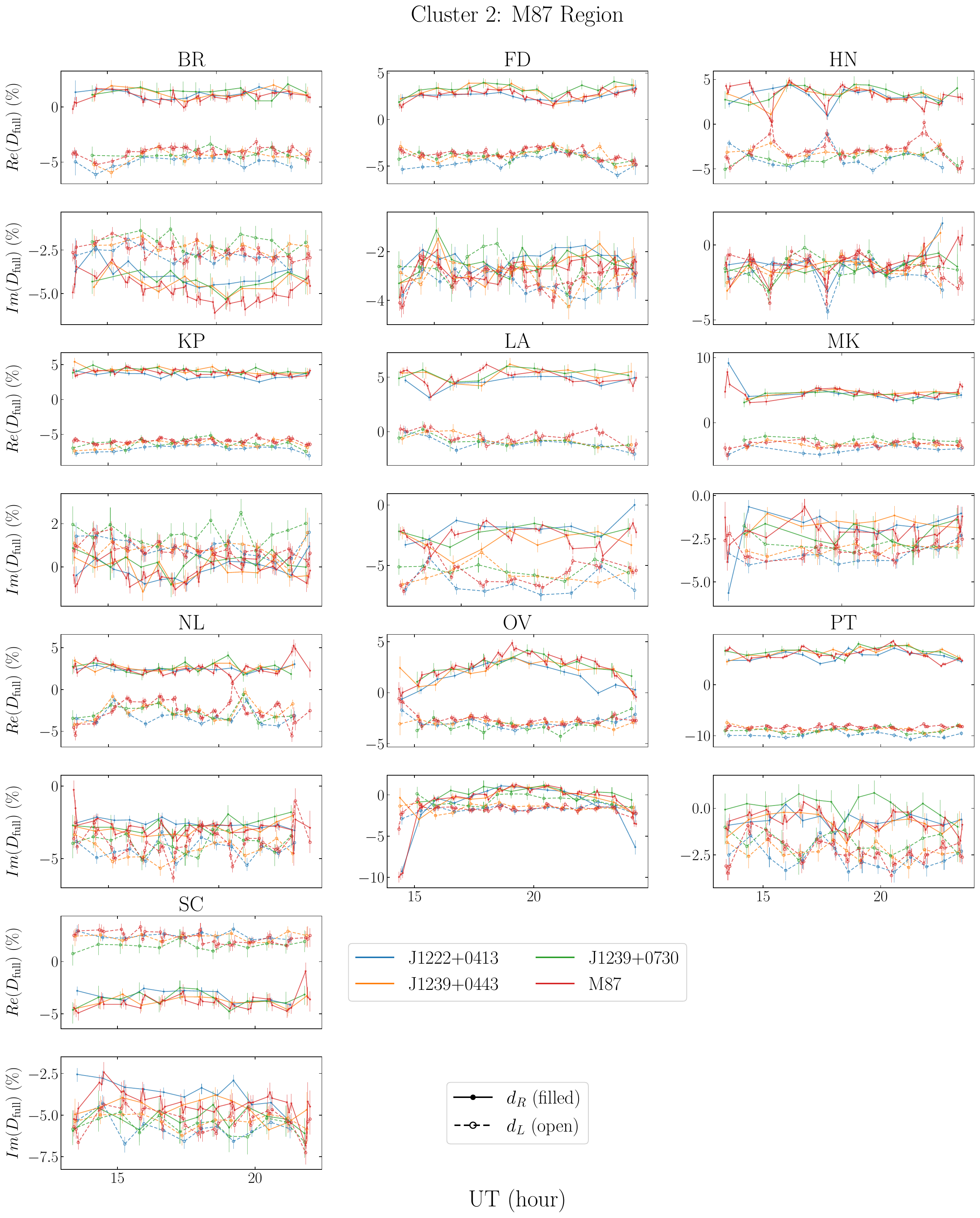}
    \caption{Same format as Figure~\ref{fig:dterm_comrade_cluster1}, but for Cluster~2 (M87 region).}
    \label{fig:dterm_comrade_cluster2}
\end{figure*}

\begin{figure*}[t!]
    \centering
    \includegraphics[width=\textwidth]{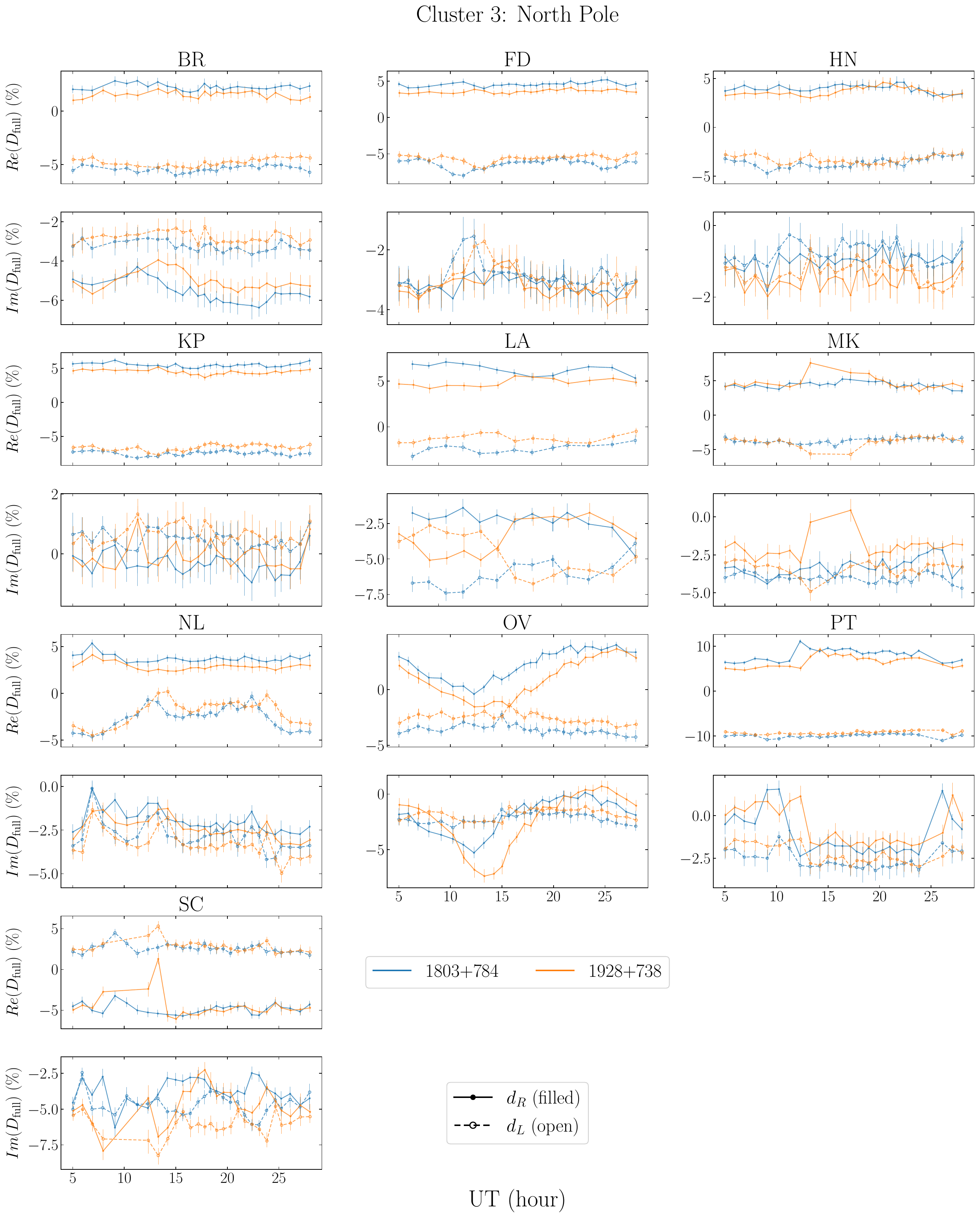}
    \caption{Same format as Figure~\ref{fig:dterm_comrade_cluster1}, but for Cluster~3 (North Pole region).}
    \label{fig:dterm_comrade_cluster3}
\end{figure*}

\begin{figure*}[t!]
    \centering
    \includegraphics[width=\textwidth]{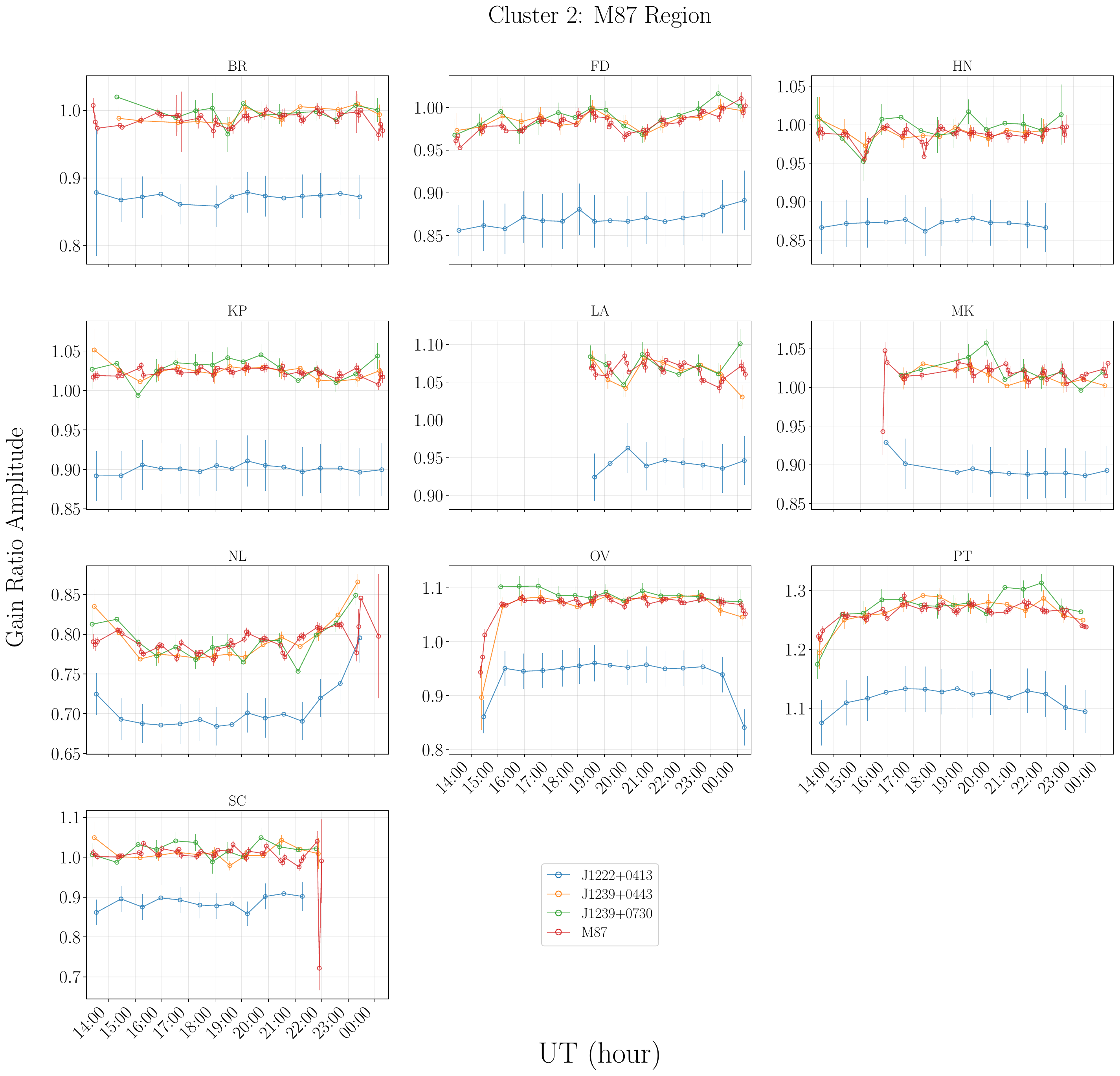}
    \caption{Same format as Figure~\ref{fig:gain_ratio_amp_cluster1} (gain ratio amplitude $|G_L/G_R|$), but for the Cluster~2 sources (M87 region).}
    \label{fig:gain_ratio_amp_cluster2}
\end{figure*}

\begin{figure*}[t!]
    \centering
    \includegraphics[width=\textwidth]{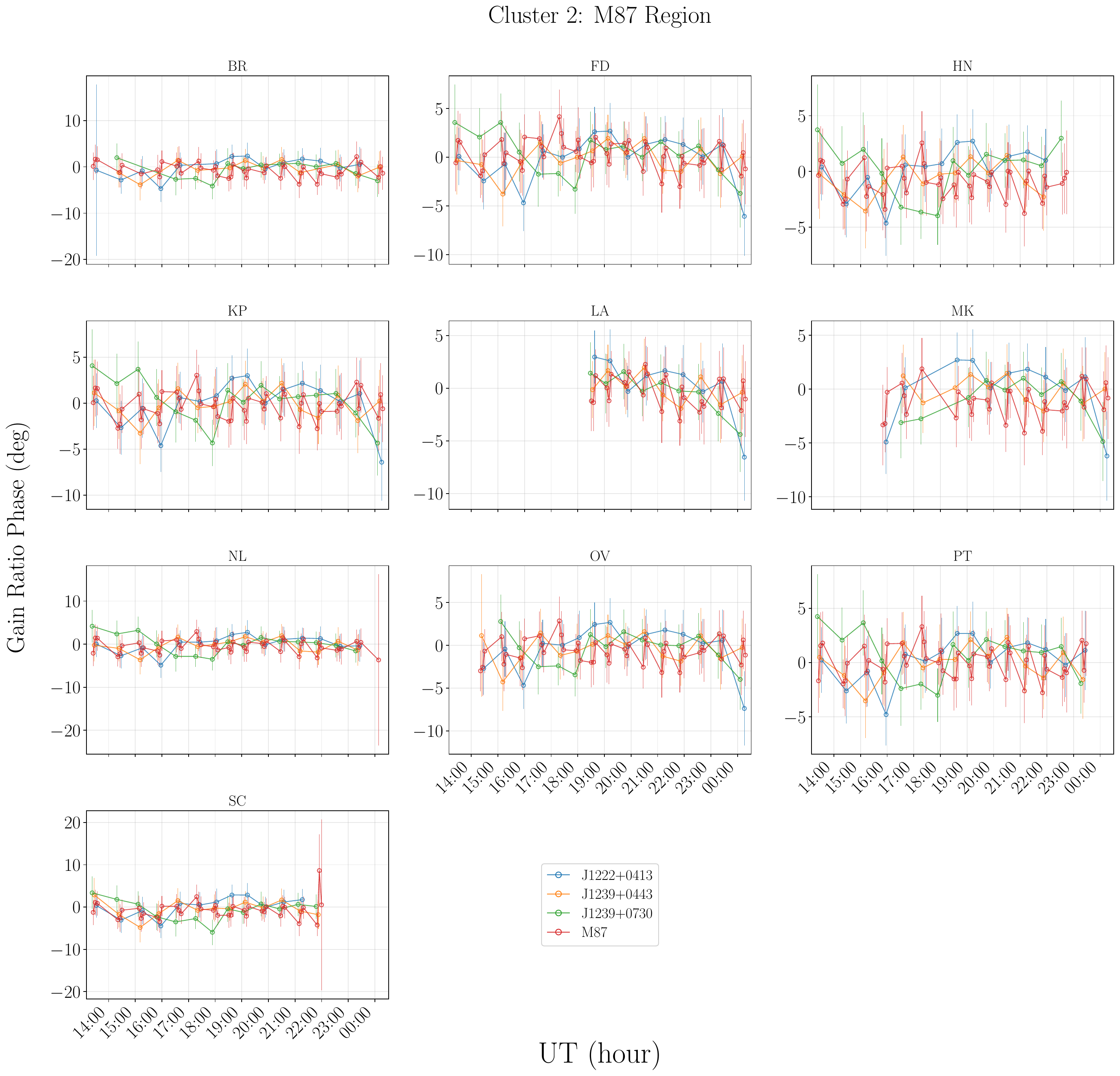}
    \caption{Same format as Figure~\ref{fig:gain_ratio_phase_cluster1} (gain ratio phase $\arg(G_L/G_R)$), but for the Cluster~2 sources (M87 region).}
    \label{fig:gain_ratio_phase_cluster2}
\end{figure*}

\begin{figure*}[t!]
    \centering
    \includegraphics[width=\textwidth]{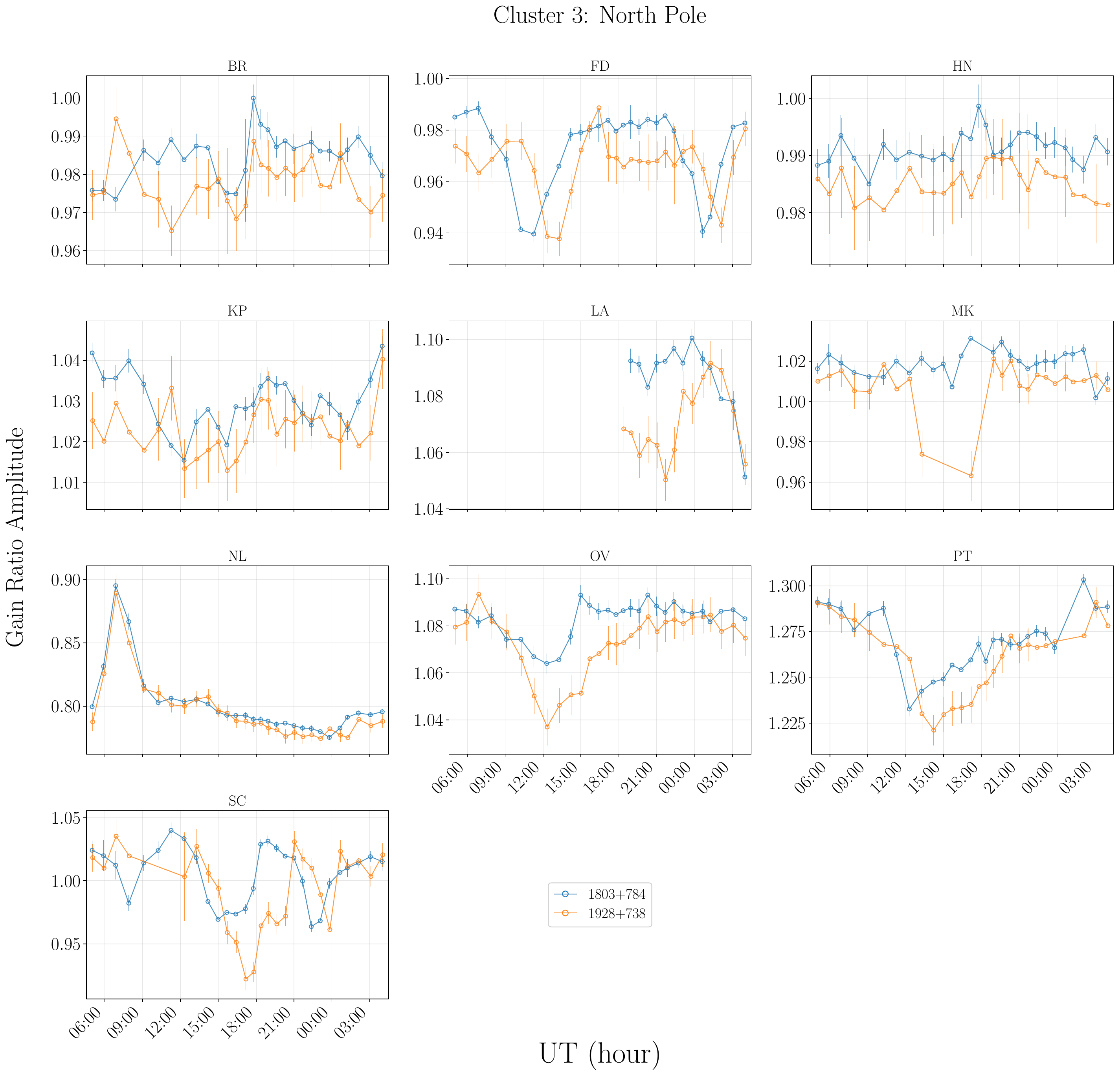}
    \caption{Same format as Figure~\ref{fig:gain_ratio_amp_cluster1} (gain ratio amplitude $|G_L/G_R|$), but for the Cluster~3 sources (North Pole region).}
    \label{fig:gain_ratio_amp_cluster3}
\end{figure*}

\begin{figure*}[t!]
    \centering
    \includegraphics[width=\textwidth]{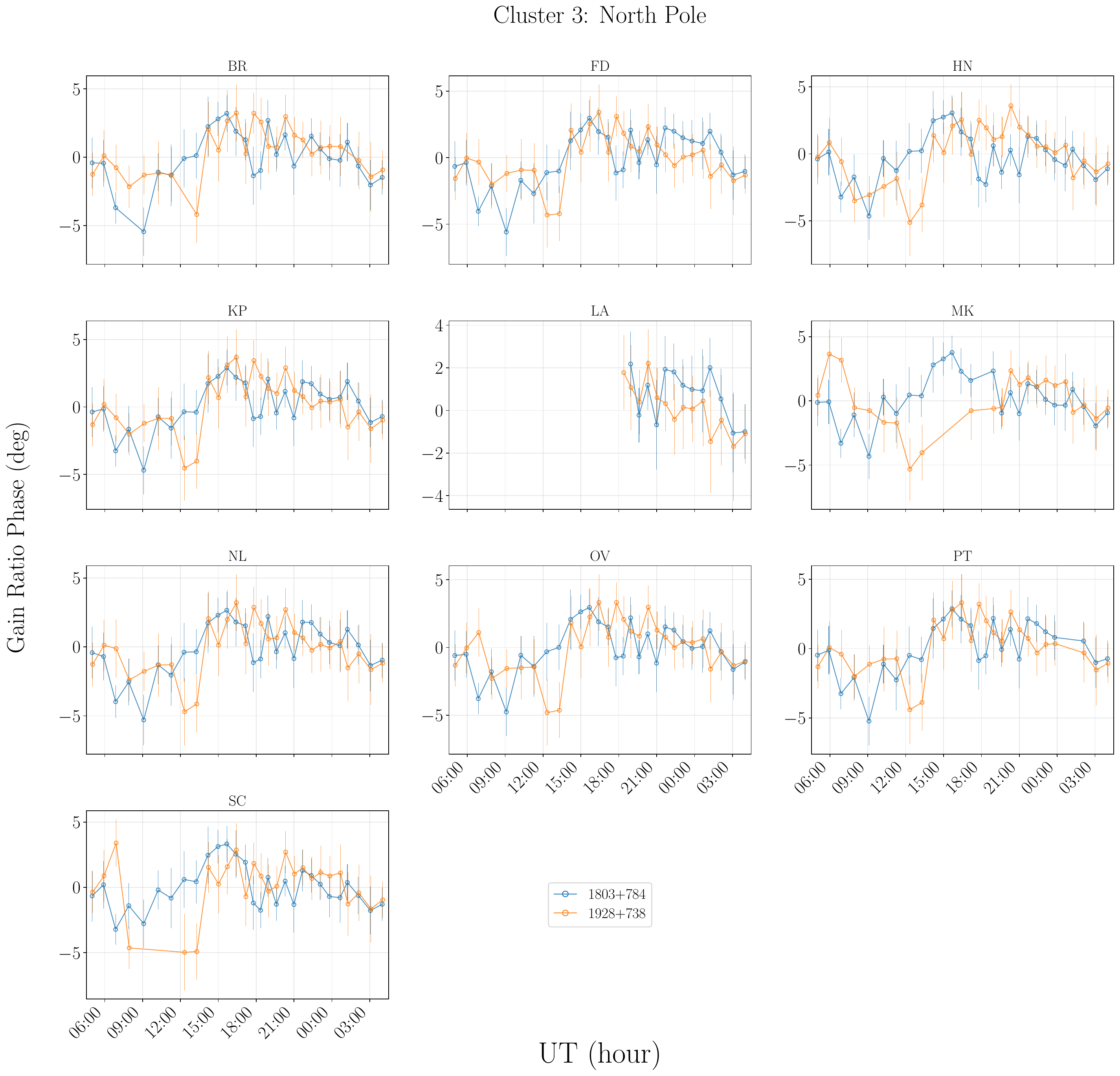}
    \caption{Same format as Figure~\ref{fig:gain_ratio_phase_cluster1} (gain ratio phase $\arg(G_L/G_R)$), but for the Cluster~3 sources (North Pole region).}
    \label{fig:gain_ratio_phase_cluster3}
\end{figure*}

\clearpage
\end{appendix}

% ============================================================
\bibliographystyle{aa}
\bibliography{references}

\begin{thebibliography}{97}
\expandafter\ifx\csname natexlab\endcsname\relax\def\natexlab#1{#1}\fi

\bibitem[{{Akiyama} {et~al.}(2017){Akiyama}, {Kuramochi}, {Ikeda}, {Fish},
  {Tazaki}, {Honma}, {Doeleman}, {Broderick}, {Dexter}, {Mo{\'s}cibrodzka},
  {Bouman}, {Chael}, \& {Zaizen}}]{Akiyama2017a}
{Akiyama}, K., {Kuramochi}, K., {Ikeda}, S., {et~al.} 2017, \apj, 838, 1

\bibitem[{{Arras} {et~al.}(2022){Arras}, {Frank}, {Haim}, {Knollm{\"u}ller},
  {Leike}, {Reinecke}, \& {En{\ss}lin}}]{Arras2022}
{Arras}, P., {Frank}, P., {Haim}, P., {et~al.} 2022, Nature Astronomy, 6, 259

\bibitem[{{Arras} {et~al.}(2025){Arras}, {Roth}, {Reinecke}, {Perley},
  {Frolov}, {Westermann}, \& {En{\ss}lin}}]{Arras2025}
{Arras}, P., {Roth}, J., {Reinecke}, M., {et~al.} 2025, arXiv e-prints,
  arXiv:2504.00227

\bibitem[{{Asada} \& {Nakamura}(2012)}]{AsadaNakamura2012}
{Asada}, K. \& {Nakamura}, M. 2012, \apjl, 745, L28

\bibitem[{{Benke} {et~al.}(2026){Benke}, {Savolainen}, {Giovannini}, {Kovalev},
  {Bruni}, {Lisakov}, {Giroletti}, \& {Ros}}]{Benke2026}
{Benke}, P., {Savolainen}, T., {Giovannini}, G., {et~al.} 2026, \aap, 709, A226

\bibitem[{{Berger}(1985)}]{Berger1985}
{Berger}, J.~O. 1985, {Statistical decision theory and Bayesian analysis}

\bibitem[{{Blandford} {et~al.}(2019){Blandford}, {Meier}, \&
  {Readhead}}]{BlandfordMeierReadhead2019}
{Blandford}, R., {Meier}, D., \& {Readhead}, A. 2019, \araa, 57, 467

\bibitem[{{Blandford} \& {Payne}(1982)}]{BlandfordPayne1982}
{Blandford}, R.~D. \& {Payne}, D.~G. 1982, \mnras, 199, 883

\bibitem[{{Blandford} \& {Znajek}(1977)}]{BlandfordZnajek1977}
{Blandford}, R.~D. \& {Znajek}, R.~L. 1977, \mnras, 179, 433

\bibitem[{{Boccardi} {et~al.}(2021){Boccardi}, {Perucho}, {Casadio}, {Grandi},
  {Macconi}, {Torresi}, {Pellegrini}, {Krichbaum}, {Kadler}, {Giovannini},
  {Karamanavis}, {Ricci}, {Madika}, {Bach}, {Ros}, {Giroletti}, \&
  {Zensus}}]{Boccardi2021}
{Boccardi}, B., {Perucho}, M., {Casadio}, C., {et~al.} 2021, \aap, 647, A67

\bibitem[{{Broderick} {et~al.}(2020){Broderick}, {Gold}, {Karami},
  {Preciado-L{\'o}pez}, {Tiede}, {Pu}, {Akiyama}, {Alberdi}, {Alef}, {Asada},
  {Azulay}, {Baczko}, {Balokovi{\'c}}, {Barrett}, {Bintley}, {Blackburn},
  {Boland}, {Bouman}, {Bower}, {Bremer}, {Brinkerink}, {Brissenden}, {Britzen},
  {Broguiere}, {Bronzwaer}, {Byun}, {Carlstrom}, {Chael}, {Chatterjee},
  {Chatterjee}, {Chen}, {Chen}, {Cho}, {Conway}, {Cordes}, {Crew}, {Cui},
  {Davelaar}, {De Laurentis}, {Deane}, {Dempsey}, {Desvignes}, {Doeleman},
  {Eatough}, {Falcke}, {Fish}, {Fomalont}, {Fraga-Encinas}, {Friberg}, {Fromm},
  {Galison}, {Gammie}, {Garc{\'\i}a}, {Gentaz}, {Georgiev}, {Goddi},
  {G{\'o}mez}, {Gu}, {Gurwell}, {Hada}, {Hecht}, {Hesper}, {Ho}, {Ho}, {Honma},
  {Huang}, {Huang}, {Hughes}, {Inoue}, {Issaoun}, {James}, {Janssen}, {Jeter},
  {Jiang}, {Jim{\'e}nez-Rosales}, {Johnson}, {Jorstad}, {Jung}, {Karuppusamy},
  {Kawashima}, {Keating}, {Kettenis}, {Kim}, {Kim}, {Kino}, {Koay}, {Koch},
  {Koyama}, {Kramer}, {Kramer}, {Krichbaum}, {Kuo}, {Lee}, {Li}, {Li},
  {Lindqvist}, {Lico}, {Liu}, {Liuzzo}, {Lo}, {Lobanov}, {Loinard}, {Lonsdale},
  {Lu}, {MacDonald}, {Mao}, {Marscher}, {Mart{\'\i}-Vidal}, {Matsushita},
  {Matthews}, {Menten}, {Mizuno}, {Mizuno}, {Moran}, {Moriyama},
  {Moscibrodzka}, {M{\"u}ller}, {Nagai}, {Nagar}, {Nakamura}, {Narayan},
  {Narayanan}, {Natarajan}, {Neri}, {Ni}, {Noutsos}, {Okino}, {Olivares},
  {Ortiz-Le{\'o}n}, {Oyama}, {Palumbo}, {Park}, {Pen}, {Pesce}, {Pi{\'e}tu},
  {Plambeck}, {PopStefanija}, {Porth}, {Prather}, {Ramakrishnan}, {Rao},
  {Rawlings}, {Raymond}, {Rezzolla}, {Ripperda}, {Roelofs}, {Rogers}, {Ros},
  {Rose}, {Rottmann}, {Ruszczyk}, {Ryan}, {Rygl}, {S{\'a}nchez},
  {S{\'a}nchez-Arguelles}, {Sasada}, {Savolainen}, {Schloerb}, {Schuster},
  {Shao}, {Shen}, {Small}, {Sohn}, {SooHoo}, {Tazaki}, {Tilanus}, {Titus},
  {Toma}, {Torne}, {Traianou}, {Trippe}, {Tsuda}, {van Bemmel}, {van
  Langevelde}, {van Rossum}, {Wagner}, {Wardle}, {Weintroub}, {Wex}, {Wharton},
  {Wielgus}, {Wong}, {Wu}, {Yoon}, {Young}, {Young}, {Younsi}, {Yuan}, {Yuan},
  {Zensus}, {Zhao}, {Zhao}, {Zhu}, \& {Event Horizon Telescope
  Collaboration}}]{Broderick2020}
{Broderick}, A.~E., {Gold}, R., {Karami}, M., {et~al.} 2020, \apj, 897, 139

\bibitem[{{Chael} {et~al.}(2018){Chael}, {Johnson}, {Bouman}, {Blackburn},
  {Akiyama}, \& {Narayan}}]{Chael2018}
{Chael}, A.~A., {Johnson}, M.~D., {Bouman}, K.~L., {et~al.} 2018, \apj, 857, 23

\bibitem[{{Chael} {et~al.}(2016){Chael}, {Johnson}, {Narayan}, {Doeleman},
  {Wardle}, \& {Bouman}}]{Chael2016}
{Chael}, A.~A., {Johnson}, M.~D., {Narayan}, R., {et~al.} 2016, \apj, 829, 11

\bibitem[{{Chavez} {et~al.}(2026){Chavez}, {Tiede}, {Issaoun}, {Johnson},
  {Pesce}, {Tsunetoe}, \& {Palumbo}}]{Chavez2026}
{Chavez}, E., {Tiede}, P., {Issaoun}, S., {et~al.} 2026, \apj, 1005, 132

\bibitem[{{Cook} {et~al.}(2006){Cook}, {Gelman}, \& {Rubin}}]{Cook2006}
{Cook}, S.~R., {Gelman}, A., \& {Rubin}, D.~B. 2006, Journal of Computational
  and Graphical Statistics, 15, 675

\bibitem[{{Debbrecht} {et~al.}(2026){Debbrecht}, {Paraschos}, {Ros},
  {Krichbaum}, {Bach}, {Gurwell}, {Hodgson}, {Janssen}, {Kim}, {Lisakov},
  {MacDonald}, {Nair}, {Oh}, \& {Zensus}}]{Debbrecht2026}
{Debbrecht}, L.~C., {Paraschos}, G.~F., {Ros}, E., {et~al.} 2026, \aap, 709,
  A192

\bibitem[{{Deller} {et~al.}(2011){Deller}, {Brisken}, {Phillips}, {Morgan},
  {Alef}, {Cappallo}, {Middelberg}, {Romney}, {Rottmann}, {Tingay}, \&
  {Wayth}}]{Deller2011}
{Deller}, A.~T., {Brisken}, W.~F., {Phillips}, C.~J., {et~al.} 2011, \pasp,
  123, 275

\bibitem[{{Event Horizon Telescope Collaboration} {et~al.}(2025){Event Horizon
  Telescope Collaboration}, {Akiyama}, {Albentosa-Ru{\'\i}z}, {Alberdi},
  {Alef}, {Algaba}, {Anantua}, {Asada}, {Azulay}, {Bach}, {Baczko}, {Ball},
  {Balokovi{\'c}}, {Bandyopadhyay}, {Barrett}, {Baub{\"o}ck}, {Benson},
  {Bintley}, {Blackburn}, {Blundell}, {Bouman}, {Bower}, {Bremer},
  {Brissenden}, {Britzen}, {Broderick}, {Broguiere}, {Bronzwaer}, {Bustamante},
  {Carlos}, {Carlstrom}, {Chael}, {Chan}, {Chang}, {Chavez}, {Chatterjee},
  {Chatterjee}, {Chen}, {Chen}, {Cheng}, {Chichura}, {Cho}, {Christian},
  {Conroy}, {Conway}, {Crawford}, {Crew}, {Cruz-Osorio}, {Cui}, {Curd},
  {Dahale}, {Davelaar}, {De Laurentis}, {Deane}, {Desvignes}, {Dexter},
  {Dhruv}, {Dihingia}, {Doeleman}, {Dzib}, {Eatough}, {Emami}, {Falcke},
  {Farah}, {Fish}, {Fomalont}, {Alyson Ford}, {Foschi}, {Fraga-Encinas},
  {Freeman}, {Friberg}, {Fromm}, {Fuentes}, {Galison}, {Gammie}, {Garc{\'\i}a},
  {Gentaz}, {Georgiev}, {Goddi}, {Gold}, {G{\'o}mez-Ruiz}, {G{\'o}mez}, {Gu},
  {Gurwell}, {Hada}, {Haggard}, {Hesper}, {Heumann}, {Ho}, {Ho}, {Hoak},
  {Honma}, {Huang}, {Huang}, {Hughes}, {Ikeda}, {Violette Impellizzeri},
  {Inoue}, {Issaoun}, {James}, {Jannuzi}, {Janssen}, {Jeter}, {Jiang},
  {Jim{\'e}nez-Rosales}, {Johnson}, {Jorstad}, {Jones}, {Joshi}, {Jung},
  {Karuppusamy}, {Kawashima}, {Keating}, {Kettenis}, {Kim}, {Kim}, {Kim},
  {Kim}, {Kino}, {Koay}, {Kocherlakota}, {Kofuji}, {Koch}, {Koyama}, {Kramer},
  {Kramer}, {Kramer}, {Krichbaum}, {Kuo}, {La Bella}, {Lee}, {Lee}, {Levis},
  {Li}, {Lico}, {Lindahl}, {Lindqvist}, {Lisakov}, {Liu}, {Liu}, {Liuzzo},
  {Lo}, {Lobanov}, {Loinard}, {Lonsdale}, {Lowitz}, {Lu}, {MacDonald}, {Mao},
  {Marchili}, {Markoff}, {Marrone}, {Marscher}, {Mart{\'\i}-Vidal},
  {Matsushita}, {Matthews}, {Medeiros}, {Menten}, {Mizuno}, {Mizuno},
  {Montgomery}, {Moran}, {Moriyama}, {Moscibrodzka}, {Mulaudzi}, {M{\"u}ller},
  {M{\"u}ller}, {Mus}, {Musoke}, {Myserlis}, {Nagai}, {Nagar}, {Nair},
  {Nakamura}, {Narayanan}, {Natarajan}, {Nathanail}, {Fuentes}, {Neilsen},
  {Ni}, {Nowak}, {Oh}, {Okino}, {S{\'a}nchez}, {Oyama}, {{\"O}zel}, {Palumbo},
  {Paraschos}, {Park}, {Parsons}, {Patel}, {Pen}, {Pesce}, {Pi{\'e}tu},
  {Plavin}, {PopStefanija}, {Porth}, {Prather}, {Principe}, \&
  {Psaltis}}]{EHTC2025_M87pol}
{Event Horizon Telescope Collaboration}, {Akiyama}, K., {Albentosa-Ru{\'\i}z},
  E., {et~al.} 2025, \aap, 704, A91

\bibitem[{{Event Horizon Telescope Collaboration}
  {et~al.}(2024{\natexlab{a}}){Event Horizon Telescope Collaboration},
  {Akiyama}, {Alberdi}, {Alef}, {Algaba}, {Anantua}, {Asada}, {Azulay}, {Bach},
  {Baczko}, {Ball}, {Balokovic}, {Bandyopadhyay}, {Barrett}, {Baub{\"o}ck},
  {Benson}, {Bintley}, {Blackburn}, {Blundell}, {Bouman}, {Bower}, {Boyce},
  {Bremer}, {Brinkerink}, {Brissenden}, {Britzen}, {Broderick}, {Broguiere},
  {Bronzwaer}, {Bustamante}, {Byun}, {Carlstrom}, {Ceccobello}, {Chael},
  {Chan}, {Chang}, {Chatterjee}, {Chatterjee}, {Chen}, {Chen}, {Cheng}, {Cho},
  {Christian}, {Conroy}, {Conway}, {Cordes}, {Crawford}, {Crew}, {Cruz-Osorio},
  {Cui}, {Dahale}, {Davelaar}, {De Laurentis}, {Deane}, {Dempsey}, {Desvignes},
  {Dexter}, {Dhruv}, {Dihingia}, {Doeleman}, {Dougal}, {Dzib}, {Eatough},
  {Emami}, {Falcke}, {Farah}, {Fish}, {Fomalont}, {Ford}, {Foschi},
  {Fraga-Encinas}, {Freeman}, {Friberg}, {Fromm}, {Fuentes}, {Galison},
  {Gammie}, {Garc{\'\i}a}, {Gentaz}, {Georgiev}, {Goddi}, {Gold},
  {G{\'o}mez-Ruiz}, {G{\'o}mez}, {Gu}, {Gurwell}, {Hada}, {Haggard}, {Haworth},
  {Hecht}, {Hesper}, {Heumann}, {Ho}, {Ho}, {Honma}, {Huang}, {Huang},
  {Hughes}, {Ikeda}, {Impellizzeri}, {Inoue}, {Issaoun}, {James}, {Jannuzi},
  {Janssen}, {Jeter}, {Jiang}, {Jim{\'e}nez-Rosales}, {Johnson}, {Jorstad},
  {Joshi}, {Jung}, {Karami}, {Karuppusamy}, {Kawashima}, {Keating}, {Kettenis},
  {Kim}, {Kim}, {Kim}, {Kim}, {Kino}, {Koay}, {Kocherlakota}, {Kofuji}, {Koch},
  {Koyama}, {Kramer}, {Kramer}, {Kramer}, {Krichbaum}, {Kuo}, {La Bella},
  {Lauer}, {Lee}, {Lee}, {Leung}, {Levis}, {Li}, {Lico}, {Lindahl},
  {Lindqvist}, {Lisakov}, {Liu}, {Liu}, {Liuzzo}, {Lo}, {Lobanov}, {Loinard},
  {Lonsdale}, {Lowitz}, {Lu}, {MacDonald}, {Mao}, {Marchili}, {Markoff},
  {Marrone}, {Marscher}, {Mart{\'\i}-Vidal}, {Matsushita}, {Matthews},
  {Medeiros}, {Menten}, {Michalik}, {Mizuno}, {Mizuno}, {Moran}, {Moriyama},
  {Moscibrodzka}, {Mulaudzi}, {M{\"u}ller}, {M{\"u}ller}, {Mus}, {Musoke},
  {Myserlis}, {Nadolski}, {Nagai}, {Nagar}, {Nakamura}, {Narayanan},
  {Natarajan}, {Nathanail}, {Fuentes}, {Neilsen}, {Neri}, {Ni}, {Noutsos},
  {Nowak}, {Oh}, {Okino}, {Olivares}, {Ortiz-Le{\'o}n}, {Oyama}, {{\"O}zel},
  {Palumbo}, {Paraschos}, {Park}, {Parsons}, {Patel}, \&
  {Pen}}]{EHTC2024_SgrApol}
{Event Horizon Telescope Collaboration}, {Akiyama}, K., {Alberdi}, A., {et~al.}
  2024{\natexlab{a}}, \apjl, 964, L25

\bibitem[{{Event Horizon Telescope Collaboration}
  {et~al.}(2024{\natexlab{b}}){Event Horizon Telescope Collaboration},
  {Akiyama}, {Alberdi}, {Alef}, {Algaba}, {Anantua}, {Asada}, {Azulay}, {Bach},
  {Baczko}, {Ball}, {Balokovi{\'c}}, {Bandyopadhyay}, {Barrett}, {Baub{\"o}ck},
  {Benson}, {Bintley}, {Blackburn}, {Blundell}, {Bouman}, {Bower}, {Boyce},
  {Bremer}, {Brinkerink}, {Brissenden}, {Britzen}, {Broderick}, {Broguiere},
  {Bronzwaer}, {Bustamante}, {Byun}, {Carlstrom}, {Ceccobello}, {Chael},
  {Chan}, {Chang}, {Chatterjee}, {Chatterjee}, {Chen}, {Chen}, {Cheng}, {Cho},
  {Christian}, {Conroy}, {Conway}, {Cordes}, {Crawford}, {Crew}, {Cruz-Osorio},
  {Cui}, {Dahale}, {Davelaar}, {De Laurentis}, {Deane}, {Dempsey}, {Desvignes},
  {Dexter}, {Dhruv}, {Dihingia}, {Doeleman}, {Dougall}, {Dzib}, {Eatough},
  {Emami}, {Falcke}, {Farah}, {Fish}, {Fomalont}, {Ford}, {Foschi},
  {Fraga-Encinas}, {Freeman}, {Friberg}, {Fromm}, {Fuentes}, {Galison},
  {Gammie}, {Garc{\'\i}a}, {Gentaz}, {Georgiev}, {Goddi}, {Gold},
  {G{\'o}mez-Ruiz}, {G{\'o}mez}, {Gu}, {Gurwell}, {Hada}, {Haggard}, {Haworth},
  {Hecht}, {Hesper}, {Heumann}, {Ho}, {Ho}, {Honma}, {Huang}, {Huang},
  {Hughes}, {Ikeda}, {Impellizzeri}, {Inoue}, {Issaoun}, {James}, {Jannuzi},
  {Janssen}, {Jeter}, {Jiang}, {Jim{\'e}nez-Rosales}, {Johnson}, {Jorstad},
  {Joshi}, {Jung}, {Karami}, {Karuppusamy}, {Kawashima}, {Keating}, {Kettenis},
  {Kim}, {Kim}, {Kim}, {Kim}, {Kino}, {Koay}, {Kocherlakota}, {Kofuji}, {Koch},
  {Koyama}, {Kramer}, {Kramer}, {Kramer}, {Krichbaum}, {Kuo}, {La Bella},
  {Lauer}, {Lee}, {Lee}, {Leung}, {Levis}, {Li}, {Lico}, {Lindahl},
  {Lindqvist}, {Lisakov}, {Liu}, {Liu}, {Liuzzo}, {Lo}, {Lobanov}, {Loinard},
  {Lonsdale}, {Lowitz}, {Lu}, {MacDonald}, {Mao}, {Marchili}, {Markoff},
  {Marrone}, {Marscher}, {Mart{\'\i}-Vidal}, {Matsushita}, {Matthews},
  {Medeiros}, {Menten}, {Michalik}, {Mizuno}, {Mizuno}, {Moran}, {Moriyama},
  {Moscibrodzka}, {Mulaudzi}, {M{\"u}ller}, {M{\"u}ller}, {Mus}, {Musoke},
  {Myserlis}, {Nadolski}, {Nagai}, {Nagar}, {Nakamura}, {Narayanan},
  {Natarajan}, {Nathanail}, {Fuentes}, {Neilsen}, {Neri}, {Ni}, {Noutsos},
  {Nowak}, {Oh}, {Okino}, {Olivares}, {Ortiz-Le{\'o}n}, {Oyama}, {{\"O}zel},
  {Palumbo}, {Paraschos}, {Park}, {Parsons}, {Patel}, \&
  {Pen}}]{EHTC2024_SgrApol_VIII}
{Event Horizon Telescope Collaboration}, {Akiyama}, K., {Alberdi}, A., {et~al.}
  2024{\natexlab{b}}, \apjl, 964, L26

\bibitem[{{Event Horizon Telescope Collaboration}
  {et~al.}(2024{\natexlab{c}}){Event Horizon Telescope Collaboration},
  {Akiyama}, {Alberdi}, {Alef}, {Algaba}, {Anantua}, {Asada}, {Azulay}, {Bach},
  {Baczko}, {Ball}, {Balokovi{\'c}}, {Bandyopadhyay}, {Barrett}, {Baub{\"o}ck},
  {Benson}, {Bintley}, {Blackburn}, {Blundell}, {Bouman}, {Bower}, {Boyce},
  {Bremer}, {Brissenden}, {Britzen}, {Broderick}, {Broguiere}, {Bronzwaer},
  {Bustamante}, {Carlstrom}, {Chael}, {Chan}, {Chang}, {Chatterjee},
  {Chatterjee}, {Chen}, {Chen}, {Cheng}, {Cho}, {Christian}, {Conroy},
  {Conway}, {Crawford}, {Crew}, {Cruz-Osorio}, {Cui}, {Dahale}, {Davelaar}, {De
  Laurentis}, {Deane}, {Dempsey}, {Desvignes}, {Dexter}, {Dhruv}, {Dihingia},
  {Doeleman}, {Dzib}, {Eatough}, {Emami}, {Falcke}, {Farah}, {Fish},
  {Fomalont}, {Ford}, {Foschi}, {Fraga-Encinas}, {Freeman}, {Friberg}, {Fromm},
  {Fuentes}, {Galison}, {Gammie}, {Garc{\'\i}a}, {Gentaz}, {Georgiev}, {Goddi},
  {Gold}, {G{\'o}mez-Ruiz}, {G{\'o}mez}, {Gu}, {Gurwell}, {Hada}, {Haggard},
  {Hesper}, {Heumann}, {Ho}, {Ho}, {Honma}, {Huang}, {Huang}, {Hughes},
  {Ikeda}, {Violette Impellizzeri}, {Inoue}, {Issaoun}, {James}, {Jannuzi},
  {Janssen}, {Jeter}, {Jiang}, {Jim{\'e}nez-Rosales}, {Johnson}, {Jorstad},
  {Jones}, {Joshi}, {Jung}, {Karuppusamy}, {Kawashima}, {Keating}, {Kettenis},
  {Kim}, {Kim}, {Kim}, {Kim}, {Kino}, {Koay}, {Kocherlakota}, {Kofuji}, {Koch},
  {Koyama}, {Kramer}, {Kramer}, {Kramer}, {Krichbaum}, {Kuo}, {La Bella},
  {Lee}, {Levis}, {Li}, {Lico}, {Lindahl}, {Lindqvist}, {Lisakov}, {Liu},
  {Liu}, {Liuzzo}, {Lo}, {Lobanov}, {Loinard}, {Lonsdale}, {Lowitz}, {Lu},
  {MacDonald}, {Mao}, {Marchili}, {Markoff}, {Marrone}, {Marscher},
  {Mart{\'\i}-Vidal}, {Matsushita}, {Matthews}, {Medeiros}, {Menten}, {Mizuno},
  {Mizuno}, {Montgomery}, {Moran}, {Moriyama}, {Moscibrodzka}, {Mulaudzi},
  {M{\"u}ller}, {M{\"u}ller}, {Mus}, {Musoke}, {Myserlis}, {Nagai}, {Nagar},
  {Nakamura}, {Narayanan}, {Natarajan}, {Nathanail}, {Fuentes}, {Neilsen},
  {Ni}, {Nowak}, {Oh}, {Okino}, {Olivares}, {Oyama}, {{\"O}zel}, {Palumbo},
  {Paraschos}, {Park}, {Parsons}, {Patel}, {Pen}, {Pesce}, {Pi{\'e}tu},
  {PopStefanija}, {Porth}, {Prather}, {Psaltis}, {Pu}, {Ramakrishnan}, {Rao},
  {Rawlings}, {Raymond}, {Rezzolla}, {Ricarte}, \&
  {Ripperda}}]{EHTC2024_M87shadow}
{Event Horizon Telescope Collaboration}, {Akiyama}, K., {Alberdi}, A., {et~al.}
  2024{\natexlab{c}}, \aap, 681, A79

\bibitem[{{Event Horizon Telescope Collaboration} {et~al.}(2023){Event Horizon
  Telescope Collaboration}, {Akiyama}, {Alberdi}, {Alef}, {Algaba}, {Anantua},
  {Asada}, {Azulay}, {Bach}, {Baczko}, {Ball}, {Balokovi{\'c}}, {Barrett},
  {Baub{\"o}ck}, {Benson}, {Bintley}, {Blackburn}, {Blundell}, {Bouman},
  {Bower}, {Boyce}, {Bremer}, {Brinkerink}, {Brissenden}, {Britzen},
  {Broderick}, {Broguiere}, {Bronzwaer}, {Bustamante}, {Byun}, {Carlstrom},
  {Ceccobello}, {Chael}, {Chan}, {Chang}, {Chatterjee}, {Chatterjee}, {Chen},
  {Chen}, {Cheng}, {Cho}, {Christian}, {Conroy}, {Conway}, {Cordes},
  {Crawford}, {Crew}, {Cruz-Osorio}, {Cui}, {Dahale}, {Davelaar}, {De
  Laurentis}, {Deane}, {Dempsey}, {Desvignes}, {Dexter}, {Dhruv}, {Doeleman},
  {Dougal}, {Dzib}, {Eatough}, {Emami}, {Falcke}, {Farah}, {Fish}, {Fomalont},
  {Ford}, {Foschi}, {Fraga-Encinas}, {Freeman}, {Friberg}, {Fromm}, {Fuentes},
  {Galison}, {Gammie}, {Garc{\'\i}a}, {Gentaz}, {Georgiev}, {Goddi}, {Gold},
  {G{\'o}mez-Ruiz}, {G{\'o}mez}, {Gu}, {Gurwell}, {Hada}, {Haggard}, {Haworth},
  {Hecht}, {Hesper}, {Heumann}, {Ho}, {Ho}, {Honma}, {Huang}, {Huang},
  {Hughes}, {Ikeda}, {Impellizzeri}, {Inoue}, {Issaoun}, {James}, {Jannuzi},
  {Janssen}, {Jeter}, {Jiang}, {Jim{\'e}nez-Rosales}, {Johnson}, {Jorstad},
  {Joshi}, {Jung}, {Karami}, {Karuppusamy}, {Kawashima}, {Keating}, {Kettenis},
  {Kim}, {Kim}, {Kim}, {Kim}, {Kino}, {Koay}, {Kocherlakota}, {Kofuji}, {Koch},
  {Koyama}, {Kramer}, {Kramer}, {Kramer}, {Krichbaum}, {Kuo}, {La Bella},
  {Lauer}, {Lee}, {Lee}, {Leung}, {Levis}, {Li}, {Lico}, {Lindahl},
  {Lindqvist}, {Lisakov}, {Liu}, {Liu}, {Liuzzo}, {Lo}, {Lobanov}, {Loinard},
  {Lonsdale}, {Lowitz}, {Lu}, {MacDonald}, {Mao}, {Marchili}, {Markoff},
  {Marrone}, {Marscher}, {Mart{\'\i}-Vidal}, {Matsushita}, {Matthews},
  {Medeiros}, {Menten}, {Michalik}, {Mizuno}, {Mizuno}, {Moran}, {Moriyama},
  {Moscibrodzka}, {Mulaudzi}, {M{\"u}ller}, {M{\"u}ller}, {Mus}, {Musoke},
  {Myserlis}, {Nadolski}, {Nagai}, {Nagar}, {Nakamura}, {Narayan}, {Narayanan},
  {Natarajan}, {Nathanail}, {Fuentes}, {Neilsen}, {Neri}, {Ni}, {Noutsos},
  {Nowak}, {Oh}, {Okino}, {Olivares}, {Ortiz-Le{\'o}n}, {Oyama}, {{\"O}zel},
  {Palumbo}, {Paraschos}, {Park}, {Parsons}, {Patel}, {Pen}, \&
  {Pesce}}]{EHTC2023_M87_StokesV}
{Event Horizon Telescope Collaboration}, {Akiyama}, K., {Alberdi}, A., {et~al.}
  2023, \apjl, 957, L20

\bibitem[{{Event Horizon Telescope Collaboration} {et~al.}(2022){Event Horizon
  Telescope Collaboration}, {Akiyama}, {Alberdi}, {Alef}, {Algaba}, {Anantua},
  {Asada}, {Azulay}, {Bach}, {Baczko}, {Ball}, {Balokovi{\'c}}, {Barrett},
  {Baub{\"o}ck}, {Benson}, {Bintley}, {Blackburn}, {Blundell}, {Bouman},
  {Bower}, {Boyce}, {Bremer}, {Brinkerink}, {Brissenden}, {Britzen},
  {Broderick}, {Broguiere}, {Bronzwaer}, {Bustamante}, {Byun}, {Carlstrom},
  {Ceccobello}, {Chael}, {Chan}, {Chatterjee}, {Chatterjee}, {Chen}, {Chen},
  {Cheng}, {Cho}, {Christian}, {Conroy}, {Conway}, {Cordes}, {Crawford},
  {Crew}, {Cruz-Osorio}, {Cui}, {Davelaar}, {De Laurentis}, {Deane}, {Dempsey},
  {Desvignes}, {Dexter}, {Dhruv}, {Doeleman}, {Dougal}, {Dzib}, {Eatough},
  {Emami}, {Falcke}, {Farah}, {Fish}, {Fomalont}, {Ford}, {Fraga-Encinas},
  {Freeman}, {Friberg}, {Fromm}, {Fuentes}, {Galison}, {Gammie}, {Garc{\'\i}a},
  {Gentaz}, {Georgiev}, {Goddi}, {Gold}, {G{\'o}mez-Ruiz}, {G{\'o}mez}, {Gu},
  {Gurwell}, {Hada}, {Haggard}, {Haworth}, {Hecht}, {Hesper}, {Heumann}, {Ho},
  {Ho}, {Honma}, {Huang}, {Huang}, {Hughes}, {Ikeda}, {Impellizzeri}, {Inoue},
  {Issaoun}, {James}, {Jannuzi}, {Janssen}, {Jeter}, {Jiang},
  {Jim{\'e}nez-Rosales}, {Johnson}, {Jorstad}, {Joshi}, {Jung}, {Karami},
  {Karuppusamy}, {Kawashima}, {Keating}, {Kettenis}, {Kim}, {Kim}, {Kim},
  {Kim}, {Kino}, {Koay}, {Kocherlakota}, {Kofuji}, {Koch}, {Koyama}, {Kramer},
  {Kramer}, {Krichbaum}, {Kuo}, {La Bella}, {Lauer}, {Lee}, {Lee}, {Leung},
  {Levis}, {Li}, {Lico}, {Lindahl}, {Lindqvist}, {Lisakov}, {Liu}, {Liu},
  {Liuzzo}, {Lo}, {Lobanov}, {Loinard}, {Lonsdale}, {Lu}, {Mao}, {Marchili},
  {Markoff}, {Marrone}, {Marscher}, {Mart{\'\i}-Vidal}, {Matsushita},
  {Matthews}, {Medeiros}, {Menten}, {Michalik}, {Mizuno}, {Mizuno}, {Moran},
  {Moriyama}, {Moscibrodzka}, {M{\"u}ller}, {Mus}, {Musoke}, {Myserlis},
  {Nadolski}, {Nagai}, {Nagar}, {Nakamura}, {Narayan}, {Narayanan},
  {Natarajan}, {Nathanail}, {Fuentes}, {Neilsen}, {Neri}, {Ni}, {Noutsos},
  {Nowak}, {Oh}, {Okino}, {Olivares}, {Ortiz-Le{\'o}n}, {Oyama}, {{\"O}zel},
  {Palumbo}, {Paraschos}, {Park}, {Parsons}, {Patel}, {Pen}, {Pesce},
  {Pi{\'e}tu}, {Plambeck}, {PopStefanija}, {Porth}, {P{\"o}tzl}, {Prather},
  {Preciado-L{\'o}pez}, \& {Psaltis}}]{EHT2022a}
{Event Horizon Telescope Collaboration}, {Akiyama}, K., {Alberdi}, A., {et~al.}
  2022, \apjl, 930, L12

\bibitem[{{Event Horizon Telescope Collaboration} {et~al.}(2019){Event Horizon
  Telescope Collaboration}, {Akiyama}, {Alberdi}, {Alef}, {Asada}, {Azulay},
  {Baczko}, {Ball}, {Balokovi{\'c}}, {Barrett}, {Bintley}, {Blackburn},
  {Boland}, {Bouman}, {Bower}, {Bremer}, {Brinkerink}, {Brissenden}, {Britzen},
  {Broderick}, {Broguiere}, {Bronzwaer}, {Byun}, {Carlstrom}, {Chael}, {Chan},
  {Chatterjee}, {Chatterjee}, {Chen}, {Chen}, {Cho}, {Christian}, {Conway},
  {Cordes}, {Crew}, {Cui}, {Davelaar}, {De Laurentis}, {Deane}, {Dempsey},
  {Desvignes}, {Dexter}, {Doeleman}, {Eatough}, {Falcke}, {Fish}, {Fomalont},
  {Fraga-Encinas}, {Freeman}, {Friberg}, {Fromm}, {G{\'o}mez}, {Galison},
  {Gammie}, {Garc{\'\i}a}, {Gentaz}, {Georgiev}, {Goddi}, {Gold}, {Gu},
  {Gurwell}, {Hada}, {Hecht}, {Hesper}, {Ho}, {Ho}, {Honma}, {Huang}, {Huang},
  {Hughes}, {Ikeda}, {Inoue}, {Issaoun}, {James}, {Jannuzi}, {Janssen},
  {Jeter}, {Jiang}, {Johnson}, {Jorstad}, {Jung}, {Karami}, {Karuppusamy},
  {Kawashima}, {Keating}, {Kettenis}, {Kim}, {Kim}, {Kim}, {Kino}, {Koay},
  {Koch}, {Koyama}, {Kramer}, {Kramer}, {Krichbaum}, {Kuo}, {Lauer}, {Lee},
  {Li}, {Li}, {Lindqvist}, {Liu}, {Liuzzo}, {Lo}, {Lobanov}, {Loinard},
  {Lonsdale}, {Lu}, {MacDonald}, {Mao}, {Markoff}, {Marrone}, {Marscher},
  {Mart{\'\i}-Vidal}, {Matsushita}, {Matthews}, {Medeiros}, {Menten}, {Mizuno},
  {Mizuno}, {Moran}, {Moriyama}, {Moscibrodzka}, {M{\"u}ller}, {Nagai},
  {Nagar}, {Nakamura}, {Narayan}, {Narayanan}, {Natarajan}, {Neri}, {Ni},
  {Noutsos}, {Okino}, {Olivares}, {Oyama}, {{\"O}zel}, {Palumbo}, {Patel},
  {Pen}, {Pesce}, {Pi{\'e}tu}, {Plambeck}, {PopStefanija}, {Porth}, {Prather},
  {Preciado-L{\'o}pez}, {Psaltis}, {Pu}, {Ramakrishnan}, {Rao}, {Rawlings},
  {Raymond}, {Rezzolla}, {Ripperda}, {Roelofs}, {Rogers}, {Ros}, {Rose},
  {Roshanineshat}, {Rottmann}, {Roy}, {Ruszczyk}, {Ryan}, {Rygl},
  {S{\'a}nchez}, {S{\'a}nchez-Arguelles}, {Sasada}, {Savolainen}, {Schloerb},
  {Schuster}, {Shao}, {Shen}, {Small}, {Sohn}, {SooHoo}, {Tazaki}, {Tiede},
  {Tilanus}, {Titus}, {Toma}, {Torne}, {Trent}, {Trippe}, {Tsuda}, {van
  Bemmel}, {van Langevelde}, {van Rossum}, {Wagner}, {Wardle}, {Weintroub},
  {Wex}, {Wharton}, {Wielgus}, {Wong}, {Wu}, {Young}, {Young}, \&
  {Younsi}}]{EHTC2019d}
{Event Horizon Telescope Collaboration}, {Akiyama}, K., {Alberdi}, A., {et~al.}
  2019, \apjl, 875, L4

\bibitem[{{Event Horizon Telescope Collaboration}
  {et~al.}(2021{\natexlab{a}}){Event Horizon Telescope Collaboration},
  {Akiyama}, {Algaba}, {Alberdi}, {Alef}, {Anantua}, {Asada}, {Azulay},
  {Baczko}, {Ball}, {Balokovi{\'c}}, {Barrett}, {Benson}, {Bintley},
  {Blackburn}, {Blundell}, {Boland}, {Bouman}, {Bower}, {Boyce}, {Bremer},
  {Brinkerink}, {Brissenden}, {Britzen}, {Broderick}, {Broguiere}, {Bronzwaer},
  {Byun}, {Carlstrom}, {Chael}, {Chan}, {Chatterjee}, {Chatterjee}, {Chen},
  {Chen}, {Chesler}, {Cho}, {Christian}, {Conway}, {Cordes}, {Crawford},
  {Crew}, {Cruz-Osorio}, {Cui}, {Davelaar}, {De Laurentis}, {Deane}, {Dempsey},
  {Desvignes}, {Dexter}, {Doeleman}, {Eatough}, {Falcke}, {Farah}, {Fish},
  {Fomalont}, {Ford}, {Fraga-Encinas}, {Freeman}, {Friberg}, {Fromm},
  {Fuentes}, {Galison}, {Gammie}, {Garc{\'\i}a}, {Gentaz}, {Georgiev}, {Goddi},
  {Gold}, {G{\'o}mez}, {G{\'o}mez-Ruiz}, {Gu}, {Gurwell}, {Hada}, {Haggard},
  {Hecht}, {Hesper}, {Ho}, {Ho}, {Honma}, {Huang}, {Huang}, {Hughes}, {Ikeda},
  {Inoue}, {Issaoun}, {James}, {Jannuzi}, {Janssen}, {Jeter}, {Jiang},
  {Jimenez-Rosales}, {Johnson}, {Jorstad}, {Jung}, {Karami}, {Karuppusamy},
  {Kawashima}, {Keating}, {Kettenis}, {Kim}, {Kim}, {Kim}, {Kim}, {Kino},
  {Koay}, {Kofuji}, {Koch}, {Koyama}, {Kramer}, {Kramer}, {Krichbaum}, {Kuo},
  {Lauer}, {Lee}, {Levis}, {Li}, {Li}, {Lindqvist}, {Lico}, {Lindahl}, {Liu},
  {Liu}, {Liuzzo}, {Lo}, {Lobanov}, {Loinard}, {Lonsdale}, {Lu}, {MacDonald},
  {Mao}, {Marchili}, {Markoff}, {Marrone}, {Marscher}, {Mart{\'\i}-Vidal},
  {Matsushita}, {Matthews}, {Medeiros}, {Menten}, {Mizuno}, {Mizuno}, {Moran},
  {Moriyama}, {Moscibrodzka}, {M{\"u}ller}, {Musoke}, {Mej{\'\i}as},
  {Michalik}, {Nadolski}, {Nagai}, {Nagar}, {Nakamura}, {Narayan}, {Narayanan},
  {Natarajan}, {Nathanail}, {Neilsen}, {Neri}, {Ni}, {Noutsos}, {Nowak},
  {Okino}, {Olivares}, {Ortiz-Le{\'o}n}, {Oyama}, {{\"O}zel}, {Palumbo},
  {Park}, {Patel}, {Pen}, {Pesce}, {Pi{\'e}tu}, {Plambeck}, {PopStefanija},
  {Porth}, {P{\"o}tzl}, {Prather}, {Preciado-L{\'o}pez}, {Psaltis}, {Pu},
  {Ramakrishnan}, {Rao}, {Rawlings}, {Raymond}, {Rezzolla}, {Ricarte},
  {Ripperda}, {Roelofs}, {Rogers}, {Ros}, {Rose}, {Roshanineshat}, {Rottmann},
  {Roy}, {Ruszczyk}, {Rygl}, {S{\'a}nchez}, {S{\'a}nchez-Arguelles}, \&
  {Sasada}}]{EHTC2021a}
{Event Horizon Telescope Collaboration}, {Akiyama}, K., {Algaba}, J.~C.,
  {et~al.} 2021{\natexlab{a}}, \apjl, 910, L12

\bibitem[{{Event Horizon Telescope Collaboration}
  {et~al.}(2021{\natexlab{b}}){Event Horizon Telescope Collaboration},
  {Akiyama}, {Algaba}, {Alberdi}, {Alef}, {Anantua}, {Asada}, {Azulay},
  {Baczko}, {Ball}, {Balokovi{\'c}}, {Barrett}, {Benson}, {Bintley},
  {Blackburn}, {Blundell}, {Boland}, {Bouman}, {Bower}, {Boyce}, {Bremer},
  {Brinkerink}, {Brissenden}, {Britzen}, {Broderick}, {Broguiere}, {Bronzwaer},
  {Byun}, {Carlstrom}, {Chael}, {Chan}, {Chatterjee}, {Chatterjee}, {Chen},
  {Chen}, {Chesler}, {Cho}, {Christian}, {Conway}, {Cordes}, {Crawford},
  {Crew}, {Cruz-Osorio}, {Cui}, {Davelaar}, {De Laurentis}, {Deane}, {Dempsey},
  {Desvignes}, {Dexter}, {Doeleman}, {Eatough}, {Falcke}, {Farah}, {Fish},
  {Fomalont}, {Ford}, {Fraga-Encinas}, {Friberg}, {Fromm}, {Fuentes},
  {Galison}, {Gammie}, {Garc{\'\i}a}, {Gelles}, {Gentaz}, {Georgiev}, {Goddi},
  {Gold}, {G{\'o}mez}, {G{\'o}mez-Ruiz}, {Gu}, {Gurwell}, {Hada}, {Haggard},
  {Hecht}, {Hesper}, {Himwich}, {Ho}, {Ho}, {Honma}, {Huang}, {Huang},
  {Hughes}, {Ikeda}, {Inoue}, {Issaoun}, {James}, {Jannuzi}, {Janssen},
  {Jeter}, {Jiang}, {Jimenez-Rosales}, {Johnson}, {Jorstad}, {Jung}, {Karami},
  {Karuppusamy}, {Kawashima}, {Keating}, {Kettenis}, {Kim}, {Kim}, {Kim},
  {Kim}, {Kino}, {Koay}, {Kofuji}, {Koch}, {Koyama}, {Kramer}, {Kramer},
  {Krichbaum}, {Kuo}, {Lauer}, {Lee}, {Levis}, {Li}, {Li}, {Lindqvist}, {Lico},
  {Lindahl}, {Liu}, {Liu}, {Liuzzo}, {Lo}, {Lobanov}, {Loinard}, {Lonsdale},
  {Lu}, {MacDonald}, {Mao}, {Marchili}, {Markoff}, {Marrone}, {Marscher},
  {Mart{\'\i}-Vidal}, {Matsushita}, {Matthews}, {Medeiros}, {Menten}, {Mizuno},
  {Mizuno}, {Moran}, {Moriyama}, {Moscibrodzka}, {M{\"u}ller}, {Musoke}, {Mus
  Mej{\'\i}as}, {Michalik}, {Nadolski}, {Nagai}, {Nagar}, {Nakamura},
  {Narayan}, {Narayanan}, {Natarajan}, {Nathanail}, {Neilsen}, {Neri}, {Ni},
  {Noutsos}, {Nowak}, {Okino}, {Olivares}, {Ortiz-Le{\'o}n}, {Oyama},
  {{\"O}zel}, {Palumbo}, {Park}, {Patel}, {Pen}, {Pesce}, {Pi{\'e}tu},
  {Plambeck}, {PopStefanija}, {Porth}, {P{\"o}tzl}, {Prather},
  {Preciado-L{\'o}pez}, {Psaltis}, {Pu}, {Ramakrishnan}, {Rao}, {Rawlings},
  {Raymond}, {Rezzolla}, {Ricarte}, {Ripperda}, {Roelofs}, {Rogers}, {Ros},
  {Rose}, {Roshanineshat}, {Rottmann}, {Roy}, {Ruszczyk}, {Rygl},
  {S{\'a}nchez}, \& {S{\'a}nchez-Arguelles}}]{EHTC2021b}
{Event Horizon Telescope Collaboration}, {Akiyama}, K., {Algaba}, J.~C.,
  {et~al.} 2021{\natexlab{b}}, \apjl, 910, L13

\bibitem[{{Foschi} {et~al.}(2025){Foschi}, {G{\'o}mez}, {Fuentes}, {Cho},
  {Marscher}, \& {Jorstad}}]{Foschi2025}
{Foschi}, M., {G{\'o}mez}, J.~L., {Fuentes}, A., {et~al.} 2025, \aap, 696, A17

\bibitem[{{Gelman} {et~al.}(2014){Gelman}, {Carlin}, {Stern}, {Dunson},
  {Vehtari}, \& {Rubin}}]{Gelman2013}
{Gelman}, A., {Carlin}, J.~B., {Stern}, H.~S., {et~al.} 2014, {Bayesian Data
  Analysis}

\bibitem[{{Giovannini} {et~al.}(2018){Giovannini}, {Savolainen}, {Orienti},
  {Nakamura}, {Nagai}, {Kino}, {Giroletti}, {Hada}, {Bruni}, {Kovalev},
  {Anderson}, {D'Ammando}, {Hodgson}, {Honma}, {Krichbaum}, {Lee}, {Lico},
  {Lisakov}, {Lobanov}, {Petrov}, {Sohn}, {Sokolovsky}, {Voitsik}, {Zensus}, \&
  {Tingay}}]{Giovannini2018}
{Giovannini}, G., {Savolainen}, T., {Orienti}, M., {et~al.} 2018, Nature
  Astronomy, 2, 472

\bibitem[{{Greisen}(2003)}]{Greisen2003}
{Greisen}, E.~W. 2003, in Astrophysics and Space Science Library, Vol. 285,
  Information Handling in Astronomy - Historical Vistas, ed. A.~{Heck}, 109

\bibitem[{{Hada} {et~al.}(2018){Hada}, {Doi}, {Wajima}, {D'Ammando}, {Orienti},
  {Giroletti}, {Giovannini}, {Nakamura}, \& {Asada}}]{Hada2018}
{Hada}, K., {Doi}, A., {Wajima}, K., {et~al.} 2018, \apj, 860, 141

\bibitem[{{Hada} {et~al.}(2016){Hada}, {Kino}, {Doi}, {Nagai}, {Honma},
  {Akiyama}, {Tazaki}, {Lico}, {Giroletti}, {Giovannini}, {Orienti}, \&
  {Hagiwara}}]{Hada2016}
{Hada}, K., {Kino}, M., {Doi}, A., {et~al.} 2016, \apj, 817, 131

\bibitem[{{Hada} {et~al.}(2013){Hada}, {Kino}, {Doi}, {Nagai}, {Honma},
  {Hagiwara}, {Giroletti}, {Giovannini}, \& {Kawaguchi}}]{Hada2013}
{Hada}, K., {Kino}, M., {Doi}, A., {et~al.} 2013, \apj, 775, 70

\bibitem[{{Hamaker} {et~al.}(1996){Hamaker}, {Bregman}, \&
  {Sault}}]{Hamaker1996}
{Hamaker}, J.~P., {Bregman}, J.~D., \& {Sault}, R.~J. 1996, \aaps, 117, 137

\bibitem[{Hoffman \& Gelman(2014)}]{HoffmanGelman2014}
Hoffman, M.~D. \& Gelman, A. 2014, Journal of Machine Learning Research, 15,
  1593

\bibitem[{{H{\"o}gbom}(1974)}]{Hogbom1974}
{H{\"o}gbom}, J.~A. 1974, \aaps, 15, 417

\bibitem[{{Homan} {et~al.}(2001){Homan}, {Attridge}, \& {Wardle}}]{Homan2001}
{Homan}, D.~C., {Attridge}, J.~M., \& {Wardle}, J. F.~C. 2001, \apj, 556, 113

\bibitem[{{Homan} \& {Lister}(2006)}]{HomanLister2006}
{Homan}, D.~C. \& {Lister}, M.~L. 2006, \aj, 131, 1262

\bibitem[{{Homan} \& {Wardle}(1999)}]{HomanWardle1999}
{Homan}, D.~C. \& {Wardle}, J.~F.~C. 1999, \aj, 118, 1942

\bibitem[{{Honma} {et~al.}(2014){Honma}, {Akiyama}, {Uemura}, \&
  {Ikeda}}]{Honma2014}
{Honma}, M., {Akiyama}, K., {Uemura}, M., \& {Ikeda}, S. 2014, \pasj, 66, 95

\bibitem[{{Jorstad} {et~al.}(2017){Jorstad}, {Marscher}, {Morozova},
  {Troitsky}, {Agudo}, {Casadio}, {Foord}, {G{\'o}mez}, {MacDonald}, {Molina},
  {L{\"a}hteenm{\"a}ki}, {Tammi}, \& {Tornikoski}}]{Jorstad2017}
{Jorstad}, S.~G., {Marscher}, A.~P., {Morozova}, D.~A., {et~al.} 2017, \apj,
  846, 98

\bibitem[{{Kam} {et~al.}(2024){Kam}, {Hodgson}, {Park}, {Kino}, {Nagai},
  {Trippe}, \& {Wagner}}]{Kam2024}
{Kam}, M., {Hodgson}, J.~A., {Park}, J., {et~al.} 2024, \apj, 970, 176

\bibitem[{{Kam} {et~al.}(2026){Kam}, {Park}, {Trippe}, {Byun}, {Asada},
  {Nagai}, {Kino}, {Ho}, {Fabian}, {Hodgson}, {Yi}, {Toma}, {Oh}, \&
  {Jung}}]{Kam2026}
{Kam}, M., {Park}, J., {Trippe}, S., {et~al.} 2026, \apj, 1003, 51

\bibitem[{{Kim} {et~al.}(2025){Kim}, {M{\"u}ller}, {Nikonov}, {Lu},
  {Knollm{\"u}ller}, {En{\ss}lin}, {Wielgus}, \& {Lobanov}}]{Kim2025_M87_86GHz}
{Kim}, J.-S., {M{\"u}ller}, H., {Nikonov}, A.~S., {et~al.} 2025, \aap, 696,
  A169

\bibitem[{{Kim} {et~al.}(2024){Kim}, {Nikonov}, {Roth}, {En{\ss}lin},
  {Janssen}, {Arras}, {M{\"u}ller}, \& {Lobanov}}]{Kim2024}
{Kim}, J.-S., {Nikonov}, A.~S., {Roth}, J., {et~al.} 2024, \aap, 690, A129

\bibitem[{{Kim} {et~al.}(2026){Kim}, {Roth}, {Park}, {Livingston}, {Arras},
  {En{\ss}lin}, {Janssen}, {Zensus}, \& {Lobanov}}]{Kim2025}
{Kim}, J.-S., {Roth}, J., {Park}, J., {et~al.} 2026, \aap, 711, A247

\bibitem[{{Komissarov} {et~al.}(2007){Komissarov}, {Barkov}, {Vlahakis}, \&
  {K{\"o}nigl}}]{Komissarov2007}
{Komissarov}, S.~S., {Barkov}, M.~V., {Vlahakis}, N., \& {K{\"o}nigl}, A. 2007,
  \mnras, 380, 51

\bibitem[{{Kovalev} {et~al.}(2020){Kovalev}, {Pushkarev}, {Nokhrina}, {Plavin},
  {Beskin}, {Chernoglazov}, {Lister}, \& {Savolainen}}]{Kovalev2020}
{Kovalev}, Y.~Y., {Pushkarev}, A.~B., {Nokhrina}, E.~E., {et~al.} 2020, \mnras,
  495, 3576

\bibitem[{{Leppanen} {et~al.}(1995){Leppanen}, {Zensus}, \&
  {Diamond}}]{Leppanen1995}
{Leppanen}, K.~J., {Zensus}, J.~A., \& {Diamond}, P.~J. 1995, \aj, 110, 2479

\bibitem[{{Lister} {et~al.}(2018){Lister}, {Aller}, {Aller}, {Hodge}, {Homan},
  {Kovalev}, {Pushkarev}, \& {Savolainen}}]{Lister2018}
{Lister}, M.~L., {Aller}, M.~F., {Aller}, H.~D., {et~al.} 2018, \apjs, 234, 12

\bibitem[{{Lobanov}(2005)}]{Lobanov2005}
{Lobanov}, A.~P. 2005, arXiv e-prints, astro

\bibitem[{{Lu} {et~al.}(2023){Lu}, {Asada}, {Krichbaum}, {Park}, {Tazaki},
  {Pu}, {Nakamura}, {Lobanov}, {Hada}, {Akiyama}, {Kim}, {Marti-Vidal},
  {G{\'o}mez}, {Kawashima}, {Yuan}, {Ros}, {Alef}, {Britzen}, {Bremer},
  {Broderick}, {Doi}, {Giovannini}, {Giroletti}, {Ho}, {Honma}, {Hughes},
  {Inoue}, {Jiang}, {Kino}, {Koyama}, {Lindqvist}, {Liu}, {Marscher},
  {Matsushita}, {Nagai}, {Rottmann}, {Savolainen}, {Schuster}, {Shen}, {de
  Vicente}, {Walker}, {Yang}, {Zensus}, {Algaba}, {Allardi}, {Bach},
  {Berthold}, {Bintley}, {Byun}, {Casadio}, {Chang}, {Chang}, {Chang}, {Chen},
  {Chen}, {Chilson}, {Chuter}, {Conway}, {Crew}, {Dempsey}, {Dornbusch},
  {Faber}, {Friberg}, {Garc{\'\i}a}, {Garrido}, {Han}, {Han}, {Hasegawa},
  {Herrero-Illana}, {Huang}, {Huang}, {Impellizzeri}, {Jiang}, {Jinchi},
  {Jung}, {Kallunki}, {Kirves}, {Kimura}, {Koay}, {Koch}, {Kramer}, {Kraus},
  {Kubo}, {Kuo}, {Li}, {Lin}, {Liu}, {Liu}, {Lo}, {Lu}, {MacDonald},
  {Martin-Cocher}, {Messias}, {Meyer-Zhao}, {Minter}, {Nair}, {Nishioka},
  {Norton}, {Nystrom}, {Ogawa}, {Oshiro}, {Patel}, {Pen}, {Pidopryhora},
  {Pradel}, {Raffin}, {Rao}, {Ruiz}, {Sanchez}, {Shaw}, {Snow}, {Sridharan},
  {Srinivasan}, {Tercero}, {Torne}, {Traianou}, {Wagner}, {Walther}, {Wei},
  {Yang}, \& {Yu}}]{Lu2023}
{Lu}, R.-S., {Asada}, K., {Krichbaum}, T.~P., {et~al.} 2023, \nat, 616, 686

\bibitem[{{Lyubarsky}(2009)}]{Lyubarsky2009}
{Lyubarsky}, Y. 2009, \apj, 698, 1570

\bibitem[{{Mart{\'\i}-Vidal} {et~al.}(2021){Mart{\'\i}-Vidal}, {Mus},
  {Janssen}, {de Vicente}, \& {Gonz{\'a}lez}}]{MartiVidal2021}
{Mart{\'\i}-Vidal}, I., {Mus}, A., {Janssen}, M., {de Vicente}, P., \&
  {Gonz{\'a}lez}, J. 2021, \aap, 646, A52

\bibitem[{{McKinney}(2006)}]{McKinney2006}
{McKinney}, J.~C. 2006, \mnras, 368, 1561

\bibitem[{{Mertens} {et~al.}(2016){Mertens}, {Lobanov}, {Walker}, \&
  {Hardee}}]{Mertens2016}
{Mertens}, F., {Lobanov}, A.~P., {Walker}, R.~C., \& {Hardee}, P.~E. 2016,
  \aap, 595, A54

\bibitem[{Moses \& Churavy(2020)}]{Moses2020}
Moses, W. \& Churavy, V. 2020, in Advances in Neural Information Processing
  Systems, ed. H.~Larochelle, M.~Ranzato, R.~Hadsell, M.~F. Balcan, \& H.~Lin,
  Vol.~33 (Curran Associates, Inc.), 12472--12485

\bibitem[{{Moses} {et~al.}(2021){Moses}, {Churavy}, {Paehler},
  {H{\"u}ckelheim}, {Narayanan}, {Schanen}, \& {Doerfert}}]{Moses2021}
{Moses}, W.~S., {Churavy}, V., {Paehler}, L., {et~al.} 2021, in SC21:
  International Conference for High Performance Computing, Networking, Storage
  and Analysis (ACM), 1--16

\bibitem[{{Mueller} {et~al.}(2026){Mueller}, {von Fellenberg}, {Zeng}, {Tiede},
  {Krichbaum}, {Gold}, {Savolainen}, {Kim}, {Peng}, {Toscano}, {Janssen},
  {Georgiev}, {Nair}, {Natarajan}, {Blackburn}, {Akiyama}, {Albentosa-Ruiz},
  {Alberdi}, {Alef}, {Algaba}, {Amanaganti}, {Anantua}, {Antonopoulou},
  {Asada}, {Azulay}, {Baczko}, {Ball}, {Bandyopadhyay}, {Barrett},
  {Baub{\"o}ck}, {Benson}, {Bintley}, {Blundell}, {Bouman}, {Bower}, {Bremer},
  {Brissenden}, {Britzen}, {Broderick}, {Broguiere}, {Bronzwaer}, {Bustamante},
  {Carlos}, {Carlstrom}, {Chael}, {Chan}, {Chang}, {Chang}, {Chatterjee},
  {Chavez}, {Chen}, {Chen}, {Cheng}, {Chichura}, {Cho}, {Conroy}, {Conway},
  {Crawford}, {Crew}, {Cruz-Osorio}, {Cui}, {Curd}, {Dahale}, {Davelaar}, {de
  Kleuver}, {De Laurentis}, {Deane}, {Dexter}, {Dhruv}, {Dihingia}, {Doeleman},
  {Dzib}, {Emami}, {Falcke}, {Farah}, {Fish}, {Fomalont}, {Ford}, {Foschi},
  {Fraga-Encinas}, {Freeman}, {Friberg}, {Fromm}, {Fuentes}, {Galison},
  {Gammie}, {Garcia}, {Gentaz}, {Goddi}, {Gomez-Ruiz}, {Gomez Miller}, {Gomez},
  {Gu}, {Gurwell}, {Hada}, {Haggard}, {Hesper}, {Heumann}, {Ho}, {Ho}, {Honma},
  {Huang}, {Huang}, {Hughes}, {Ikeda}, {Impellizzeri}, {Inoue}, {Issaoun},
  {Iwata}, {James}, {Jannuzi}, {Jeter}, {Jiang}, {Jim{\'e}nez-Rosales},
  {Johnson}, {Jorstad}, {Jones}, {Joshi}, {Jung}, {Kawashima}, {Keating},
  {Kettenis}, {Kim}, {Kim}, {Kim}, {Kino}, {Knollm{\"u}ller}, {Koay},
  {Kocherlakota}, {Kofuji}, {Koch}, {Koyama}, {Kramer}, {Kramer}, {Kramer},
  {Kuo}, {La Bella}, {Lee}, {Lee}, {Levis}, {Li}, {Li}, {Lico}, {Lindahl},
  {Lindqvist}, {Lisakov}, {Liu}, {Liu}, {Liuzzo}, {Lo}, {Lobanov}, {Loinard},
  {Lonsdale}, {Lowitz}, {Lu}, {MacDonald}, {Mao}, {Marchili}, {Markoff},
  {Marrone}, {Marscher}, {Marti-Vidal}, {Matsushita}, {Matthews}, {Medeiros},
  {Menten}, {Messias}, {Mizuno}, {Mizuno}, {Montgomery}, {Moriyama},
  {Moscibrodzka}, {Mulaudzi}, {M{\"u}ller}, {Mus}, {Musoke}, {Myserlis},
  {Nagai}, {Nagar}, {Nakamura}, {Narayanan}, {Nathanail}, {Navarro Fuentes},
  {Neilsen}, {Ni}, {Nilipour}, {Nowak}, {Okino}, {Olivares Sanchez},
  {{\"O}zel}, {Palumbo}, {Paraschos}, {Park}, {Parsons}, {Patel}, {Pen},
  {Pesce}, {Pietu}, {Plavin}, \& {PopStefanija}}]{Mueller2026}
{Mueller}, H., {von Fellenberg}, S.~D., {Zeng}, A.-L., {et~al.} 2026, arXiv
  e-prints, arXiv:2607.16104

\bibitem[{{Nagai} {et~al.}(2014){Nagai}, {Haga}, {Giovannini}, {Doi},
  {Orienti}, {D'Ammando}, {Kino}, {Nakamura}, {Asada}, {Hada}, \&
  {Giroletti}}]{Nagai2014}
{Nagai}, H., {Haga}, T., {Giovannini}, G., {et~al.} 2014, \apj, 785, 53

\bibitem[{{Nakahara} {et~al.}(2018){Nakahara}, {Doi}, {Murata}, {Hada},
  {Nakamura}, \& {Asada}}]{Nakahara2018}
{Nakahara}, S., {Doi}, A., {Murata}, Y., {et~al.} 2018, \apj, 854, 148

\bibitem[{{Nakamura} \& {Asada}(2013)}]{NakamuraAsada2013}
{Nakamura}, M. \& {Asada}, K. 2013, \apj, 775, 118

\bibitem[{{Nakamura} {et~al.}(2018){Nakamura}, {Asada}, {Hada}, {Pu}, {Noble},
  {Tseng}, {Toma}, {Kino}, {Nagai}, {Takahashi}, {Algaba}, {Orienti},
  {Akiyama}, {Doi}, {Giovannini}, {Giroletti}, {Honma}, {Koyama}, {Lico},
  {Niinuma}, \& {Tazaki}}]{Nakamura2018}
{Nakamura}, M., {Asada}, K., {Hada}, K., {et~al.} 2018, \apj, 868, 146

\bibitem[{{Nikonov} {et~al.}(2023){Nikonov}, {Kovalev}, {Kravchenko},
  {Pashchenko}, \& {Lobanov}}]{Nikonov2023}
{Nikonov}, A.~S., {Kovalev}, Y.~Y., {Kravchenko}, E.~V., {Pashchenko}, I.~N.,
  \& {Lobanov}, A.~P. 2023, \mnras, 526, 5949

\bibitem[{{Nokhrina} {et~al.}(2020){Nokhrina}, {Kovalev}, \&
  {Pushkarev}}]{Nokhrina2020}
{Nokhrina}, E.~E., {Kovalev}, Y.~Y., \& {Pushkarev}, A.~B. 2020, \mnras, 498,
  2532

\bibitem[{{Okino} {et~al.}(2022){Okino}, {Akiyama}, {Asada}, {G{\'o}mez},
  {Hada}, {Honma}, {Krichbaum}, {Kino}, {Nagai}, {Bach}, {Blackburn}, {Bouman},
  {Chael}, {Crew}, {Doeleman}, {Fish}, {Goddi}, {Issaoun}, {Johnson},
  {Jorstad}, {Koyama}, {Lonsdale}, {Lu}, {Mart{\'\i}-Vidal}, {Matthews},
  {Mizuno}, {Moriyama}, {Nakamura}, {Pu}, {Ros}, {Savolainen}, {Tazaki},
  {Wagner}, {Wielgus}, \& {Zensus}}]{Okino2022}
{Okino}, H., {Akiyama}, K., {Asada}, K., {et~al.} 2022, \apj, 940, 65

\bibitem[{{Paraschos} {et~al.}(2024){Paraschos}, {Kim}, {Wielgus}, {R{\"o}der},
  {Krichbaum}, {Ros}, {Agudo}, {Myserlis}, {Moscibrodzka}, {Traianou},
  {Zensus}, {Blackburn}, {Chan}, {Issaoun}, {Janssen}, {Johnson}, {Fish},
  {Akiyama}, {Alberdi}, {Alef}, {Algaba}, {Anantua}, {Asada}, {Azulay}, {Bach},
  {Baczko}, {Ball}, {Balokovi{\'c}}, {Barrett}, {Baub{\"o}ck}, {Benson},
  {Bintley}, {Blundell}, {Bouman}, {Bower}, {Boyce}, {Bremer}, {Brinkerink},
  {Brissenden}, {Britzen}, {Broderick}, {Broguiere}, {Bronzwaer}, {Bustamante},
  {Byun}, {Carlstrom}, {Ceccobello}, {Chael}, {Chang}, {Chatterjee},
  {Chatterjee}, {Chen}, {Chen}, {Cheng}, {Cho}, {Christian}, {Conroy},
  {Conway}, {Cordes}, {Crawford}, {Crew}, {Cruz-Osorio}, {Cui}, {Dahale},
  {Davelaar}, {De Laurentis}, {Deane}, {Dempsey}, {Desvignes}, {Dexter},
  {Dhruv}, {Doeleman}, {Dougal}, {Dzib}, {Eatough}, {Emami}, {Falcke}, {Farah},
  {Fomalont}, {Ford}, {Foschi}, {Fraga-Encinas}, {Freeman}, {Friberg}, {Fromm},
  {Fuentes}, {Galison}, {Gammie}, {Garc{\'\i}a}, {Gentaz}, {Georgiev}, {Goddi},
  {Gold}, {G{\'o}mez-Ruiz}, {G{\'o}mez}, {Gu}, {Gurwell}, {Hada}, {Haggard},
  {Haworth}, {Hecht}, {Hesper}, {Heumann}, {Ho}, {Ho}, {Honma}, {Huang},
  {Huang}, {Hughes}, {Ikeda}, {Impellizzeri}, {Inoue}, {James}, {Jannuzi},
  {Jeter}, {Jaing}, {Jim{\'e}nez-Rosales}, {Jorstad}, {Joshi}, {Jung},
  {Karami}, {Karuppusamy}, {Kawashima}, {Keating}, {Kettenis}, {Kim}, {Kim},
  {Kim}, {Kino}, {Koay}, {Kocherlakota}, {Kofuji}, {Koch}, {Koyama}, {Kramer},
  {Kramer}, {Kramer}, {Kuo}, {La Bella}, {Lauer}, {Lee}, {Lee}, {Leung},
  {Levis}, {Li}, {Lico}, {Lindahl}, {Lindqvist}, {Lisakov}, {Liu}, {Liu},
  {Liuzzo}, {Lo}, {Lobanov}, {Loinard}, {Lonsdale}, {Lowitz}, {Lu},
  {MacDonald}, {Mao}, {Marchili}, {Markoff}, {Marrone}, {Marscher},
  {Mart{\'\i}-Vidal}, {Matsushita}, {Matthews}, {Medeiros}, {Menten},
  {Michalik}, {Mizuno}, {Mizuno}, {Moran}, {Moriyama}, {Mulaudzi},
  {M{\"u}ller}, {M{\"u}ller}, {Mus}, {Musoke}, {Nadolski}, {Nagai}, {Nagar},
  {Nakamura}, {Narayanan}, {Natarajan}, {Nathanail}, {Navarro Fuentes},
  {Neilsen}, {Neri}, {Ni}, {Noutsos}, {Nowak}, {Oh}, {Okino}, {Olivares},
  {Ortiz-Le{\'o}n}, {Oyama}, {{\"O}zel}, {Palumbo}, \& {Park}}]{Paraschos2024}
{Paraschos}, G.~F., {Kim}, J.-Y., {Wielgus}, M., {et~al.} 2024, \aap, 682, L3

\bibitem[{{Park} {et~al.}(2023{\natexlab{a}}){Park}, {Asada}, \&
  {Byun}}]{Park2023b}
{Park}, J., {Asada}, K., \& {Byun}, D.-Y. 2023{\natexlab{a}}, \apj, 958, 27

\bibitem[{{Park} {et~al.}(2023{\natexlab{b}}){Park}, {Asada}, \&
  {Byun}}]{Park2023a}
{Park}, J., {Asada}, K., \& {Byun}, D.-Y. 2023{\natexlab{b}}, \apj, 958, 28

\bibitem[{{Park} {et~al.}(2021{\natexlab{a}}){Park}, {Asada}, {Nakamura},
  {Kino}, {Pu}, {Hada}, {Kravchenko}, \& {Giroletti}}]{Park2021c}
{Park}, J., {Asada}, K., {Nakamura}, M., {et~al.} 2021{\natexlab{a}}, \apj,
  922, 180

\bibitem[{{Park} {et~al.}(2021{\natexlab{b}}){Park}, {Byun}, {Asada}, \&
  {Yun}}]{Park2021}
{Park}, J., {Byun}, D.-Y., {Asada}, K., \& {Yun}, Y. 2021{\natexlab{b}}, \apj,
  906, 85

\bibitem[{{Park} {et~al.}(2019{\natexlab{a}}){Park}, {Hada}, {Kino},
  {Nakamura}, {Hodgson}, {Ro}, {Cui}, {Asada}, {Algaba}, {Sawada-Satoh}, {Lee},
  {Cho}, {Shen}, {Jiang}, {Trippe}, {Niinuma}, {Sohn}, {Jung}, {Zhao},
  {Wajima}, {Tazaki}, {Honma}, {An}, {Akiyama}, {Byun}, {Kim}, {Zhang},
  {Cheng}, {Kobayashi}, {Shibata}, {Lee}, {Roh}, {Oh}, {Yeom}, {Jung}, {Oh},
  {Kim}, {Hwang}, \& {Hagiwara}}]{Park2019a}
{Park}, J., {Hada}, K., {Kino}, M., {et~al.} 2019{\natexlab{a}}, \apj, 887, 147

\bibitem[{{Park} {et~al.}(2019{\natexlab{b}}){Park}, {Hada}, {Kino},
  {Nakamura}, {Ro}, \& {Trippe}}]{Park2019b}
{Park}, J., {Hada}, K., {Kino}, M., {et~al.} 2019{\natexlab{b}}, \apj, 871, 257

\bibitem[{{Park} {et~al.}(2021{\natexlab{c}}){Park}, {Hada}, {Nakamura},
  {Asada}, {Zhao}, \& {Kino}}]{Park2021b}
{Park}, J., {Hada}, K., {Nakamura}, M., {et~al.} 2021{\natexlab{c}}, \apj, 909,
  76

\bibitem[{{Park} {et~al.}(2024{\natexlab{a}}){Park}, {Kino}, {Nagai},
  {Nakamura}, {Asada}, {Kam}, \& {Hodgson}}]{Park2024b}
{Park}, J., {Kino}, M., {Nagai}, H., {et~al.} 2024{\natexlab{a}}, \aap, 685,
  A115

\bibitem[{{Park} {et~al.}(2026){Park}, {Takahashi}, {Toma}, {Hada}, {Nakamura},
  {Pu}, {Asada}, {Ho}, {Kino}, {Kawashima}, {Kam}, {Yi}, \& {Cho}}]{Park2026}
{Park}, J., {Takahashi}, K., {Toma}, K., {et~al.} 2026, \apjl, 996, L22

\bibitem[{{Park} {et~al.}(2024{\natexlab{b}}){Park}, {Zhao}, {Nakamura},
  {Mizuno}, {Pu}, {Asada}, {Takahashi}, {Toma}, {Kino}, {Cho}, {Hada},
  {Edwards}, {Ro}, {Kam}, {Yi}, {Lee}, {Koyama}, {Byun}, {Phillips},
  {Reynolds}, {Hodgson}, \& {Lee}}]{Park2024a}
{Park}, J., {Zhao}, G.-Y., {Nakamura}, M., {et~al.} 2024{\natexlab{b}}, \apjl,
  973, L45

\bibitem[{{Pesce}(2021)}]{Pesce2021}
{Pesce}, D.~W. 2021, \aj, 161, 178

\bibitem[{{Pushkarev} {et~al.}(2017){Pushkarev}, {Kovalev}, {Lister}, \&
  {Savolainen}}]{Pushkarev2017}
{Pushkarev}, A.~B., {Kovalev}, Y.~Y., {Lister}, M.~L., \& {Savolainen}, T.
  2017, \mnras, 468, 4992

\bibitem[{{Ricci} {et~al.}(2022){Ricci}, {Boccardi}, {Nokhrina}, {Perucho},
  {MacDonald}, {Mattia}, {Grandi}, {Madika}, {Krichbaum}, \&
  {Zensus}}]{Ricci2022}
{Ricci}, L., {Boccardi}, B., {Nokhrina}, E., {et~al.} 2022, \aap, 664, A166

\bibitem[{{Savolainen} {et~al.}(2023){Savolainen}, {Giovannini}, {Kovalev},
  {Perucho}, {Anderson}, {Bruni}, {Edwards}, {Fuentes}, {Giroletti},
  {G{\'o}mez}, {Hada}, {Lee}, {Lisakov}, {Lobanov}, {L{\'o}pez-Miralles},
  {Orienti}, {Petrov}, {Plavin}, {Sohn}, {Sokolovsky}, {Voitsik}, \&
  {Zensus}}]{Savolainen2023}
{Savolainen}, T., {Giovannini}, G., {Kovalev}, Y.~Y., {et~al.} 2023, \aap, 676,
  A114

\bibitem[{{Shepherd}(1997)}]{Shepherd1997}
{Shepherd}, M.~C. 1997, in Astronomical Society of the Pacific Conference
  Series, Vol. 125, Astronomical Data Analysis Software and Systems VI, ed.
  G.~{Hunt} \& H.~{Payne}, 77

\bibitem[{{Smirnov}(2011{\natexlab{a}})}]{Smirnov2011}
{Smirnov}, O.~M. 2011{\natexlab{a}}, \aap, 527, A106

\bibitem[{{Smirnov}(2011{\natexlab{b}})}]{Smirnov2011b}
{Smirnov}, O.~M. 2011{\natexlab{b}}, \aap, 527, A107

\bibitem[{{Tchekhovskoy} {et~al.}(2008){Tchekhovskoy}, {McKinney}, \&
  {Narayan}}]{Tchekhovskoy2008}
{Tchekhovskoy}, A., {McKinney}, J.~C., \& {Narayan}, R. 2008, \mnras, 388, 551

\bibitem[{{Tiede}(2022)}]{Tiede2022}
{Tiede}, P. 2022, The Journal of Open Source Software, 7, 4457

\bibitem[{{Tiede} {et~al.}(2026){Tiede}, {Moses}, {Churavy}, {Johnson},
  {Pesce}, {Blackburn}, \& {Galison}}]{Tiede2026}
{Tiede}, P., {Moses}, W., {Churavy}, V., {et~al.} 2026, \apj, 997, 262

\bibitem[{{Tseng} {et~al.}(2016){Tseng}, {Asada}, {Nakamura}, {Pu}, {Algaba},
  \& {Lo}}]{Tseng2016}
{Tseng}, C.-Y., {Asada}, K., {Nakamura}, M., {et~al.} 2016, \apj, 833, 288

\bibitem[{{Virtanen} {et~al.}(2020){Virtanen}, {Gommers}, {Oliphant},
  {Haberland}, {Reddy}, {Cournapeau}, {Burovski}, {Peterson}, {Weckesser},
  {Bright}, {van der Walt}, {Brett}, {Wilson}, {Millman}, {Mayorov}, {Nelson},
  {Jones}, {Kern}, {Larson}, {Carey}, {Polat}, {Feng}, {Moore}, {VanderPlas},
  {Laxalde}, {Perktold}, {Cimrman}, {Henriksen}, {Quintero}, {Harris},
  {Archibald}, {Ribeiro}, {Pedregosa}, {van Mulbregt}, \& {SciPy 1. 0
  Contributors}}]{Virtanen2020}
{Virtanen}, P., {Gommers}, R., {Oliphant}, T.~E., {et~al.} 2020, Nature
  Methods, 17, 261

\bibitem[{{Vlahakis} \& {K{\"o}nigl}(2004)}]{VlahakisKonigl2004}
{Vlahakis}, N. \& {K{\"o}nigl}, A. 2004, \apj, 605, 656

\bibitem[{{Walker} {et~al.}(2018){Walker}, {Hardee}, {Davies}, {Ly}, \&
  {Junor}}]{Walker2018}
{Walker}, R.~C., {Hardee}, P.~E., {Davies}, F.~B., {Ly}, C., \& {Junor}, W.
  2018, \apj, 855, 128

\bibitem[{{Wardle} \& {Homan}(2003)}]{WardleHoman2003}
{Wardle}, J. F.~C. \& {Homan}, D.~C. 2003, \apss, 288, 143

\bibitem[{{Wardle} {et~al.}(1998){Wardle}, {Homan}, {Ojha}, \&
  {Roberts}}]{Wardle1998}
{Wardle}, J.~F.~C., {Homan}, D.~C., {Ojha}, R., \& {Roberts}, D.~H. 1998, \nat,
  395, 457

\bibitem[{{Yan} {et~al.}(2025){Yan}, {Cui}, {Hada}, {Frey}, {Lu}, {Chen}, {Xu},
  {Fariyanto}, \& {Ho}}]{Yan2025}
{Yan}, X., {Cui}, L., {Hada}, K., {et~al.} 2025, \apj, 991, 75

\bibitem[{{Yan} {et~al.}(2023){Yan}, {Lu}, {Jiang}, {Krichbaum}, \&
  {Shen}}]{Yan2023}
{Yan}, X., {Lu}, R.-S., {Jiang}, W., {Krichbaum}, T.~P., \& {Shen}, Z.-Q. 2023,
  \apj, 957, 32

\bibitem[{{Yi} {et~al.}(2024){Yi}, {Park}, {Nakamura}, {Hada}, \&
  {Trippe}}]{Yi2024}
{Yi}, K., {Park}, J., {Nakamura}, M., {Hada}, K., \& {Trippe}, S. 2024, \aap,
  688, A94

\bibitem[{{Zavala} \& {Taylor}(2002)}]{ZT2002}
{Zavala}, R.~T. \& {Taylor}, G.~B. 2002, \apjl, 566, L9

\end{thebibliography}

\end{document}